\documentclass[journal]{IEEEtran}
\usepackage{fancyhdr}
\usepackage{cite}
\usepackage{amsmath,amssymb,amsfonts}
\usepackage{algorithmic}
\usepackage{graphicx}
\usepackage{textcomp}
\usepackage{xcolor}
\usepackage{braket}
\usepackage{cases}
\usepackage{algorithm}
\usepackage{subfigure}
\usepackage{multirow}

\usepackage{cancel}
\usepackage{makecell}
\ifCLASSINFOpdf
\else
\fi
\DeclareMathOperator*{\argmin}{arg\,min}

\newcommand\mc[1]{\mathcal{#1}}

\newcommand{\norm}[1]{\lVert#1\rVert}

\begin{document}
%

\title{Global 4D Ionospheric STEC Prediction based on DeepONet for GNSS Rays}

%\title{ Predicting 4D Global Ionospheric STEC based on DeepONet for arbitrary GNSS Rays}

%\author{Zenghui~Shi,~\IEEEmembership{Student Member,~IEEE,}

\author{Dijia~Cai,~\IEEEmembership{}
      Zenghui~Shi,
      Haiyang~Fu,~\IEEEmembership{Member,~IEEE,} 
      Huan~Liu,
      Hongyi~Qian, 
      Yun~Sui,
      Feng~Xu,
      and Ya-Qiu~Jin,~\IEEEmembership{Life~Fellow,~IEEE}\\

      % <-this % stops a space
% \thanks{M. Shell is with the Department
% of Electrical and Computer Engineering, Georgia Institute of Technology, Atlanta,
% GA, 30332 USA 

% e-mail: (see http://www.michaelshell.org/contact.html).}% <-this % stops a space
\thanks{Manuscript received: 
% April 19, 2005
%; revised
% September 17, 2014
Corresponding authors: Haiyang~Fu (e-mail: haiyang\_fu@fudan.edu.cn)}
\thanks{Dijia Cai, Zenghui~Shi, Haiyang Fu, Hongyi Qian, Yun~Sui, Yaqiu Jin are with the Key Laboratory for Information Science of Electromagnetic Waves (MoE), school of information science and engineering, Fudan University, Shanghai 200433, China. Huan Liu is in the school of mathematics, Shanghai University of Finance and Economics, Shanghai 200433, China. This work is supported by National Key Research and Development Program of China (2021YFA0717300). This work is also supported by the National Science Foundation (42074189; 62231010).}% <-this % stops a space
}


\markboth{IEEE TRANSACTIONS ON GEOSCIENCE AND REMOTE SENSING}%,~Vol.~13, No.~9, September~2014}%
{Shell \MakeLowercase{\textit{et al.}}: Bare Demo of IEEEtran.cls for Journals}



\maketitle

\begin{abstract}
The ionosphere is a vitally dynamic charged particle region in the Earth's upper atmosphere, playing a crucial role in applications such as radio communication and satellite navigation. The Slant Total Electron Contents (STEC) is an important parameter for characterizing wave propagation, representing the integrated electron density along the ray of radio signals passing through the ionosphere. The accurate prediction of STEC is essential for mitigating the ionospheric impact particularly on Global Navigation Satellite Systems (GNSS). In this work, we propose a high-precision STEC prediction model named DeepONet-STEC, which learns nonlinear operators to predict the 4D temporal-spatial integrated parameter for specified ground station - satellite ray path globally. As a demonstration, we validate the performance of the model based on GNSS observation data for global and US-CORS regimes under ionospheric quiet and storm conditions. The DeepONet-STEC model results show that the three-day 72 hour prediction in quiet periods could achieve high accuracy using observation data by the Precise Point Positioning (PPP) with temporal resolution $\rm{30\,s}$. Under active solar magnetic storm periods, the DeepONet-STEC also demonstrated its robustness and superiority than traditional deep learning methods. This work presents a neural operator regression architecture for predicting the 4D temporal-spatial ionospheric parameter for satellite navigation system performance, which may be further extended for various space applications and beyond.

%We validate the performance of the model using observed STEC data retreived by the Undifferenced and Uncombined Precise Point Positioning (UCPPP) method and simulated STEC data generated by the Nequick model over 30 days. 

%The Relative Mean Square Error (RMSE) of predicted STEC for test stations in globe and the United States region are 0.4192/0.1745 in simulation data and 1.4843/1.4617 in observation data, respectively, with coefficient of determination $R^2$ (R-square) values above 0.85 for the fitted curves. The results demonstrate that our method achieves favorable performance in STEC prediction. This research presents a novel, efficient, and accurate approach for predicting STEC in the ionosphere, contributing to the improvement of satellite navigation system accuracy and performance. 
\end{abstract}


\IEEEpeerreviewmaketitle

\section{Introduction}
\IEEEPARstart
{T}{he}
ionosphere, spanning from 60 km to 1000 km above the Earth, constitutes the ionized particles of the upper atmosphere\cite{kelley2009earth}. Comprised of plasma electrons, ions and neutral components, the ionosphere exhibits complex and dynamic nature influenced by solar radiation, the Earth's magnetic field, and the atmospheric activity, which holds great significance particularly in the realms of radio communications, satellite navigation and L-band synthetic aperture radar (SAR)\cite{kintner2005ionosphere, hu2015background}. When electromagnetic (EM) signals such as Global Navigation Satellite Systems (GNSS) propagate in the ionosphere\cite{budden2009radio}, the EM wave encounters diverse effects such as dispersion, refraction, and group delay, producing a certain degree of interference which leads to amplified positioning errors\cite{kintner2005ionosphere, Hernandez,Liu202010.1007/s10291-019-0940-1} from several centimeters to hundred meters in severe ionospheric conditions \cite{shiImprovedApproachModel2012}. As global space ionospheric weather and GNSS service rapidly growing, the development of highly precise and low-latency global ionospheric sensing become imperative in order to obtain high temporal-spatial (4D) resolution accurate estimation and fast prediction of ionospheric properties for various applications.

Ionospheric sensing based on GNSS data have been developed as one of the dominant source due to global constellation, massive stations and high accuracy positioning, navigation and timing (PNT) service. The global ionospheric sensing based on GNSS data has been evolving from traditional two dimensional (2D) modeling of the Vertical Total Electron Contents(VTEC) using spherical Harmonics function fitting\cite{steinInterpolationSpatialData1999,mannucciGlobalMapping1998}, and Neural Network (NN) interpolation\cite{tecdeep} to three dimensional (3D) ionospheric tomography \cite{SuiTomo2022} . The altitude and temporal resolution of the ionospheric tomography cannot be neglected in order to capture high temporal-spatial resolution of ionospheric parameter dynamics. To satisfy high GNSS application, the Slant Total Electron Contents (STEC) estimation has been widely used in the GNSS society such as in Wide Area Augmentation System (WAAS) using the Kriging model \cite{sparks2011} and multiple kernal model \cite{shiMethodDSTECInterpolation2022} to express the spatial ionospheric correlation under ionospheric scintillation. However, the STEC interpolation \cite{shiMethodDSTECInterpolation2022} actually could not predict the future temporal state. The present ionospheric parameter estimation is actually lacking of achieving  estimation and prediction with high temporal-spatial resolution. 

Indeed, it has been pursued to predict the temporally varying ionosphere from short several hours to long term as another topic. The predicting method includes both traditional statistical method and machine learning methods. In recently years, due to its applicability of machine learning to nonlinearity, several works have been done for ionospheric VTEC prediction, including  recurrent neural network (RNN) \cite{YuanDeepLearningRecurrentNeuralNetwork2018}, long short-term memory (LSTM) neural networks\cite{HochreiterLSTM1997,SunFORECASTINGVTECLSTM2017}, LSTM with convolutional neural networks\cite{RuwaliLSTM-CNNTEC2021}, and encoder-decoder ConvLSTMcite \cite{LiED-ConvLSTM2023} ande etc. There are few methods to predict STEC directly.  Recently Zhu et al.\cite{xgboostSTECZhu2023} proposed an XGBoost-based method to predict the single-station STEC for high accuracy Precise Point Positionin(PPP)  service. Although these methods can predict VTEC/STEC, the spatial-temporal prediction cannot be simultaneously achieved for any satellite-station ray path with high accuracy and resolution.  

The existing ionospheric estimation and prediction can not satisfy the high 4D temporal-spatial resolution of global ionospheric parameter, which is possible to improve. On one hand, existing VTEC map is actually a low accurate method. The mapping function is usually used for converting STEC to VTEC. But the mapping function cannot characterize the STEC variation in different azimuths between the user and satellites. The Ionospheric VTEC models are established for all GNSS satellites as a whole and can only provide an averaged solution. But this is not the optimal solution for each individual satellite\cite{Li2021}. The conversion process from VTEC to STEC will also introduce errors by the mapping function. The tomographic modeling from electron density to calculate STEC suffers from meshing errors and expensive computational cost. These factors hinder significant improvements in the accuracy of estimating STEC. On the other hand, prevailing STEC estimation approaches primarily focus on spatial interpolation techniques, with limited emphasis on temporal prediction. The drawback of current exisiting methodologies lacks a unified framework for achieving 4D temporal-spatial STEC estimation, which encompasses constructing models based on past STEC observations to estimate STEC for the present and future, and along any ray between satellites and stations within the domain.

In recent years, with the development of deep learning, neural operators have gradually played an important for predicting the complex dynamics. The operator regression methods approximate mappings between infinite function spaces and considered as discretization invariant.  Some of the operator regression includes the Fourier neural operator (FNO)\cite{FNO:journals/corr/abs-2010-08895}, the graph kernal network (GKN) \cite{li2020neural} and the deep neural operator(DeepONet) \cite{luLearningNonlinearOperators2021}. The neural operator learns representations between functions and work on any discretization of inputs to a meshless resolution \cite{ChenBo2024},  while the performance of standard neural network such as CNN can degrade when data resolution during deployment changes from model training\cite{kovachki2023neural}. Recently, the FNO operator based prediction model such FourCastNet \cite{pathak2022fourcastnet} has achieved substantial progress for accurate medium-range global weather forecasting, which triggers several atmospheric prediction model based on machine learning \cite{bi2023accurate} with capability superior than traditional methods. In addition, prediction methods for VTEC map using deep learning methods have also been attempted \cite{KSivakrishna2022revise,QZhang2023revise}, including detecting pre-earthquake anomalies\cite{PXiong2022revise}. This motivates our attempt to further achieve the 4D ionospheric temporally-spatial prediction based on advanced machine learning methods.     
 
In this paper, we developed a 4D temporal-spatial prediction framework DeepONet-STEC for predicting ionospheric parameter for station-satellite ray path based on deep operator regression. The advantage of this model is capable of predicting the future 10 days STEC for any specified ground station-satellite rays by capturing representations between functions from data. The main contributions of this work can be summarized as follows: 1) We proposed a novel 4D ionospheric parameter prediction framework based on neural operator regression. 2) The proposed framework provides simultaneously high accuracy spatial estimation and temporal prediction of the STEC values for global GNSS data. 3) The innovative Kernel method is adopted to construct the input function from the real observation data for the DeepONet-STEC model. The structure of this paper is organized as following. After introduction in Section I, the description of the proposed 4D DeepONet-STEC model will be explained in detail in Section II. In Section III, we will present the process of dataset generation, model training, prediction results based on simulated and observation data. Section IV will draw conclusions with discussion. 

\section{Prediction Model}

In this section, we present the modeling approach for the STEC 4D prediction problem mentioned above, emphasizing two key methods used in the modeling process: the Kernel Methods (KM) and the principles of DeepONet. The construction process of the 4D DeepONet-STEC model is then detailed, along with the techniques employed for data processing.

\subsection{Problem statement}
%% Figure 1 %%%%%%%%%%%%%%%%%%%
\begin{figure*}[!ht]
    \centering
    \includegraphics[width=1\linewidth]{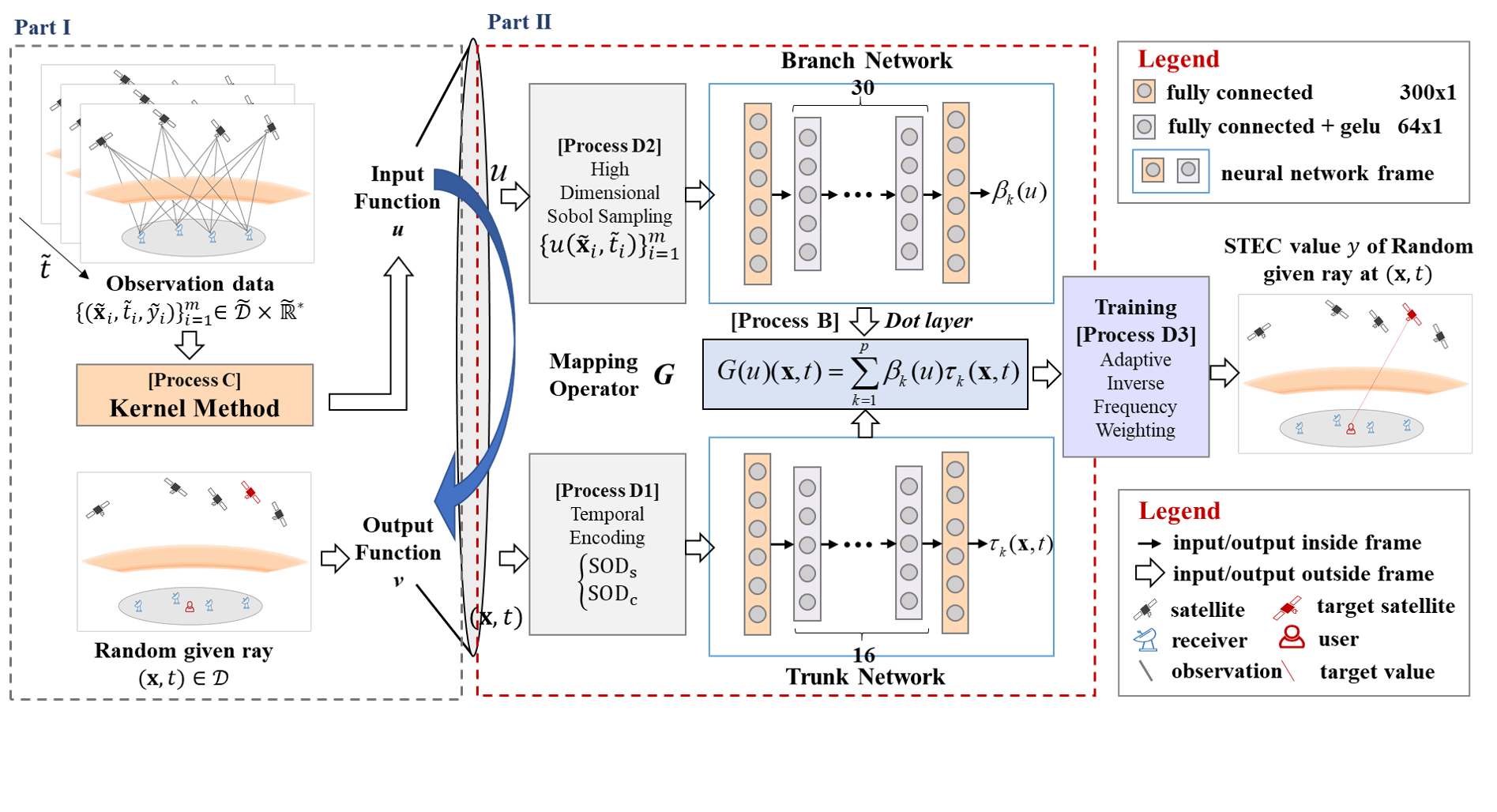}
    \vspace{-1cm}
    \caption{Illustration of the 4D DeepONet-STEC architecture. Using $\{(\tilde{\textbf{x}}_i, \tilde{t}_i, \tilde{y}_i)\}_{i=1}^M$ to train, the model can predict given ray $(\textbf{x}, t) \in \mc{D}$ to get the output STEC value $y$. The 4D DeepONet-STEC comprises two main components: the Branch network and the Trunk network. Both neural networks adopt a fully connected structure. The Branch network takes the output value of the $u$ function at fixed point location, representing the mapped STEC value, as its input. The input for the Trunk network includes the coordinates of the corresponding rays in the observed data, as well as sine and cosine encoded time. An activation function ReLu is applied to the output of each hidden layer. The branch network has the input layer with 300 neurons, 30 hidden layer with 64 neurons, and the output layer with 300 neurons.}
    \label{fig:DeepONet-STEC}
\end{figure*}

In ionospheric applications, the observations are often limited due to the limitation of observation such as GNSS. Typically, only partial observable STEC value $y$ along the $i$-th ray can be measured from a restricted number of satellite $s$ and receiver $r$ reference pairs $\textbf{x}$ at a specific time $t$, where $\textbf{x}$ represents the coordinates of the satellite and receiver. Consequently, there arises a need to predict the STEC value $y \in \mathbb{R}^*$ at any temporal-spatial point $(\textbf{x}, t) \in \mc{D}$, therefore we develop the 4D DeepONet-STEC model based on the operator neural network. 

Let all observed rays $\{(\tilde{\textbf{x}}_i, \tilde{t}_i, \tilde{y}_i)\}_{i=1}^M$, be the data set, with each observed sample $(\tilde{\textbf{x}}_i, \tilde{t}_i, \tilde{y}_i) \in \tilde{\mc{D}} \times \tilde{\mathbb{R}}^*$ and the subscript $i$ denotes the {$i$-th} ray. To evaluate the generalization performance of the 4D DeepONet-STEC model, the training set and test set are created by collecting partial data from the data set, where the training set is denoted by $\mc{T} := \{(\tilde{\textbf{x}}_i^{(0)}, \tilde{t}_i^{(0)}, \tilde{y}_i^{(0)})\}_{i=1}^{N}$.

In contrast with conventional neural networks to approximate functions, DeepONet approximates linear and nonlinear operators. The model comprises two deep neural networks: one network that encodes the discrete input function space (i.e., branch net) and one that encodes the output function space (i.e., trunk net).

Furthermore, we create two subsets from the training set $\mc{T}$, which are denoted by $\mc{U}$ and $\mc{V}$, respectively. The first subset $\mc{U} := \{(\tilde{\textbf{x}}_i^{(1)}, \tilde{t}_i^{(1)}, \tilde{y}_i^{(1)})\}_{i=1}^{m}$ consisting of $m$ observed samples is used to construct the input function $u$ of the  4D DeepONet-STEC model, and the second subset $\mc{V} := \{(\tilde{\textbf{x}}_i^{(2)}, \tilde{t}_i^{(2)}, \tilde{y}_i^{(2)})\}_{i=1}^{h}$ consisting of $h$ observed samples will be employed to train the model. Obviously, $(\tilde{\textbf{x}}_i^{(j)}, \tilde{t}_i^{(j)}) \in \tilde{\mc{D}}$ for $j = 0, 1, 2$ with $\tilde{\mc{D}}$ being a subset of $\mc{D}$.

Therefore, the problem can be reformulated as follows: (I) Constructing the input function $u:
\tilde{\mc{D}}\rightarrow \tilde{\mathbb{R}}^*$ satisfying $u(\tilde{\textbf{x}}_i^{(1)}, \tilde{t}_i^{(1)}) = \tilde{y}_i^{(1)}$ for $i=1, \ldots, m$, i.e., the STEC value at a given time on a given path can be obtained through the function $u$; (II) Define the mapping operator $G: u\rightarrow v$ using DeepONet, and learn the optimal hyper-parameters of the DeepONet by minimizing the loss $\sum_{i=1}^{h}|G(u)(\tilde{\textbf{x}}_i^{(2)}, \tilde{t}_i^{(2)}) - \tilde{y}_i^{(2)}|^2$; (III) Once the optimal hyper-parameters of the DeepONet are obtained, then the STEC value $y$ at any path and any time $(\textbf{x}, t) \in \mc{D}$ can be predicted through the network output $G(u)(\textbf{x}, t)$.

\subsection{Constructing mapping operator $G$ using DeepONet}
We use DeepONet to construct the operator $G$, which is a deep neural network-based model that can learn both linear and nonlinear operators. The DeepONet is proposed based on the universal approximation theorem of operators, which states that a neural network with a single hidden layer can accurately approximate any nonlinear continuous operator. More specifically, given two functions $u$ and $v$ defined on different spaces as follows
\begin{equation}
    u: \tilde{\mc{D}}\ni (\tilde{\textbf{x}},\tilde{t}) \rightarrow u(\tilde{\textbf{x}},\tilde{t}) \in  \tilde{\mathbb{R}}^*
\end{equation}
 and 
\begin{equation}
    v: \mc{D}\ni (\textbf{x},t) \rightarrow v(\textbf{x},t) \in \mathbb{R}^*
\end{equation}
DeepONet can train the mapping operator $G$:
 
\begin{equation}
G: u\rightarrow v
\end{equation}

In terms of concrete implementation, DeepONet consists of two sub-networks: one for encoding the input function at a fixed number of sensors (branch network), and another for encoding the locations for the output functions (trunk network)\cite{luLearningNonlinearOperators2021}.

The branch network takes the function evaluations $[u(\tilde{\textbf{x}}_1^{(1)}, \tilde{t}_1^{(1)}), u(\tilde{\textbf{x}}_2^{(1)}, \tilde{t}_2^{(1)}), \cdots, u(\tilde{\textbf{x}}_m^{(1)}, \tilde{t}_m^{(1)})]$ as the input, and the output of the branch network is $[\beta_1(u), \beta_2(u), \cdots , \beta_p(u)]$, where $p$ is the number of neurons. The trunk network takes all the rays from $\{(\tilde{\textbf{x}}_1^{(2)}, \tilde{t}_1^{(2)}), (\tilde{\textbf{x}}_2^{(2)}, \tilde{t}_2^{(2)}), \cdots, (\tilde{\textbf{x}}_h^{(2)}, \tilde{t}_h^{(2)})\}$ as the input and the outputs are  $\{[\tau_1(\tilde{\textbf{x}}_i^{(2)}, \tilde{t}_i^{(2)}), \tau_2(\tilde{\textbf{x}}_i^{(2)}, \tilde{t}_i^{(2)}), \cdots, \tau_p(\tilde{\textbf{x}}_i^{(2)}, \tilde{t}_i^{(2)})]\}^h_{i=1}$. Then, by taking the inner product of trunk and branch outputs, the output of DeepONet is:

\begin{equation}
    G(u)(\tilde{\textbf{x}}_i^{(2)}, \tilde{t}_i^{(2)}) = \sum_{k=1}^{p}\beta_k(u)\tau_k(\tilde{\textbf{x}}_i^{(2)}, \tilde{t}_i^{(2)}), ~~i = 1,\ldots, h
    \label{eq:deeponet-out}
\end{equation}

We use the following loss function to train the DeepONet, which can be minimized by many well-developed gradient-based optimization methods:
\begin{equation}
        L = \sum_{i=1}^{{h}} |{G(u)(\tilde{\textbf{x}}_i^{(2)}, \tilde{t}_i^{(2)}) - \tilde{y}_i^{(2)}}|^2 
        \label{eq:loss}
\end{equation}

For any  pair $(\textbf{x},t) \in \mc{D}$ as the input, the outputs of the trunk network is $[\tau_1(\textbf{x},t), \tau_2(\textbf{x},t), \cdots, \tau_p(\textbf{x},t)]$. Therefore, we can obtain the predicted STEC value $y =G(u)(\textbf{x},t) = \sum_{k=1}^{p}\beta_k(u)\tau_k(\textbf{x},t)$ for any temporal-spatial pair $(\textbf{x},t)$ provided that the optimal hyper-parameters of the DeepONet are obtained after training process.  

\subsection{Constructing input function $u$ using Kernel Method}
% \hl{original interpolation methods GRF vs Representer Theorem}
Kernel methods are mathematical functions that operate on vectors in the original space and yield the inner product of vectors in the feature space \cite{hofmannKernelMethodsMachine2008,1950Theory}. In this paper, we use the kernel method to construct the input function $u$. Distinct from Gaussian Random Field (GRF) and Chebyshev polynomials in DeepONet\cite{luLearningNonlinearOperators2021} due to lacking the observation data, our work uses the Representer Theorem\cite{scholkopfGeneralizedRepresenterTheorem2001} to construct the input function $u$, which can achieve preliminary fitting of the observation data.

Based on the kernel method, there exists a feature map $\Phi(\cdot):\mathcal{Z}\rightarrow\mathcal{Z}^\prime$, where $\mathcal{Z}$ is a nonempty set that the input vectors are taken from. The feature map transforms the vector in original space $\mathcal{Z}$ to the feature space $\mathcal{Z}^\prime$. Suppose $\boldsymbol{x}_i$ and $\boldsymbol{x}_j$ are two random vectors in $\mathcal{Z}$, then the kernel function corresponding to the feature map $\Phi$ is defined as a function that maps $\mathcal{Z} \times \mathcal{Z} \rightarrow \mathbb{R}$ satisfying {\cite{hofmannKernelMethodsMachine2008}}:

\begin{equation}
\kappa(\boldsymbol{x}_i,\boldsymbol{x}_j) = \langle \Phi(\boldsymbol{x}_i),\Phi(\boldsymbol{x}_j) \rangle 
\label{eq:kappa}
\end{equation}

There are several widely used kernel functions, such as the Gaussian kernel function (Radial Basis Function, RBF) \cite{10.5555/48424.48433}
\begin{equation}
    \kappa_{\text{RBF}}({\boldsymbol{x}}_{i},{\boldsymbol{x}}_{j})=\exp(-\frac{\norm{{\boldsymbol{x}}_{i}-{\boldsymbol{x}}_{j}}_2^2}{2\sigma^2})
    \label{eq:rbf}
\end{equation}
and the Mat$\acute{\text{e}}$rn kernel function\cite{steinInterpolationSpatialData1999}
\begin{equation}
    % \begin{aligned}
    \kappa_{\text{Mat$\acute{\text{e}}$rn}}({\boldsymbol{x}}_{i},{\boldsymbol{x}}_{j}) = (1+\frac{\sqrt{3}\norm{{\boldsymbol{x}}_{i}-{\boldsymbol{x}}_{j}}_1}{\sigma}) \cdot \exp(-\frac{\sqrt{3}\norm{{\boldsymbol{x}}_{i}-{\boldsymbol{x}}_{j}}_1}{\sigma}) 
    % \end{aligned}
    \label{eq:matern}
\end{equation}
where $\|\cdot\|_{2}$ and $\|\cdot\|_1$ denote the $L_2$ norm and $L_1$ norm, respectively, $\sigma$ is the width of the kernel function as the hyper-parameter describing the correlation between any two sampled ray pairs ${\boldsymbol{x}}_{i}=(\tilde{\textbf{x}}_i,\tilde{t}_i)$ and ${\boldsymbol{x}}_{j}=(\tilde{\textbf{x}}_j,\tilde{t}_j)$ in our model.
% {\color{red} In Eq. (8), what is the definition of norm $\|\cdot\|_1$?}

%\end{theorem}

Based on Representer Theorem, given a kernel function $\kappa$ defined on Reproducing Kernel Hilbert Space (RKHS) $\mathcal{H}$, the solutions of a large class of optimization problems can be expressed as kernel expansions over the sample points. In 4D STEC prediction problem, the input function $u$ can be constructed that satisfies
\begin{equation}
        u^\ast = \argmin_{u\in\mc{H}} \norm{u(\textbf{X}, \textbf{T})-\textbf{Y}}_2^2
\end{equation}
and for any $(\tilde{\textbf{x}}, \tilde{t}) \in \tilde{\mc{D}}$, $u^\ast$ can be expressed as
\begin{equation}
    \begin{aligned}
        u^\ast(\tilde{\textbf{x}}, \tilde{t}) =  \kappa\big([\tilde{\textbf{x}}, \tilde{t}],[\textbf{X}, \textbf{T}]\big) \kappa\big([\textbf{X}, \textbf{T}],[\textbf{X}, \textbf{T}]\big)^{-1}\textbf{Y}
    \end{aligned}
    \label{eq:stec-u}
\end{equation}
where $\textbf{X}=[\textbf{x}_1^{(1)}, \cdots, \textbf{x}_m^{(1)}]^\top$, $\textbf{T}=[t_1^{(1)},\cdots,t_m^{(1)}]^\top$, $\textbf{Y} = [y_1^{(1)},\cdots,y_m^{(1)}]^\top$. 
Therefore, when we specify the form of the kernel function $\kappa$ and the data characteristics, then the input function $u$ can be obtained.

\subsection{Preprocessing}
\subsubsection{\textbf{Temporal Encoding}}
In the STEC 4D estimation problem, without a suitable encoding, the variance between adjacent time $t$ is not obvious and is overwhelmed by large values. In the input domain, we specify time $t$ into year (Year, YY), day of year (Day Of Year, DOY), and second of day (Second Of Day, SOD), namely as $t=(\text{YY, DOY, SOD})$. DOY refers to the sequential count of days within a calendar year from the beginning of the year. It is commonly sine-cosine-encoded\cite{Williscroft1996}

\begin{equation}
    \begin{aligned}
        \text{SOD}_s = \sin(\text{SOD}\cdot 2\pi/P) & \\
        \text{SOD}_c = \cos(\text{SOD}\cdot 2\pi/P)
    \end{aligned}
\end{equation}
where $\text{SOD}_s$ and $\text{SOD}_c$ represent the sine and cosine encoding of SOD, respectively, and $P$ represents the period, which is taken as $3600 \times 24$ seconds for one day here. The large and non-periodic values of time data is avoided by encoding time $t$ into two orthogonal periodic components.

\subsubsection{\textbf{High Dimensional Sobol Sampling}}

When constructing the branch network input, it is necessary to obtain the input function $\{u(\tilde{\textbf{x}}_i^{(1)}, \tilde{t}_i^{(1)})\}^m_{i=1}$ at $m$ scattered temporal-spatial sample points from the observation domain $\tilde{\mc{D}}$. However, due to the fact that the domain $\tilde{\mc{D}}$ in the 4D STEC estimation problem is high-dimensional, the sampling is required to ensure the distribution of these $m$ sampling points as uniform as possible. 

To achieve high-dimensional sampling, an efficient uniform method-the Sobol sampling \cite{sobolDistributionPointsCube1967} is adopted to deal with this difficulty. The key idea behind the Sobol sequence is to use a set of primitive polynomials that exhibit desirable properties when generating each component of the points. These polynomials are carefully chosen to have a  balance between low-discrepancy and efficient computation.

%% Figure 2 %%%%%%%%%%%%%%%%%%%
\begin{figure*}[!htbp]
    \centering
    \subfigure[US station before sampling]{
        \includegraphics[width=0.4\linewidth]{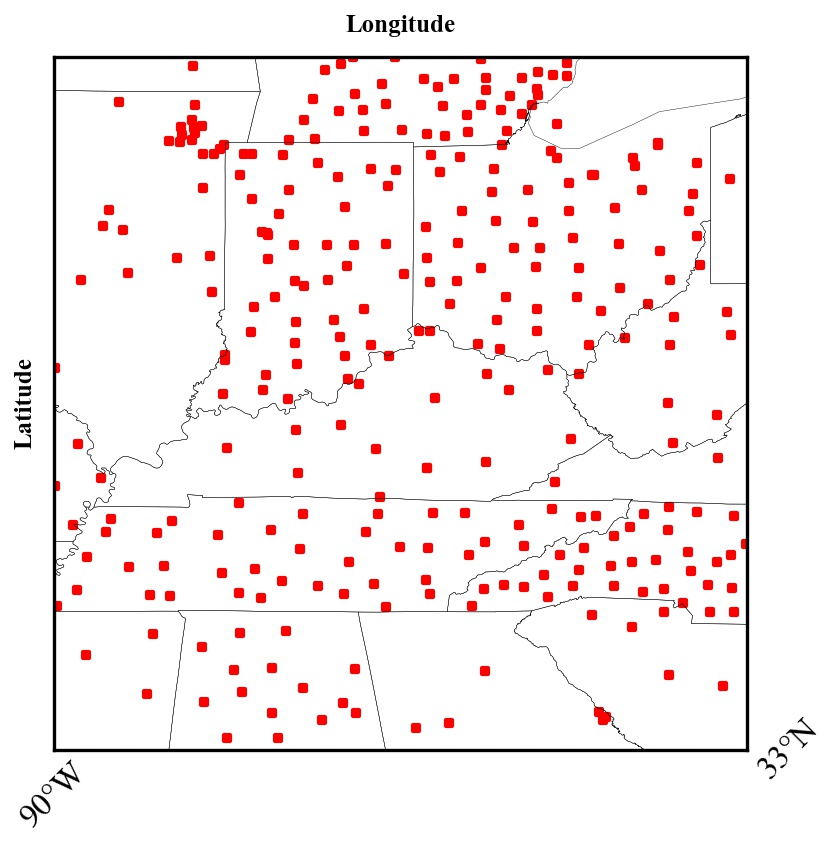}
    }
    \subfigure[US station after sampling]{
        \includegraphics[width=0.4\linewidth]{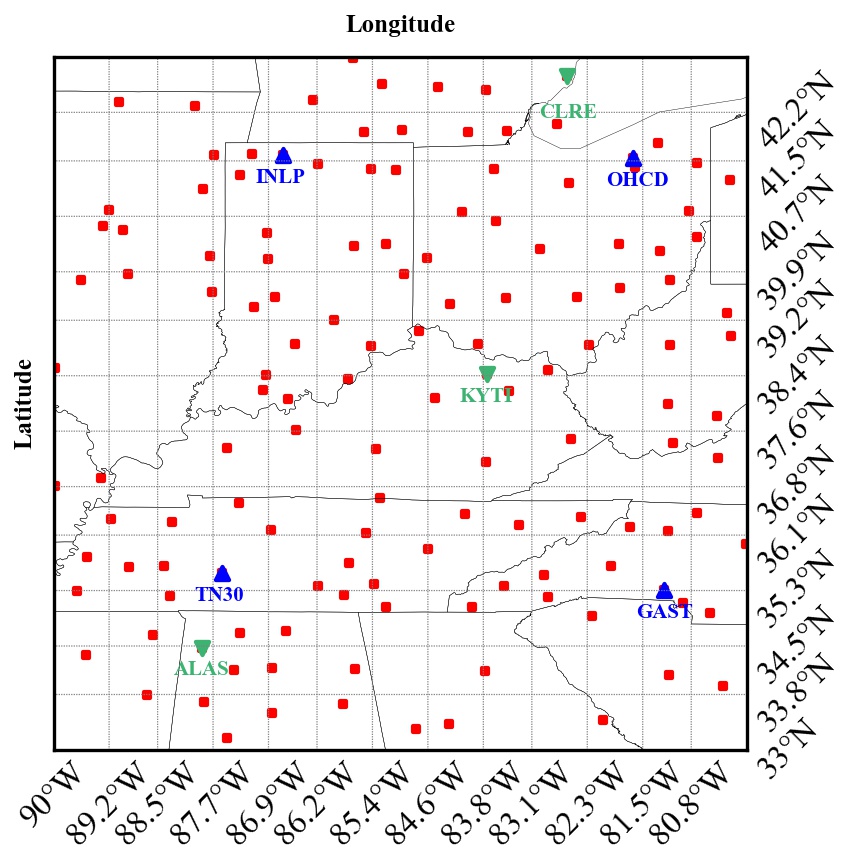}
    }
    \caption{US-CORS station distribution for the training (red), validation (blue) and test (green) data for (a) all available sites; (b) the down-sampling sites.}
    \label{fig:us_map}
    \vspace{-0.5cm}
\end{figure*}

\subsubsection{\textbf{Adaptive Inverse Frequency Weighting Optimization}}

Since the distribution of the train data is imbalanced, the Adaptive Inverse Frequency Weighting (AIFW) method is adopted to optimize \cite{yang2021delving}. The AIFW method handles the imbalanced training data and assigns weights to each data point based on the frequency of its target value, which enables models to adjust weights for sparse large value data. 

\begin{algorithm}[htb]
 \caption{4D DeepONet-STEC}
 \label{algorithm:DeepONet-STEC}
 \begin{algorithmic}[1]
 \renewcommand{\algorithmicrequire}{\textbf{Input:}}
 \renewcommand{\algorithmicensure}{\textbf{Output:}}
 \REQUIRE a training set $\mc{T}$ consisting of $N$ observation data  and a random given ray $(\textbf{x},t) \in \mc{D}$ in real space.
 \ENSURE  the estimated STEC data $G(u)(\textbf{x},t)$ at $(\textbf{x},t)$.
  
\STATE Pre-sample observation data and construct the input function $u$, i.e., in Eq. (\ref{eq:stec-u}) [Process C]
\STATE Fix $m$ sample points $(\tilde{\textbf{x}}_m^{(1)}, \tilde{t}_m^{(1)})$ in the current space [Process D2] and encode $\tilde{t}_m^{(1)}$ for the sample points $(\tilde{\textbf{x}}_m^{(1)}, \tilde{t}_m^{(1)})$ [Process D1]
\STATE Feed preliminary estimated STEC values gained into the Branch network and select input $(\textbf{x},t)$ into the Trunk network after temporal encoding
\STATE Use the Branch network output $\beta_k(u)$ and Trunk network output $\tau_k(\textbf{x},t)$ to obtain the network output in Eq. (\ref{eq:deeponet-out}). [Process B]
\STATE Train the model using weighted loss in Eq. (\ref{eq:loss-w}) to obtain the operator $G(u)(\cdot,\cdot)$. [Process D3]
% \STATE Obtain the estimation data of STEC at $(\textbf{x},t)$ using best function $G(u)(\cdot,\cdot)$ have trained
\RETURN  $G(u)(\textbf{x},t)$
\end{algorithmic}
\end{algorithm}

For the training set $\mc{T}$, we introduce the label space $\mc{Y}$, where we divide $\mc{Y}$ into $B$ groups (bins) with equal intervals, i.e., $[y_0, y_1), [y_1, y_2), \cdots, [y_{B-1}, y_B)$. $\mc{Y} = \{\textbf{y}_i\}_{i=1}^B$ with $\textbf{y}_i = [y_{i-1}, y_i)$. We denote $b\in \{0,\cdots, B-1\}$ as the group index of the target value. The defined bins reflect a minimum resolution we care for grouping data in a regression task. We define $\delta y \triangleq y_{b+1}-y_{b}=1$, showing a minimum difference of 1 TECU is of interest for our problem. Finally, we could define each interval $\textbf{y}_i$ as the target value $y_i$ in each bin, $\mc{P}(\textbf{y}_i)$ is as the number of the target value within this bin, then the weight $\omega_i$ for ${y_i}$ can be expressed as 
\begin{equation}
    \omega_i = \big(\frac{y_i}{\mc{P}(\textbf{y}_i)}\big)^{\lambda}
\end{equation}
where $\lambda$ represents the control factor of the weighting decay amplitude. In this paper, $\lambda$ was chosen as 0.05 by trial and error. By adding weights into Eq. (\ref{eq:loss}) of DeepONet, the modified loss function becomes 
\begin{equation}
        L= \sum_{i=1}^{{h}} |G(u)(\tilde{\textbf{x}}_i^{(2)}, \tilde{t}_i^{(2)}) - y_i^{(2)}|^2\cdot \frac{w_i}{\sum\limits_{i=1}^{h}w_i}
        \label{eq:loss-w}
\end{equation}

%% Figure 3 %%%%%%%%%%%%%%%%%%%
\begin{figure*}[!htbp]
    \centering
    \includegraphics[width=0.85\linewidth]{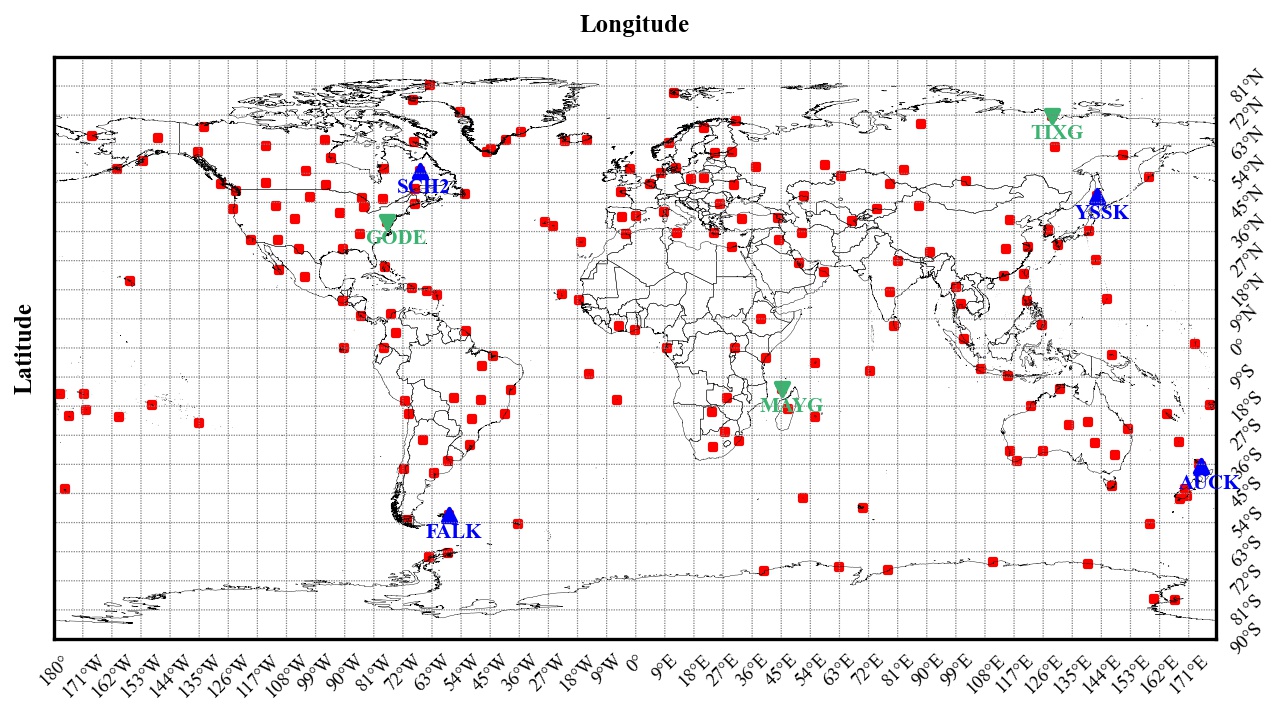}
    \caption{Global station distribution for the training, validation and test data after down-sampling. Tree test sites (green triangle) and four validation sites (blue triangle) are selected globally, respectively, while the red square refers to the training sites. }
    \label{fig:global_map}
    \vspace{-0.5cm}
\end{figure*}

The 4D DeepONet-STEC prediction model is illustrated in Figure 1. Using $\{(\tilde{\textbf{x}}_i, \tilde{t}_i, \tilde{y}_i)\}_{i=1}^M$ to train, the model can input random giving ray $(\textbf{x}, t) \in \mc{D}$ to get the output STEC value $y$. Specifically, assuming the time period to predict is $\textbf{T}_0$ and $\textbf{T}_0 \in [0,\textbf{T}]$, the input space during training is defined as $\tilde{\mc{D}}_{input} = \tilde{\mc{X}}\times[0,\textbf{T}-\textbf{T}_0]$, while the output space is defined as $\tilde{\mc{D}}_{output} = \tilde{\mc{X}}\times[\textbf{T}-\textbf{T}_0, \textbf{T}]$. The architecture consists of two key parts. The first part involves constructing the input function $u$, with the input space defined as $\tilde{\mc{D}}$ and the output space as $\mc{D}$. During the network training, the available data is limited to observations, necessitating the partitioning of $\tilde{\mc{D}}$.  The input function $u$ is constructed using the kernel function labeled \textit{Process C}, utilizing observation data to map a given ray in the observation space to STEC. After \textit{Process D1} and \textit{Process D2}, this serves as the input for the second part, which trains the operator $G$ to extend the range of applicability and accuracy of the input function $u$ as Process B and \textit{Process D3}. As the final generated function $G(u)(\cdot,\cdot)$ is the dot product of the outputs from the Branch and Trunk networks, both networks require an equal number of neurons in their output layers. Using the optimized function $G(u)(\cdot,\cdot)$ obtained from the training, the STEC estimate $G(u)(\textbf{x},t)$ can be obtained. Thus, the constructed 4D DeepONet-STEC model enables 4D STEC estimation.
The  4D DeepONet-STEC algorithmic flow is depicted in \textit{Algorithm \ref{algorithm:DeepONet-STEC}}. 
\\
%% Figure 4 %%%%%%%%%%%%%%%%%%%
\begin{figure*}[!htbp]
	\subfigure[Kp\&Dst index (Quiet Period)]{
		\begin{minipage}[c]{0.5\linewidth}
			\centering
			\includegraphics[width=3.4in]{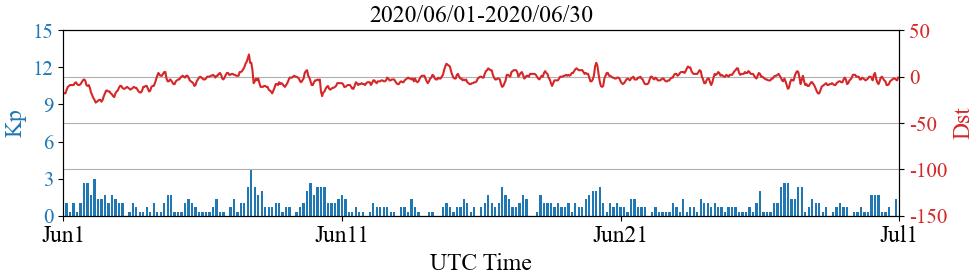}\\
		\end{minipage}%
	}%
	\subfigure[All rays at one GODE station (Quiet Period)]{
		\begin{minipage}[c]{0.5\linewidth}
			\centering
			\includegraphics[width=3.4in]{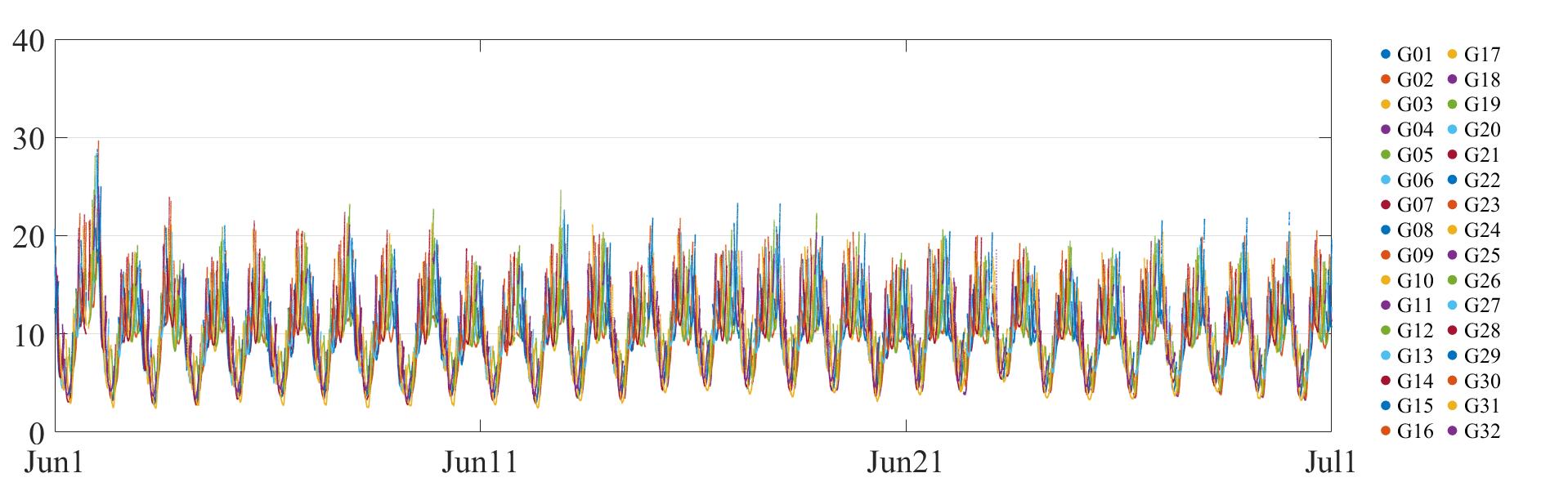}\\
		\end{minipage}%
	}%
	\centering
	\vspace{-0.2cm}
   	\centering
            \subfigure[Kp\&Dst index (Storm Period)]{
 		\includegraphics[width=6.8in]{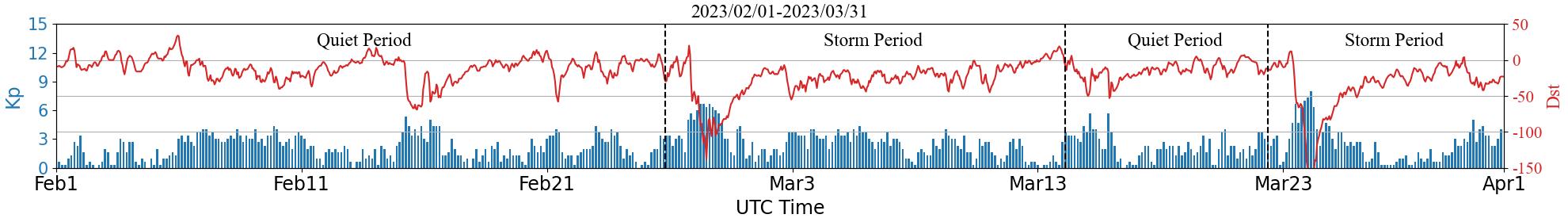}}\\
			\vspace{-0.4cm}
            \subfigure[All rays at one GODE station (Storm Period)]{
			\includegraphics[width=7.2in]{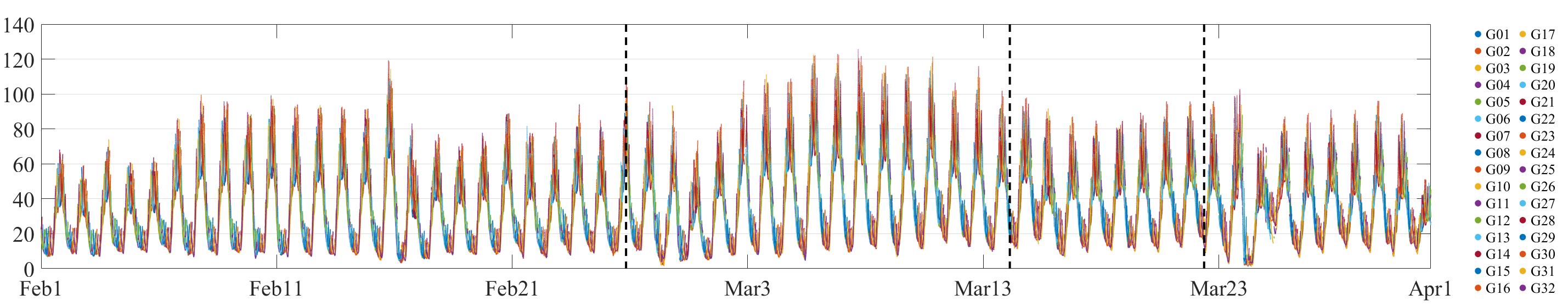}}\\
   %          \subfigure[Rays at one GODE station (Storm Period magnified)]{
			% \includegraphics[width=3in]{deeponet_fig/0228revise/59magnify.jpg}}\\
   %       \centering

    \caption{Temporal dataset of the geomagnetic index Kp\&Dst and retrieved STEC value at a single station (GODE) during solar ionospheric quiet (a-b) and storm (c-d) time periods.}
    \label{fig:kpdst30}
\end{figure*}

\section{Validation Results and Analysis}
To verify and validate the accuracy and robustness of the 4D DeepONet-STEC model, we perform a series of tests using both the simulation datasets and real observation datasets under global and regional GNSS station geometry scenarios. 

\subsection{Validation Setup}
The validation sites were selected based on US-CORS (Continuously Operating Reference Stations) GNSS stations in Fig.~\ref{fig:us_map} and global IGS (International GNSS service) as illustrated in Fig.~\ref{fig:global_map} to ensure representative coverage. To assess the performance of the proposed 4D DeepONet-STEC model, the whole number of stations is divided into the training sites (red square) , the validation sites (blue triangle) and the test sites  (green triangle) in Fig.~\ref{fig:us_map}(b) and Fig.~\ref{fig:global_map}. By employing this selection of sites and training data, this model aims to simulate real-world scenarios for users at arbitrary locations within the region of interest.

In order to verify the performance of the model at different solar activity for diverse ionospheric conditions, we choose two global dataset to test the model: the quiet period in Fig.~\ref{fig:kpdst30}(a) and the storm period to test the model in Fig.~\ref{fig:kpdst30}(c). The quiet time period spans from June 1, 2020, to June 30, 2020 and the storm time period spans from February 1, 2023 to March 31, 2023 based on the magnetic storm $\rm{Kp}$ and $\rm{Dst}$ index. %The kp\&dst index and STEC distribution figure is shown in Fig.~\ref{fig:kpdst30}.}

In quiet periods, the training dataset for the train stations were utilized during June 1 to June 27, 2020. The validation stations without training are used to optimize the train model performance by making prediction from June 25 to June 27, 2020. The test sites were selected to predict from June 28 to June 30, 2020. Similarly, in storm periods from February 1, 2023 to March 31, 2023, the test days is enhanced to 10 days from March 22 to April 1, 2023. The training and validation dataset is from February 1 to March 21, 2023. The ratio of the training and test days for quiet periods is $27:3$ and the ratio of the training and test days for magnetic storm periods is $49:10$, respectively. respectively. 

For the simulation data generation, we use the NeQuick2 model \cite{navaNewVersionNeQuick2008}. The NeQuick model is a three-dimensional and time dependent ionospheric electron density, which has been adapted for Galileo satellite navigation single-frequency ionospheric corrections. The NeQuick2 model can provide the electron density along any ground-to-satellite straight line ray-path and the corresponding slant Total Electron Content (STEC) by numerical integration. The expression for the Epstein layer function of the Nequick2 model is shown as

\begin{equation}
    N_{Epstein} = \frac{4N_{max}}{\big(1+\exp(\frac{h-h_{max}}{B})\big)^2}\exp(\frac{h-h_{max}}{B})
\end{equation}
where $h$ is the height, $h_{max}$ is the layer peak height, $B$ is the thickness parameter, and $N_{max}$ is the layer peak electron density. 

The NeQuick-STEC simulator is used to generate rays by extracting satellite coordinates for specified time periods using precision ephemeris files. The NeQuick2 model was then applied to these rays, for selected time as inputs, to generate simulated STEC values for each ray, forming the synthetic STEC dataset {\cite{shiMethodDSTECInterpolation2022}}, which can provide efficient test for evaluating the developed model.

For the real observation dataset generation, the undifferenced and uncombined precise point positioning (UC-PPP) with known fixed station coordinates is used to extract the STEC values from GNSS observables \cite{zhouGAMPOpensourceSoftware2018,BZhang2016revise}. For electromagnetic wave frequency $f_i$, the GNSS observation equations can be written as \cite{zhouGAMPOpensourceSoftware2018}:

\begin{equation}
\begin{aligned}
    P^s_{r,f_i} = \rho^s_r + (dt_r - dt^s) + T_r^s + I_{r,f_i}^s & \\ + (d_{r,f_i} - d^s_{f_i})+\epsilon^s_r&
    \label{eq:code navigation equation}
\end{aligned}
\end{equation}

\begin{equation}
\begin{aligned}
    L^s_{r,f_i} = \rho^s_r + (dt_r - dt^s) + T_r^s  -{I_{r,f_i}^s} & \\  + (b_{r,f_i} - b^s_{f_i})+ \lambda N^s_{r,f_i}  + \eta^s_r&
    \label{eq:phase navigaion equation}
\end{aligned}
\end{equation}
where the superscript $s$ and the subscript $r$ denotes a satellite and a receiver, respectively. $P^s_r$ and $L^s_r$ denotes the pseudo-range and carrier phase observable, respectively. $\rho^s_r$ is the geometric distance between the receiver $r$ to the satellite $s$. Here, $dt_r$ and $dt^s$ are the receiver and satellite clock offsets, respectively. $d_{r,f_i}$ and $b_{r,f_i}$ are the receiver code and phase biases, and $d^s_{f_i}$ and $b^s_{f_i}$ are the satellite code and phase biases (in meters). Here, $I_r^s$ is the line of sight (LOS) slant ionospheric delay and $T^s_r$ is troposphere delay, $N^s_r$ is the phase ambiguity, $\lambda$ is the wavelength and $\epsilon_p$ is the unmodelled errors for pseudo-range and carrier phase observable. The ionospheric STEC is estimated by the dual-frequency undifferenced and uncombined UCPPP method using both pseudo-range and carrier phase observables\cite{9096522}. The DCBs of the receivers and satellites are separated by ionospheric single layer modeling (SLM), yielding the final unbiased STECs. The epoch temporal resolution is $30\,\rm{s}$.  
%According to the GNSS observation equation, for receiver $r$ and satellite $s$, the raw GNSS pseudo-range observation value $P^s_{r,f_i}$ at frequency point $f_i$ can be expressed as\cite{zhouGAMPOpensourceSoftware2018}:
\begin{table}[!htbp]
\caption{List of Selected IGS Station for Test and Validation}
\label{table:station-set}
\centering
\begin{tabular}{ccccc}
\hline
Station & Region   & Name   & Latitude[$^\circ$] & Longitude[$^\circ$]  \\
\hline
\multirow{2}{*}[-5ex]{Test} & \multirow{4}{*}{USA} & CLRE & 42.72 & -82.59\\
                                            & & KYTI & 38.42 & -83.74\\
                                            & & ALAS & 34.47 & -87.86\\
                            \cline{3-5}
                      & \multirow{3}{*}{Global} & TIXG & 71.63 & 128.87\\
                                                & & GODE & 39.02 & -76.83\\
                                                & & MAYG & -12.78 & 45.26\\
\hline
\multirow{2}{*}[-8ex]{Validation} & \multirow{4}{*}{USA} & OHCD & 41.54 & -81.63\\
                                            & & GAST & 35.31 & -81.18\\
                                            & & INLP & 41.58 & -86.69\\
                                            & & TN30 & 35.55 & -87.57\\
                            \cline{3-5}
                      & \multirow{4}{*}{Global} & FALK & -51.69 & -57.87\\
                                                & & AUCK & -36.6 & 174.83\\
                                                & & SCH2 & 54.83 & -66.83\\
                                                & & YSSK & 47.03 & 142.72\\
\hline
\end{tabular}
\end{table}

\subsubsection{US region data generation}
The latitude and longitude range for the North American region was set as [33$^\circ$N, 43$^\circ$N] and [80$^\circ$W, 90$^\circ$W], consisting of 300 GNSS stations, respectively. To address the large number of stations for high computational cost, a sparse downsampling process was employed. Specifically, the entire region was divided into  a 0.77$^\circ\times$0.77$^\circ$ grid patch unit with a number of 13$\times$13 patches.  
Only one station is selected from each grid patch. This downsampling process reduces the dataset size while maintaining spatial diversity. Within the total number of stations 137 after downsampling, three stations were chosen for testing, while four stations were designated for validation. The distribution of the test, validation, and training stations in the North American region is illustrated in Fig.~\ref{fig:us_map}(b).
%% Figure 5 %%%%%%%%%%%%%%%%%%%
\begin{figure*}[!htbp]
	\centering
    \subfigure[US CORS observation STEC dataset of single-station, single-satellite between June 1-27 (Training) and June 28-30 (Test)]{
    \begin{minipage}[b]{1\linewidth}
			\includegraphics[width=7in]{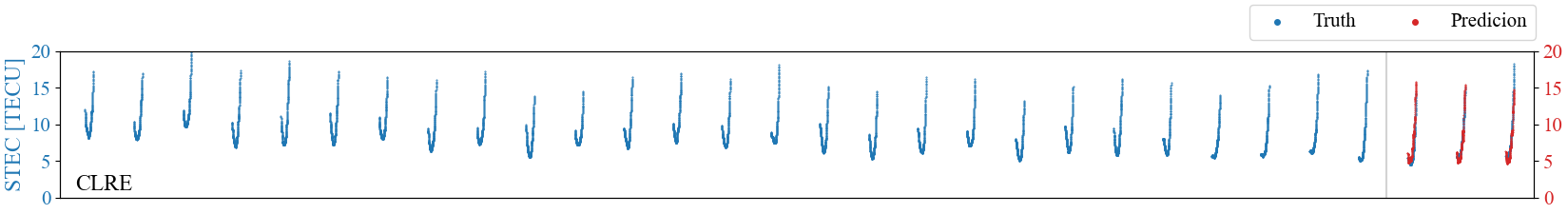}\\
			\includegraphics[width=7in]{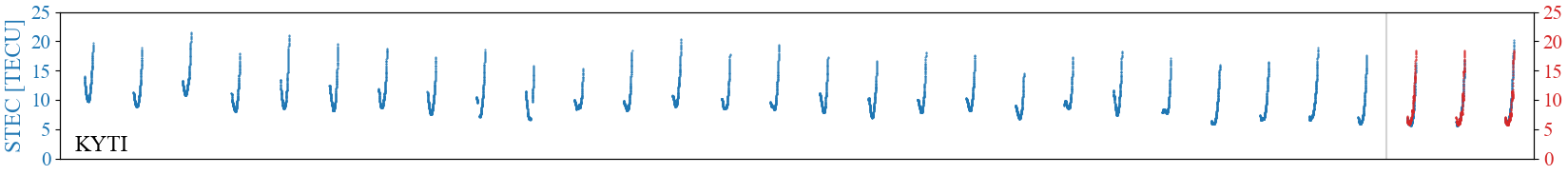}\\
			\includegraphics[width=7in]{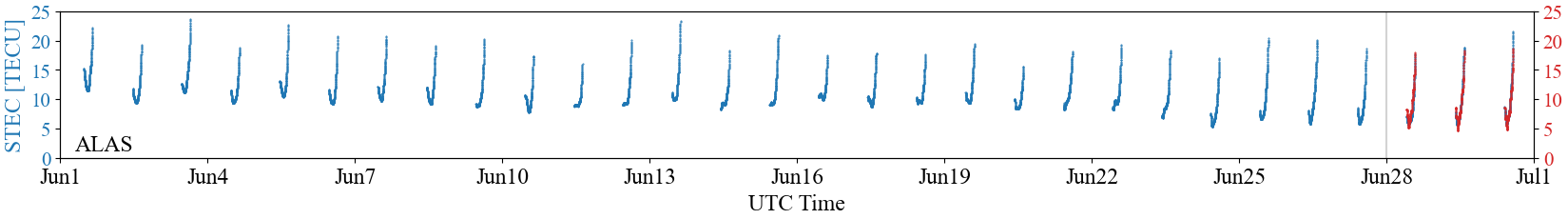}\\
			\vspace{-0.6cm}
   \end{minipage}
	\centering
    }\\
    \subfigure[US CORS predicted (red) and observed (blue) STEC between June 28-30 (only test comparison)]{
	\centering
    \begin{minipage}[b]{1\linewidth}
			\includegraphics[width=7in]{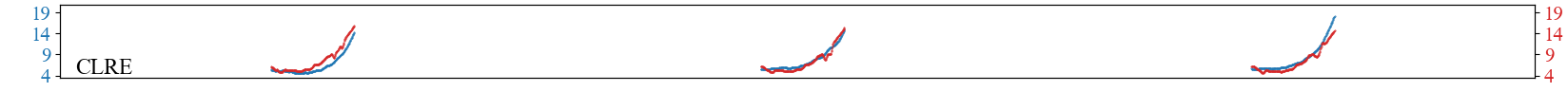}\\
			\includegraphics[width=7in]{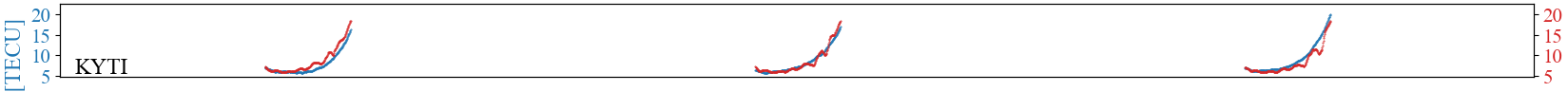}\\
			\includegraphics[width=7in]{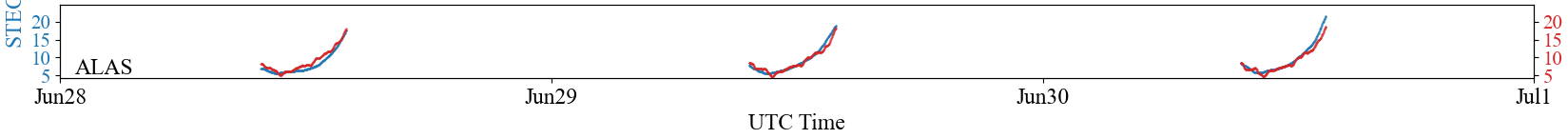}\\
			\vspace{-0.6cm}
   \end{minipage}
	\centering
	\vspace{-0.2cm}}
    \caption{US CORS observation dataset predicted STEC of single-station, single-satellite (G08) for three test stations for (a) one-month data and (b) only the three-day prediction period. The blue dots represents the observation, while the red dots represents the predicted STEC values in quiet periods.}
    \label{fig:us-obs-rays}
    \vspace{-0.6cm}
\end{figure*}

%% Figure 6 %%%%%%%%%%%%%%%%%%%
\begin{figure*}[!htbp]
	\centering
	\subfigure[CLRE]{
		\begin{minipage}[c]{0.32\linewidth}
			\centering
			\includegraphics[width=2in]{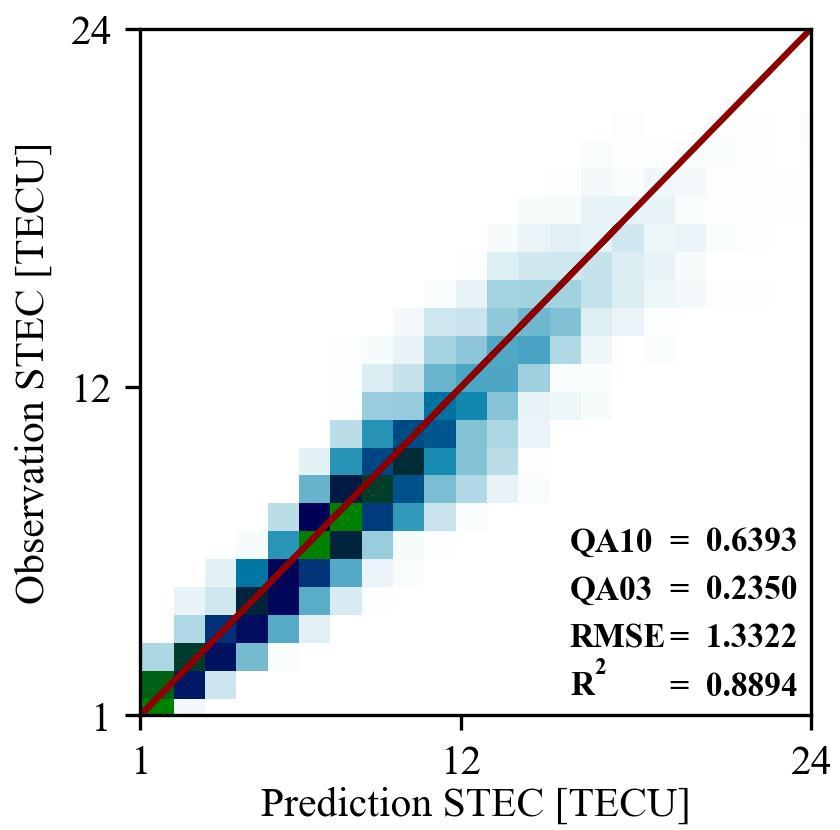}\\
			\vspace{0.02cm}
			\centering
			\includegraphics[width=2.3in]{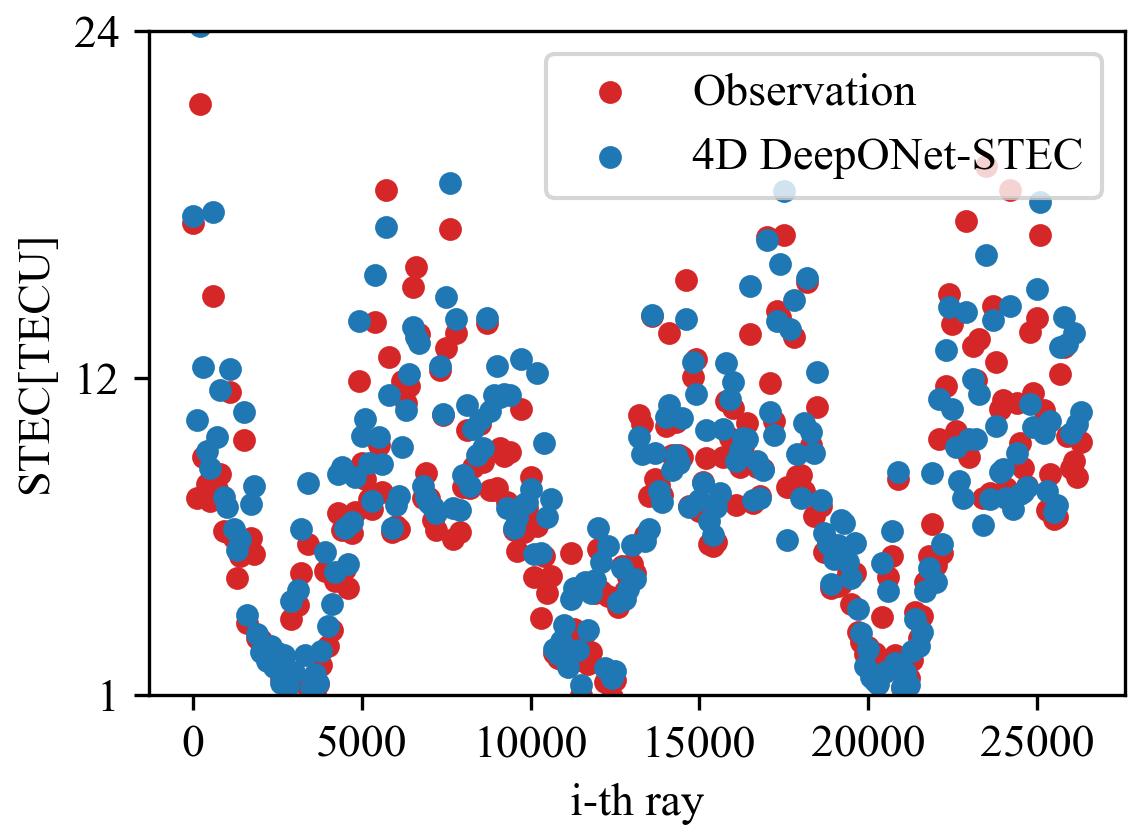}\\
			\vspace{0.02cm}
		\end{minipage}%
	}%
	\subfigure[KYTI]{
		\begin{minipage}[c]{0.32\linewidth}
			\centering
			\includegraphics[width=2in]{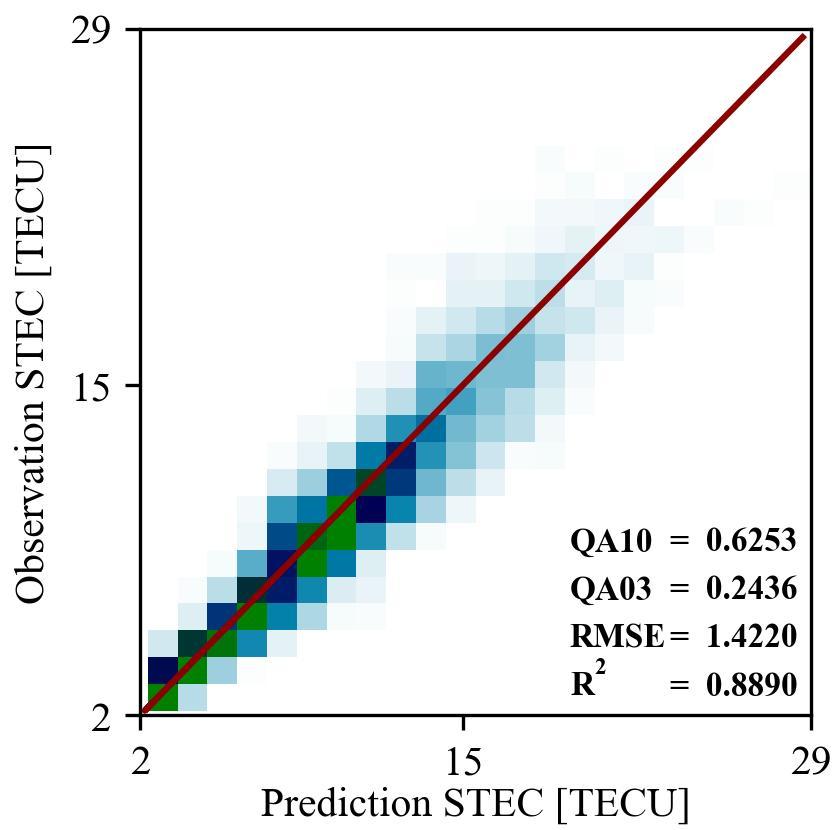}\\
			\vspace{0.02cm}
			\centering
			\includegraphics[width=2.3in]{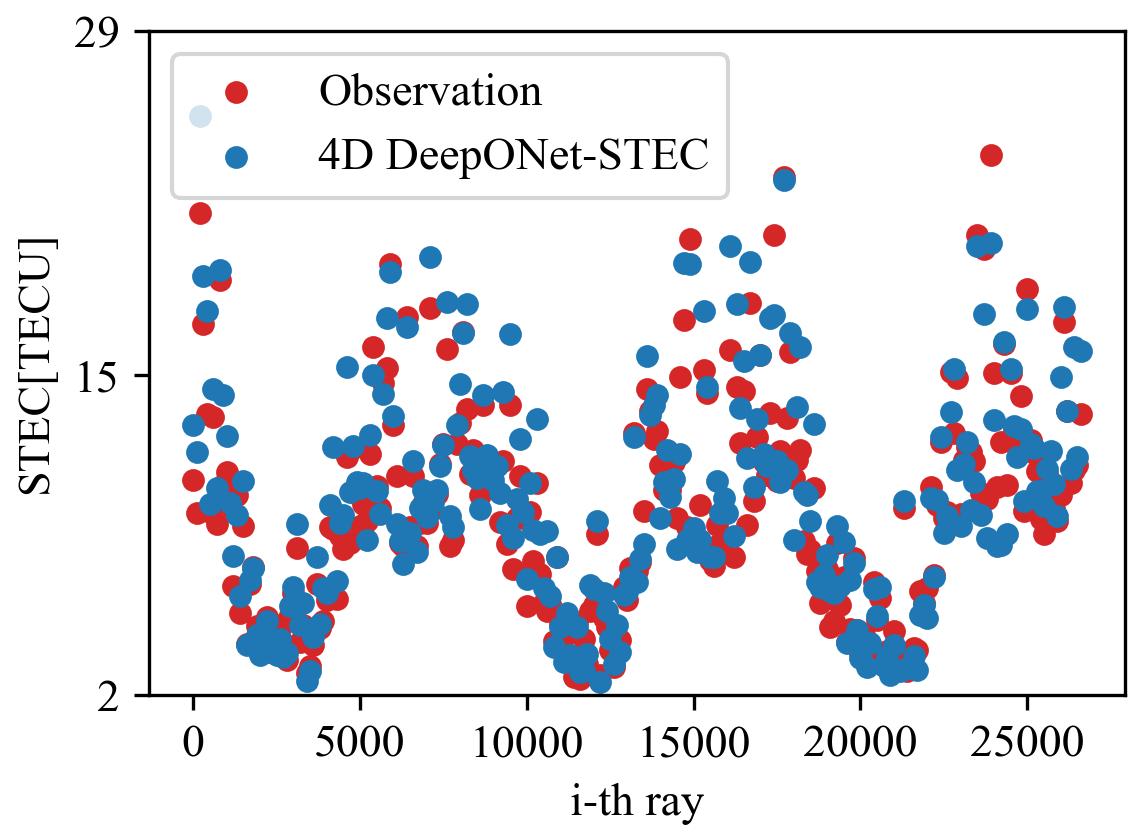}\\
			\vspace{0.02cm}
		\end{minipage}%
	}%
	\subfigure[ALAS]{
		\begin{minipage}[c]{0.34\linewidth}
			\centering
			\includegraphics[width=2.57in]{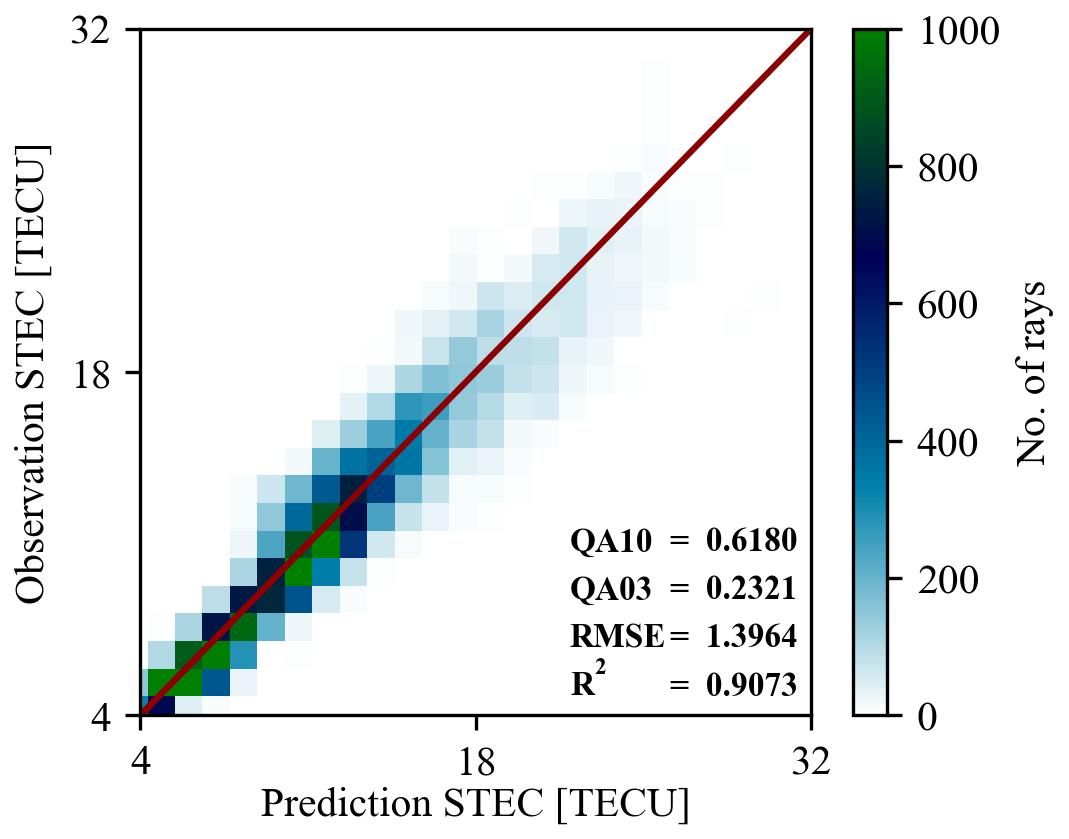}\\
			\vspace{0.02cm}
			\centering
			\includegraphics[width=2.3in]{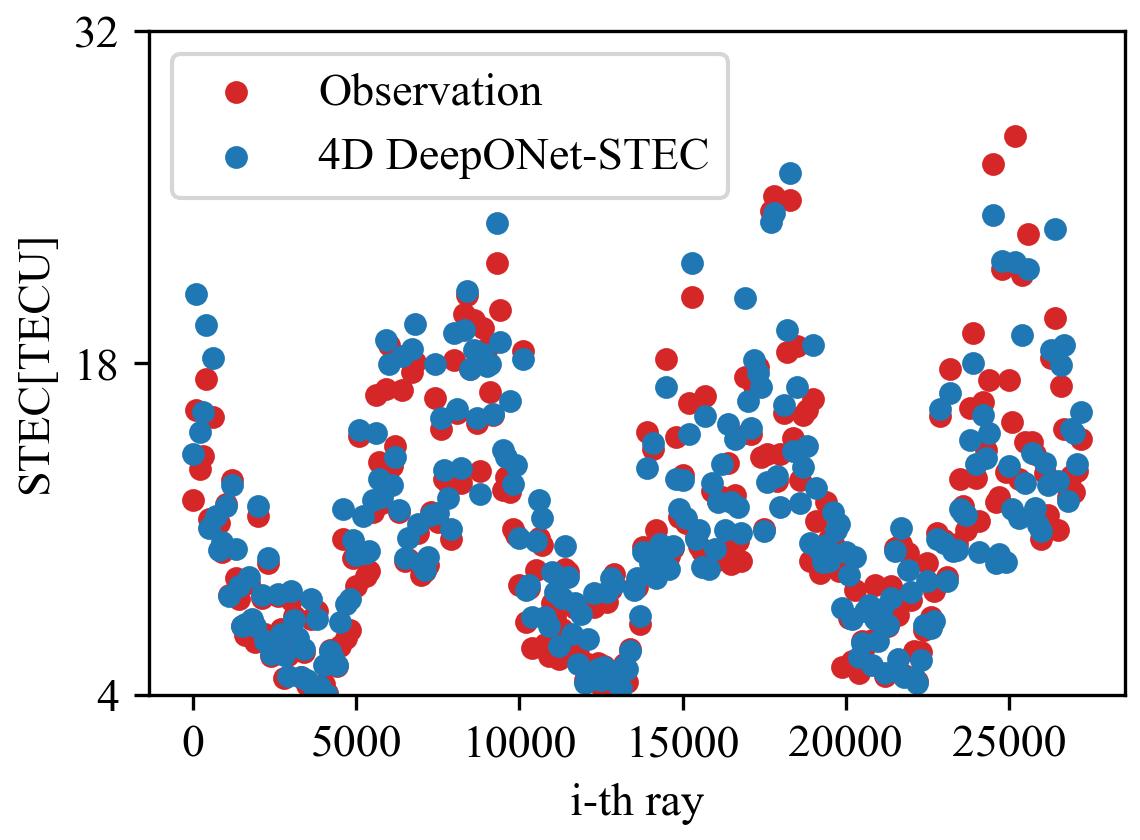}\\
			\vspace{0.02cm}
		\end{minipage}%
	}%
	\centering
	\vspace{-0.2cm}
    \caption{Up: US CORS observation data prediction result by the DeepONet-STEC model for three test station-satellite rays, where the color counts the number of rays denoting the correlation between the observation and predicted value within the prediction period 3 days. Down: the DeepONet-STEC model prediction result comparison with observation within 3 days in quiet periods.}
    \label{fig:usa-pred-obs}
    \vspace{-0.5cm}
\end{figure*}

\subsubsection{Global data generation}
To generate the global dataset, the entire Earth was considered with latitudes from -90$^\circ$ to 90$^\circ$ and longitudes from -180$^\circ$ to 180$^\circ$. In order to handle the large number of available stations worldwide, similar downsampling process was applied to divide the global stations into a 9$^\circ\times$9$^\circ$ grid patch unit with a total number of 800 patches. Then, we averagely select 114 sites from original 534 sites in total. The resulting distribution of selected stations across the global region forms the basis of the global dataset used in this study as shown in Fig.~\ref{fig:global_map}. The maximum STEC value is 74 TECU.

Fig.~\ref{fig:kpdst30}(b) shows the retrieved STEC measurement data by UCPPP for the global dataset during the quiet period for June $1-30$, 2020.  Fig.~\ref{fig:kpdst30}(d) shows the retrieved STEC measurement data by UCPPP during the storm period from Feb 1 to Mar 31, 2023 for different GPS satellites at the cutoff elevation angel $15^{\rm o}$. The time resolution for each ray data is $60s$. Based on  high temporal resolution historical data, the proposed DeepONet-STEC model can predict STEC for the next 72 hours for instance with each minute interval. The maximum STEC value in quiet periods is 42 TECU  in Fig.~\ref{fig:kpdst30}(b) and the maximum STEC value in storm periods is 243 TECU in Fig.~\ref{fig:kpdst30}d.

%\subsection{Evaluation Metrics} 
To evaluate the prediction performance of the 4D DeepONet-STEC model,  we define the Root Mean Square Error (RMSE) and $R^2$ metrics to quantify the accuracy and goodness-of-fit of the model's predictions. Furthermore, to assess the model's suitability for navigation applications, the Quality Assessment (QA) metrics are introduced {\cite{psychasAssessmentIonosphericCorrections2019}}, which provides additional metric for the model's capability to achieve high-precision and reliable navigation systems.
% Specifically, the percentage of absolute errors less than 1 TECU will be calculated, and this metric will be denoted as QA10.
\begin{equation}
        RMSE = \sqrt{\frac{1}{N} \sum_{i=1}^{N}(y_i-\hat{y_i})^2}
\end{equation}

\begin{equation}
        \it{R^{2}}=1-\frac{\sum_{i=1}^{N}\left(y_{i}-\hat{y_{i}}\right)^{2}}{\sum_{i=1}^{N}\left(y_{i}-\bar{y}\right)^{2}}
\end{equation}

\begin{equation}
        \it{QA}=\frac{num(\left|y_{i}-\hat{y_{i}}\right|<\theta)}{N}\times100\%
\end{equation}

\begin{equation}
        \it{MAPE}=\frac{1}{N}\sum_{i=1}^{N}\left|\frac{y_{i}-\hat{y_{i}}}{\hat{y_{i}}}\right|
\end{equation}

where $y_i$ represents the $i$-th prediction value, $\hat{y}_i$ represents the $i$-th true value, $N$ represents the total number of data for test stations. In QA metrics, $\theta$ represents the scale of quality assessment and $num(\mc{A})$ represents the number of individuals satisfying condition $\mc{A}$. We denoted QA value as QA03 for {$\theta=0.3$} TECU and QA10 for {$\theta=1.0$} TECU. For mean absolute percentage error (MAPE), we calculated the MAPE value  during the predicting period as shown in Table II, III and IV, respectively.

\subsection{Simulated Data Results in Quiet Periods}
%% Figure 7-8%%%%%%%%%%%%%%%%%%%
%%%%%%%%%%%%%%%%%%%%%Global Simulation Data%%%%%%%%%%%%%%%%%%%%%%%%%%%%%%%%%%%

\begin{figure*}[!htbp]
	\centering
    \subfigure[Global simulated STEC dataset of single-station, single-satellite between June 1-27 (Training) and June 28-30 (Test)]{
    \begin{minipage}[b]{1\linewidth}
			\includegraphics[width=6.8in]{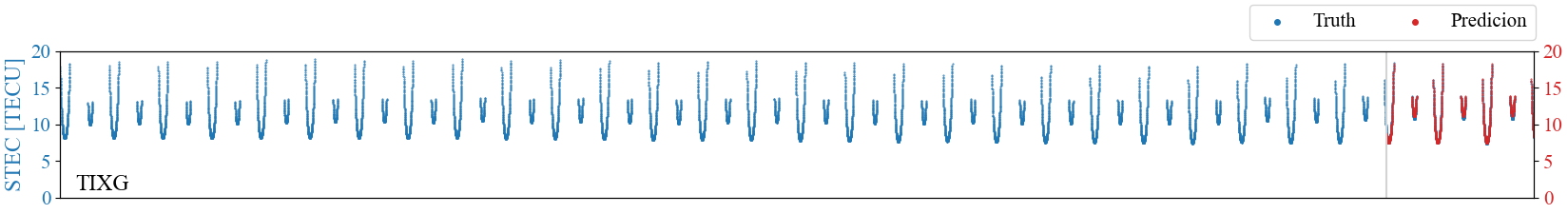}\\
			\includegraphics[width=6.8in]{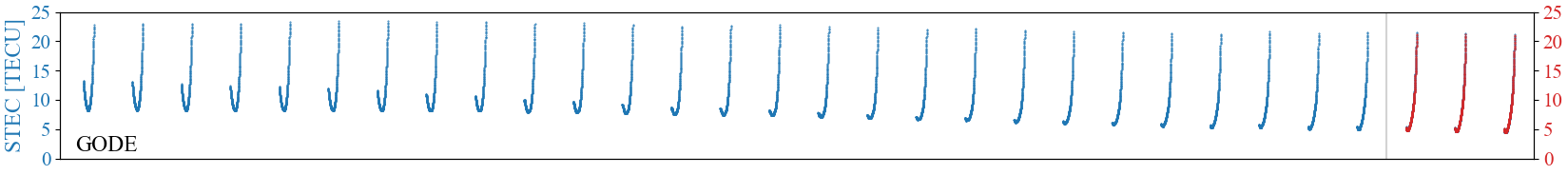}\\
			\includegraphics[width=6.8in]{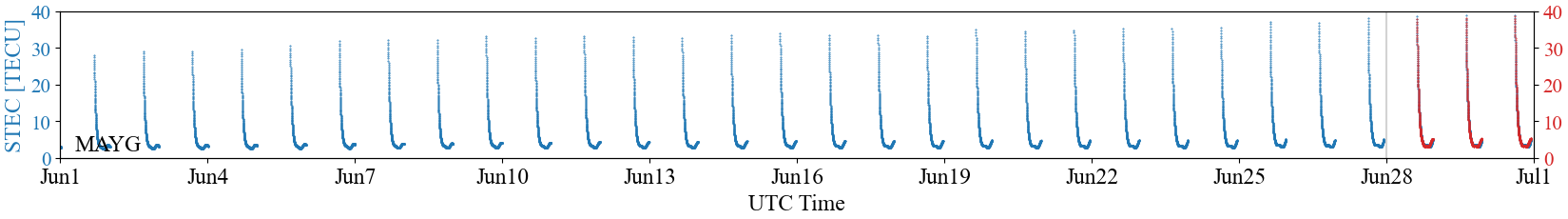}\\
			\vspace{-0.6cm}
   \end{minipage}
	\centering
    }\\  
 
    \subfigure[Global simulated dataset predicted (red) and observed (blue) STEC between June 28-30  (only test comparison)]{
	\centering
    \begin{minipage}[b]{1\linewidth}
			\includegraphics[width=6.8in]{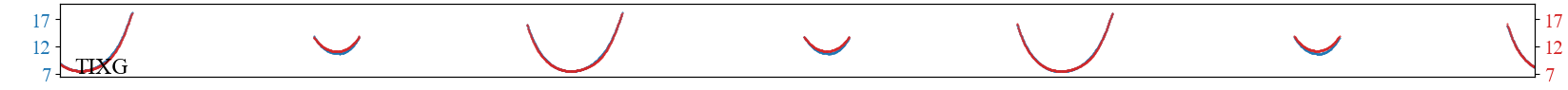}\\
			\includegraphics[width=6.8in]{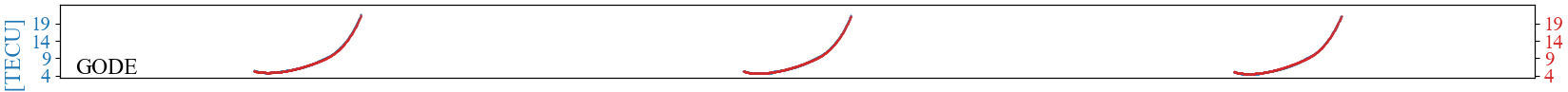}\\
			\includegraphics[width=6.8in]{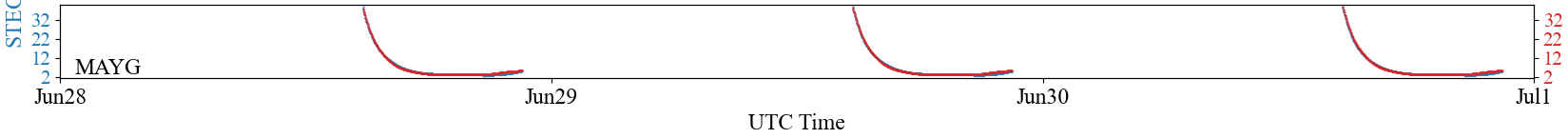}\\
			\vspace{-0.6cm}
   \end{minipage}
	\centering
	\vspace{-0.2cm}}
    \caption{Global simulated dataset predicted STEC of single-station, single-satellite (G08) for three test stations. The blue dots represents the observation, while the red dots represents the predicted STEC values in quiet periods.}
    \label{fig:Global-simu-rays}
    \vspace{-0.6cm}
\end{figure*}

US CORS observation STEC dateset of single-station, single-satellite between June 1-30; US CORS predicted (red) and observed (blue) STEC between June 28-30.

\begin{figure*}[!htbp]
	\centering
	\subfigure[TIXG]{
		\begin{minipage}[c]{0.32\linewidth}
			\centering
			\includegraphics[width=2in]{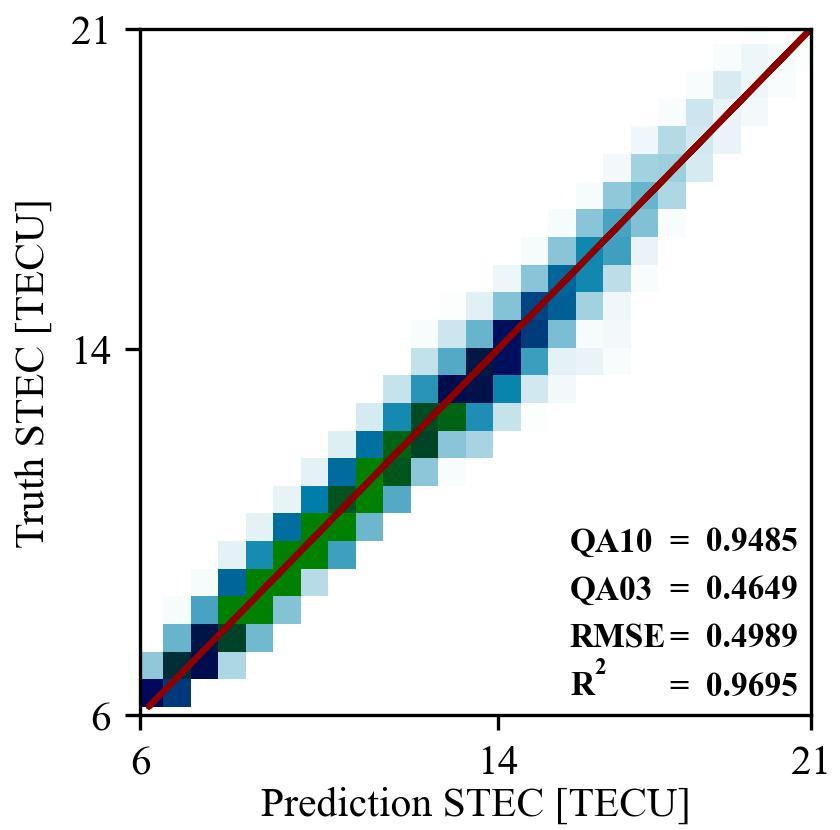}\\
			\vspace{0.02cm}
		\end{minipage}%
	}%
	\subfigure[GODE]{
		\begin{minipage}[c]{0.32\linewidth}
			\centering
			\includegraphics[width=2in]{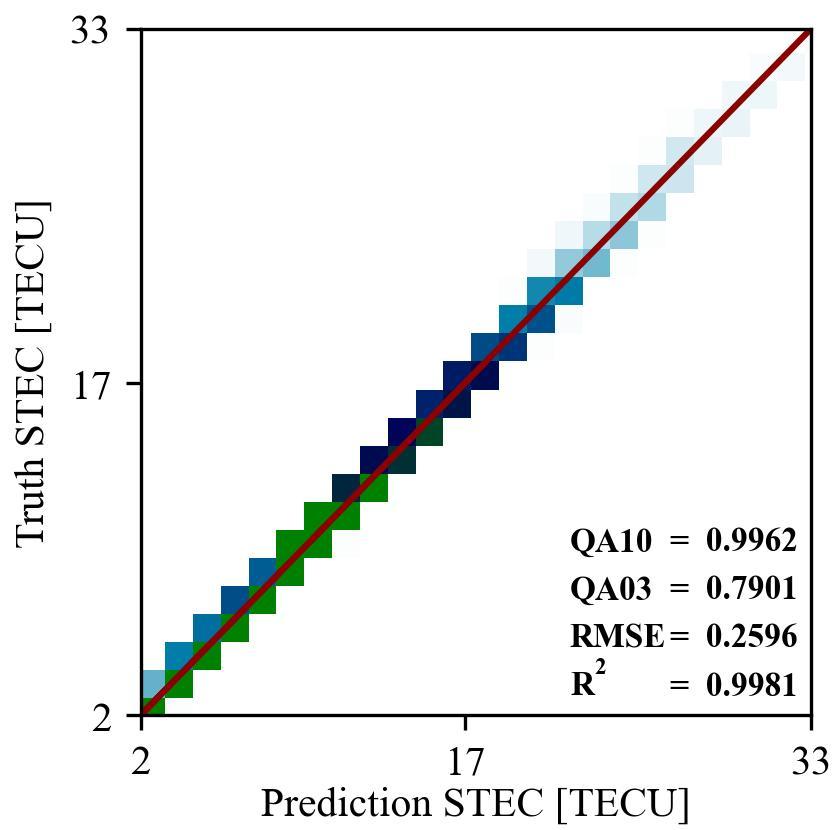}\\
			\vspace{0.02cm}
		\end{minipage}%
	}%
	\subfigure[MAYG]{
		\begin{minipage}[c]{0.34\linewidth}
			\centering
			\includegraphics[width=2.57in]{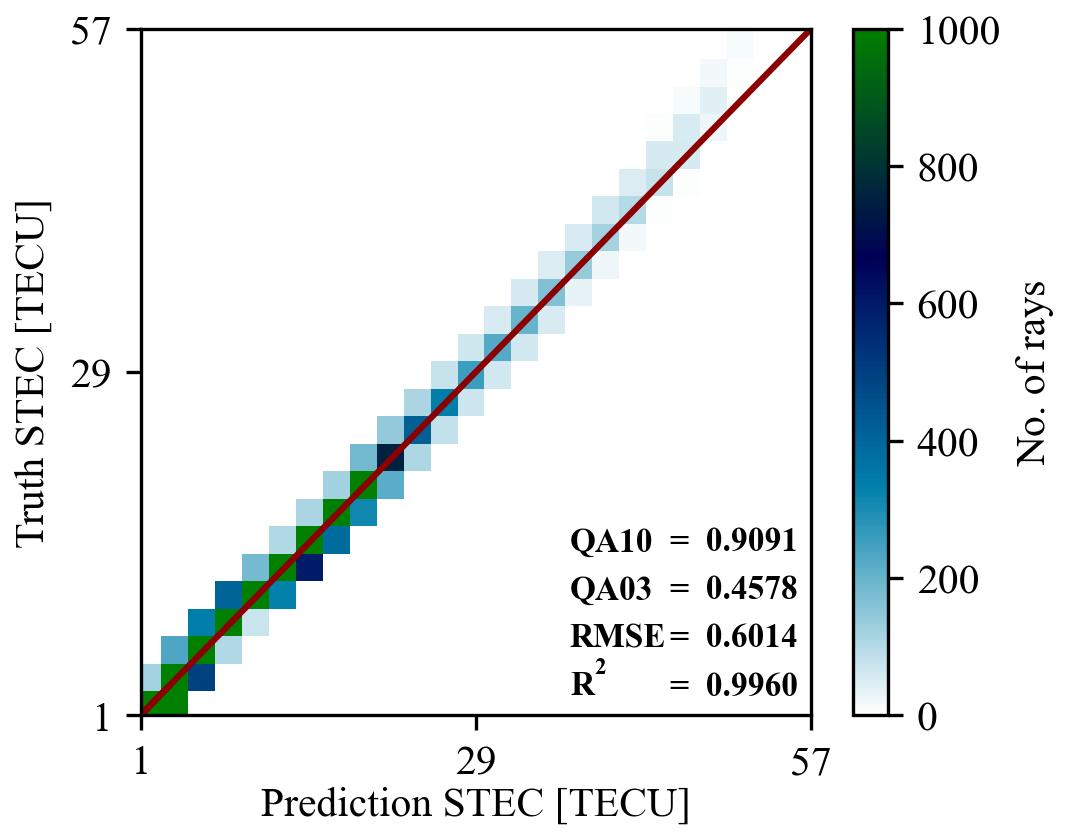}\\
			\vspace{0.02cm}
		\end{minipage}%
	}%
	\centering
	\vspace{-0.2cm}
    \caption{Global simulated dataset correlation between the predicted value and the model truth based on the global simulated 27 days dataset over all satellite-station rays for the future three days in quiet periods.}
    \label{fig:global-pred-simu}
    \vspace{-0.5cm}
\end{figure*}

\begin{table*}[!htbp]
\caption{Statistical prediction results of the simulated data based on the test stations in quiet period}
\label{table:simu-statistical-results}
\centering
\begin{tabular}{lcccccc}
\hline
Region & Station        & RMSE [TECU]   & $R^2$    & MAPE [\%]  & QA03 [\%] & QA10 [\%] \\
\hline
\multirow{3}{*}{USA}    & CLRE & 0.1347 & 0.9992   & 1.36      & 95.95  & 100.00 \\
                        & KYTI & 0.1366 & 0.9993   & 1.39      & 95.53  & 100.00 \\
                        & ALAS & 0.1559 & 0.9994   & 1.44      & 94.93  & 99.83 \\
\hline
\multirow{3}{*}{Global} & TIXG & 0.4989 & 0.9695   & 3.54      & 46.49  & 94.85 \\
                        & GODE & 0.2596 & 0.9981   & 3.45      & 79.01  & 99.62 \\
                        & MAYG & 0.6014 & 0.9960   & 5.78      & 45.78  & 90.91 \\
\hline
\end{tabular}
\end{table*}

% \begin{table}[!ht]
% \caption{
% Simulated data
% }
% \label{table:simu-statistical-results}
% \centering
% \begin{tabular}{ccccc}
% \hline
% Region   & RMSE [TECU] & R2      & QA03 [\%] & QA10 [\%] \\ \hline
% USA      & 0.1745      & 0.9990  & 92.01     & 99.88  \\
% Global   & 0.4192      & 0.9918  & 53.74     & 97.39  \\ \hline
% \end{tabular}
% \end{table}

The result shows that the model's predicted STEC RMSE error for the US region is at $\sim 0.15$ based on simulation data as shown in Table II, the disparity between the predicted STEC and true STEC is very trivial, and the differences in latitudes and longitudes of the stations in the US region are negligible that can be regarded as a quasi-baseline. Therefore, only the simulated STEC results of global stations are presented below.

Fig.~\ref{fig:Global-simu-rays} represents the predicted STEC of single-station, single-satellite (G08) within three days for the cutoff elevation angel $15^o$ based on global 30-day simulated dataset 
for three test stations (TIXG, GODE, MAYG). Based on the simulation STEC data, the overall trend of single station single satellite within 30 days of single satellite is relatively smooth and periodic. The simulated results show that the predicted results within June 28-30 in the red curve in good agreement with the simulation true STEC values in blue, which demonstrates that the model learns the variability of the STEC data with time at different regimes, indicating good generalisation based on the Nequick2 simulated data.

The simulation data prediction results for the USA and global region data are summarized in Table II. In the North American region, the CLRE station and KYTI station achieved results with RMSE less than 0.14 TECU. For the two stations, the QA03 value is higher than 95\% and QA10 of 100\%. On the other hand, the ALAS station exhibited relatively lower accuracy with an RMSE of 0.15 TECU, $R^2$ value of 0.9994, QA03 of 94.93\%, and QA10 of 99.83\%.
The reason for the degraded performance at ALAS can be attributed to its location in the southern region of the North American region with fewer stations available. The global MAPE $ 3.5\% \sim 5.7\%$ is higher than that in the USA simulated dataset $ 1.3\% \sim 1.4\%$.

%% Figure 9-10%%%%%%%%%%%%%%%%%%%

\begin{figure*}[!htbp]
	\centering
    \subfigure[Global observation dataset predicted STEC of single-station between June 1-30]{
    \begin{minipage}[b]{1\linewidth}
			\includegraphics[width=6.8in]{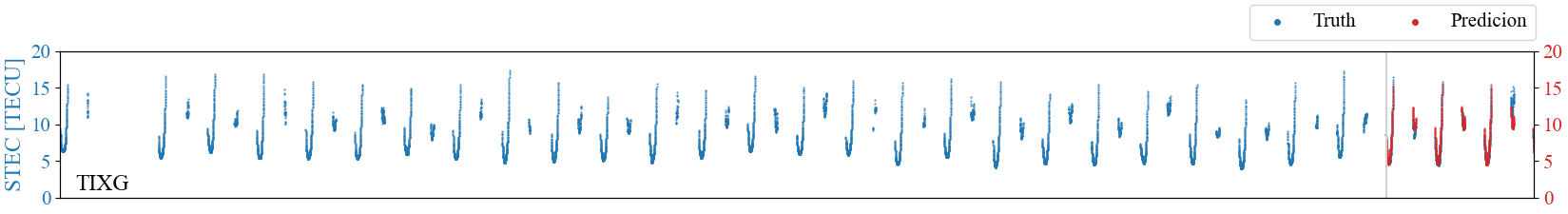}\\
			\includegraphics[width=6.8in]{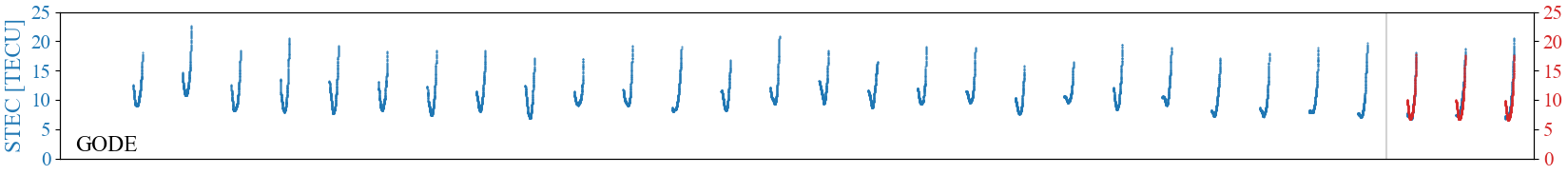}\\
			\includegraphics[width=6.8in]{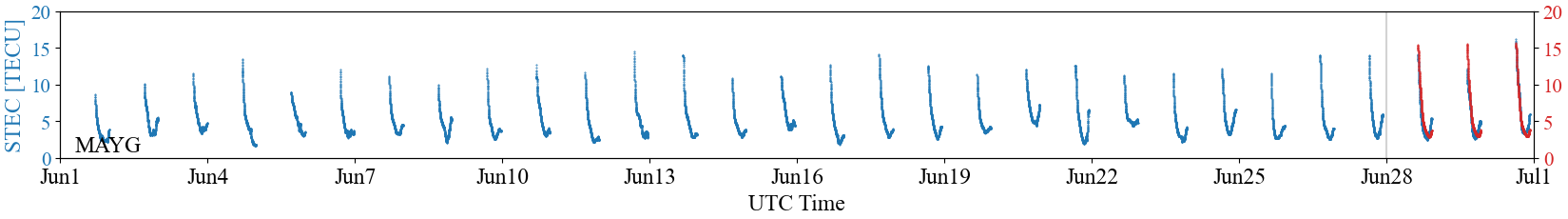}\\
			\vspace{-0.6cm}
   \end{minipage}
	\centering
    }\\
    \subfigure[Global observation dataset predicted STEC of single-station between June 28-30 (only test comparison)]{
	\centering
    \begin{minipage}[b]{1\linewidth}
			\includegraphics[width=6.8in]{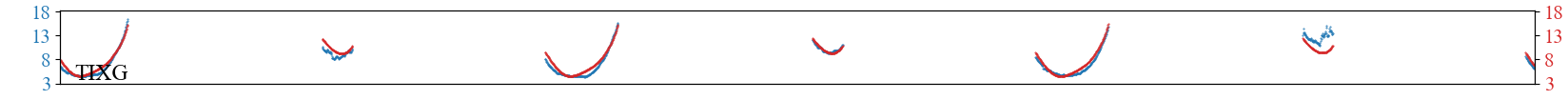}\\
			\includegraphics[width=6.8in]{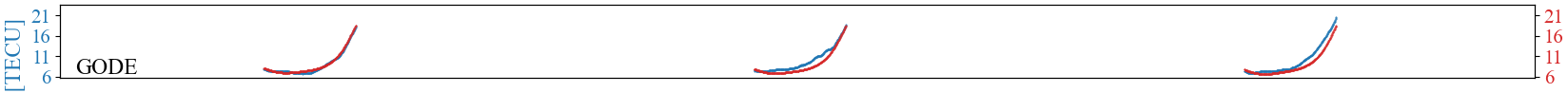}\\
			\includegraphics[width=6.8in]{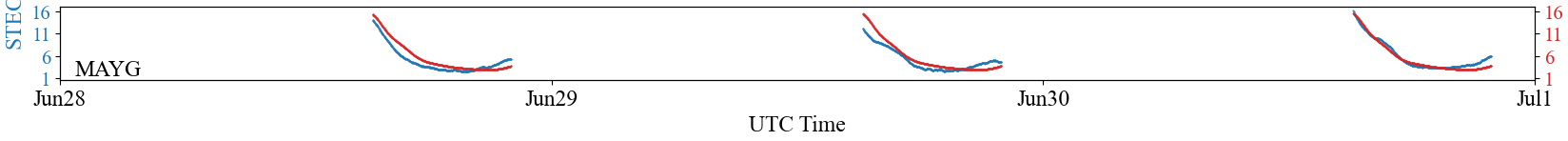}\\
			\vspace{-0.6cm}
   \end{minipage}
	\centering
	\vspace{-0.2cm}}
    \caption{Global observation dataset predicted STEC of single-station, single-satellite (G08) for three test stations. The blue dots represents the observation, while the red dots represents the predicted STEC values in quiet periods.}
    \label{fig:Global-obs-rays}
    \vspace{-0.6cm}
\end{figure*}

\begin{figure*}[!htbp]
	\centering
	\subfigure[TIXG]{
		\begin{minipage}[c]{0.32\linewidth}
			\centering
			\includegraphics[width=2in]{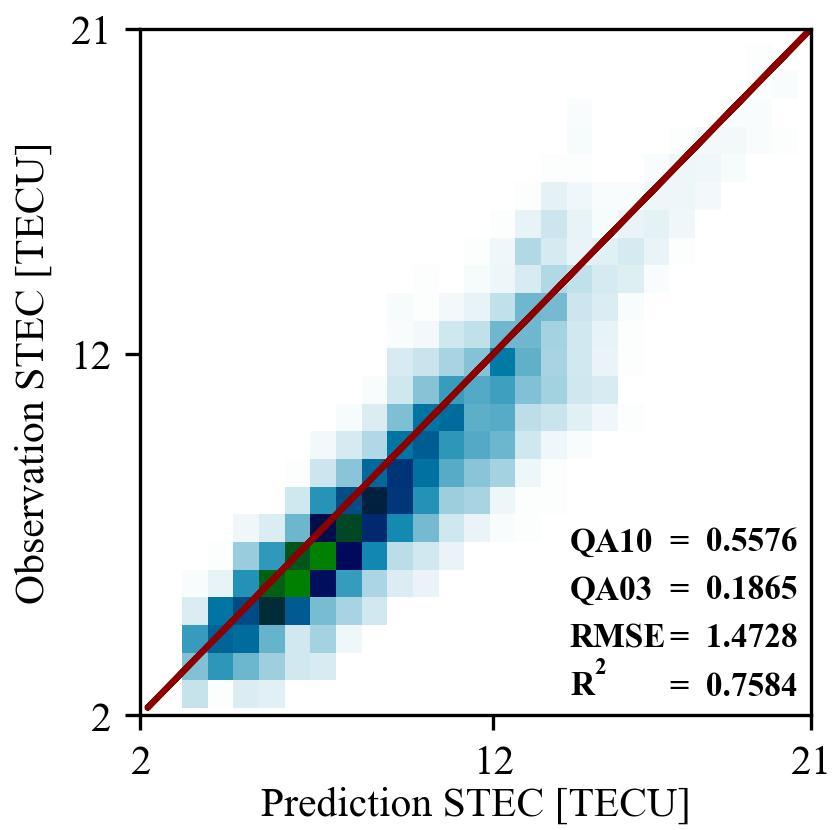}\\
			\vspace{0.02cm}
			\centering
			\includegraphics[width=2.3in]{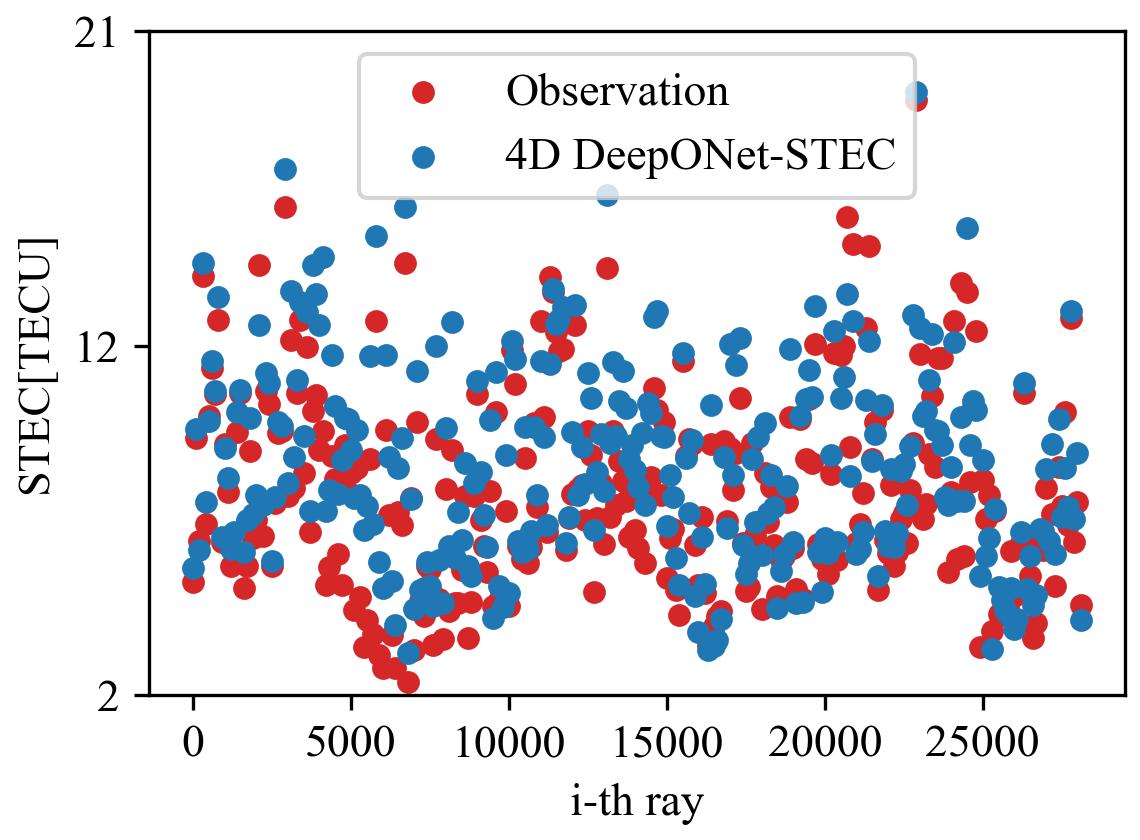}\\
			\vspace{0.02cm}
		\end{minipage}%
	}%
	\subfigure[GODE]{
		\begin{minipage}[c]{0.32\linewidth}
			\centering
			\includegraphics[width=2in]{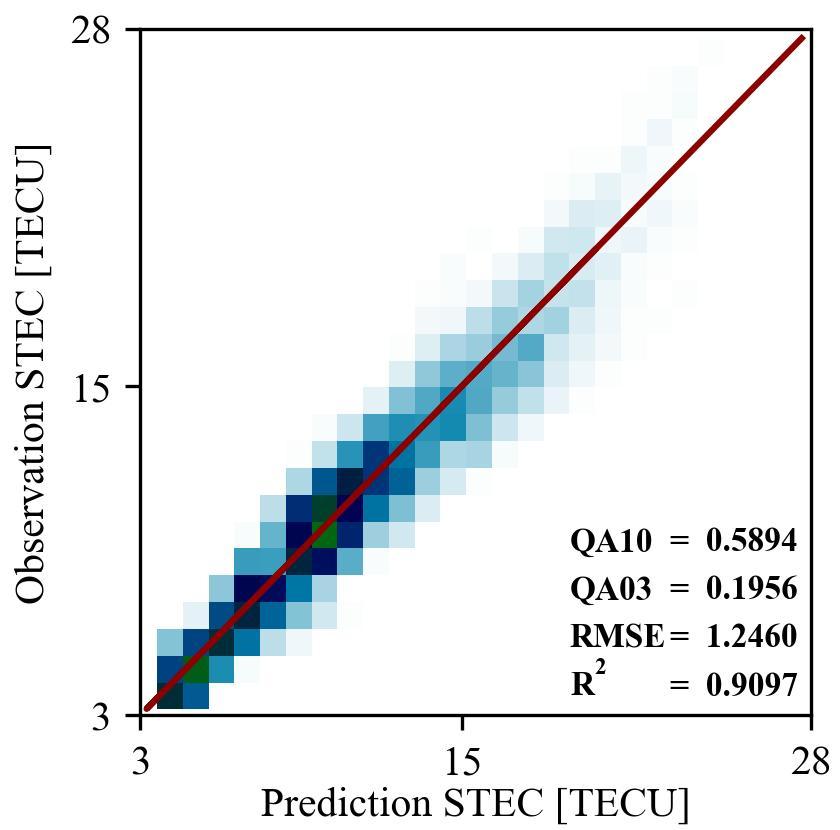}\\
			\vspace{0.02cm}
			\centering
			\includegraphics[width=2.3in]{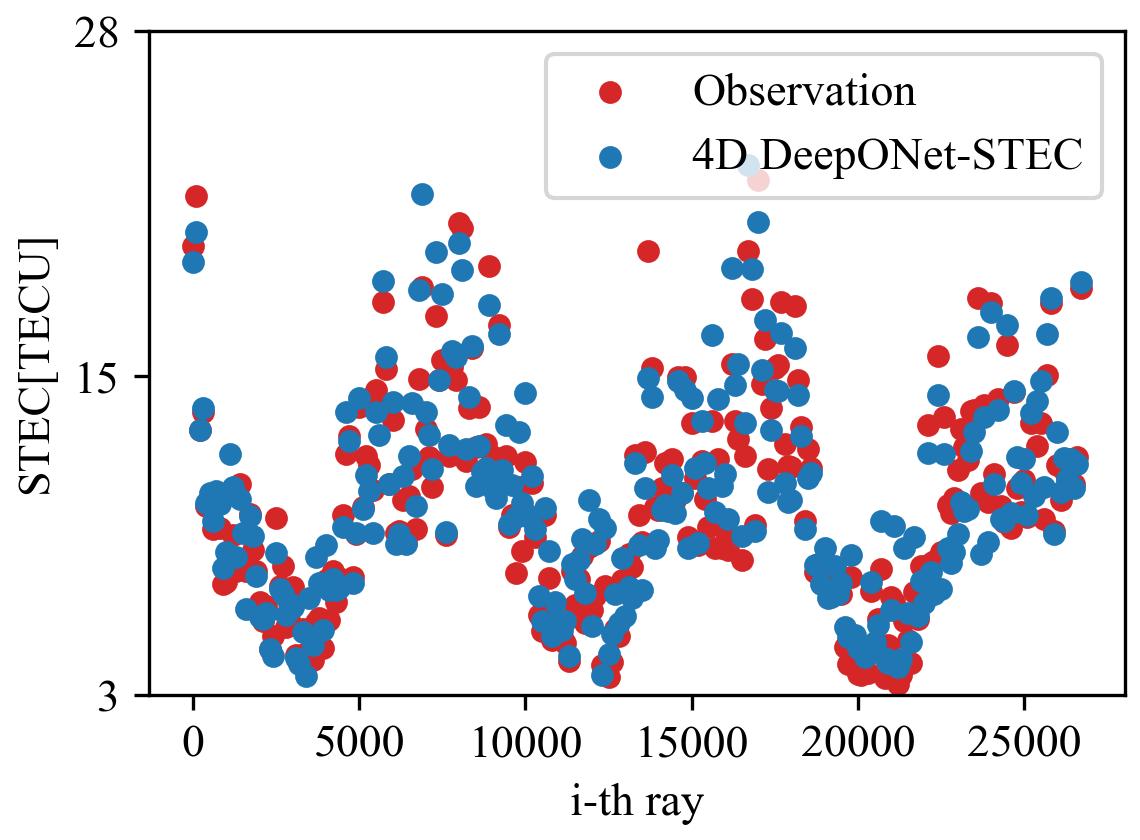}\\
			\vspace{0.02cm}
		\end{minipage}%
	}%
	\subfigure[MAYG]{
		\begin{minipage}[c]{0.34\linewidth}
			\centering
			\includegraphics[width=2.57in]{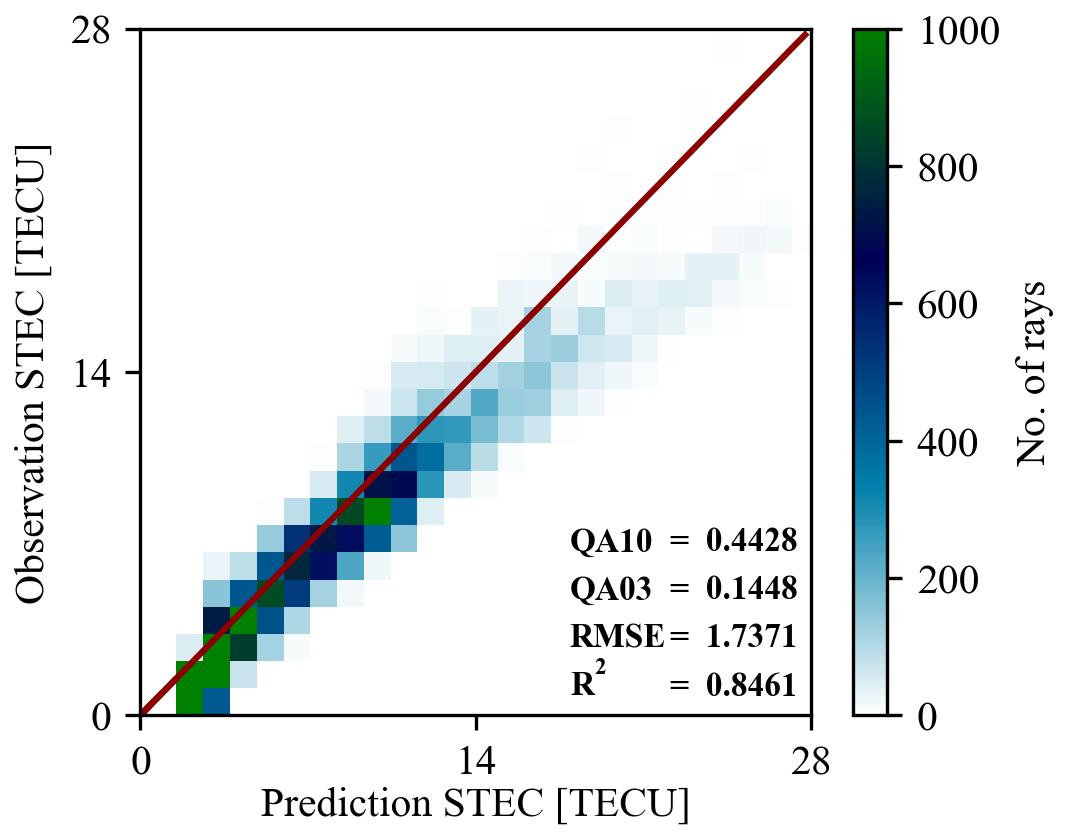}\\
			\vspace{0.02cm}
			\centering
			\includegraphics[width=2.3in]{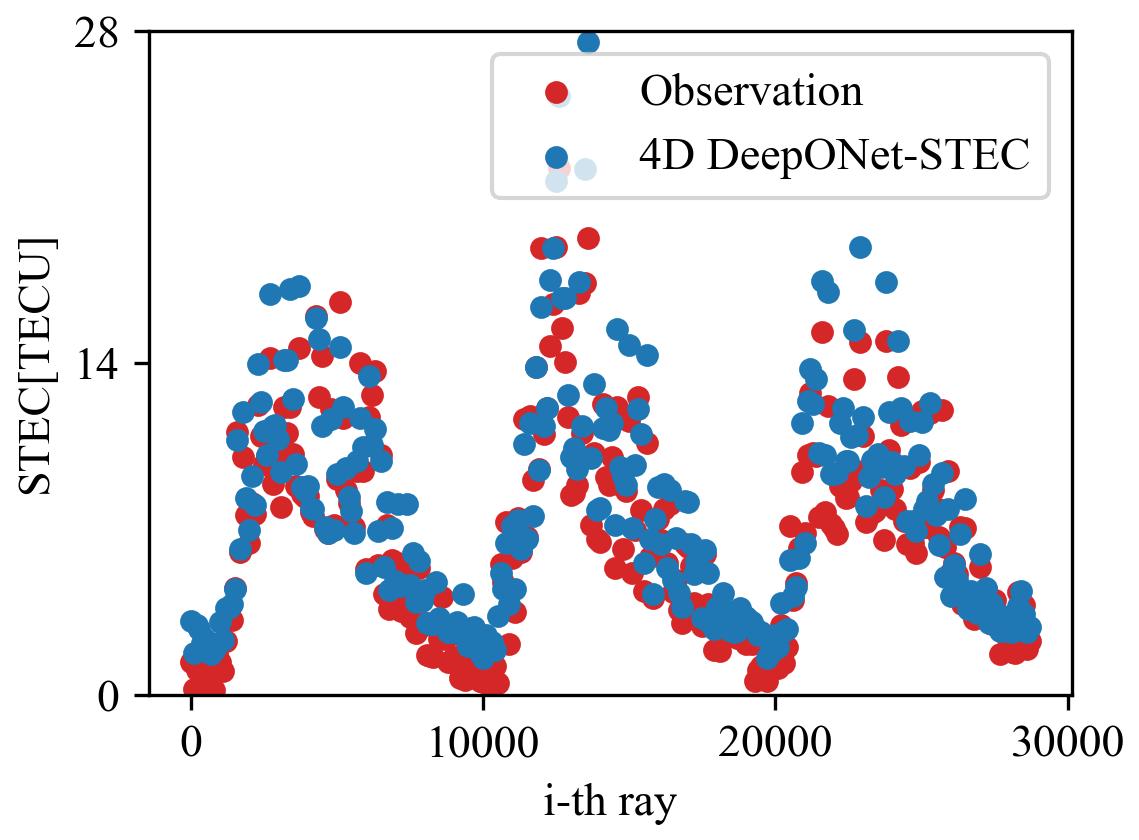}\\
			\vspace{0.02cm}
		\end{minipage}%
	}%
	\centering
	\vspace{-0.2cm}
    \caption{Up: Global observation data prediction result by the DeepONet-STEC model, where the color counts the number of rays denoting the correlation between the observation and predicted value for all rays within the prediction period 3 days. Down: the model DeepONet-STEC prediction result comparison with observation within 3 days in  periods.}
    \label{fig:global-pred-obs}
\end{figure*}

Fig.~\ref{fig:global-pred-simu}. presents a correlation between the model predicted STEC and simulated true STEC value within three days. In the Global region, the GODE station exhibited high accuracy with an RMSE of 0.2596 TECU, $R^2$ value of 0.9981, QA03 of 79.01\%, and QA10 of 99.62\%. The performance can be attributed to its location, which falls within the densely populated region of stations. The availability of nearby stations contributes to improved data coverage and more precise predictions. In contrast, the TIXG station and MAYG station displayed relatively higher RMSE values of 0.4989 TECU and 0.6014 TECU. The limited data coverage near the equator and at high latitudes leads to low accuracy and high uncertainty. For the global simulation data, with a wider area and sparser station distribution, the model still performs well, which demonstrates that the model has good generalization.

The prediction result of the 4D DeepONet-STEC model based on the simulation data clearly demonstrates its superiority in accurately predicting the STEC variations. The selected stations within the North American and global regions have consistently exhibited high accuracy, with RMSE values within 0.7 TECU and $R^2$ indices exceeding 0.96.

\subsection{Observation Data Results in Quiet Periods}
\begin{table*}[!htbp]
\caption{Statistical prediction results of the observation data based on the test stations in quiet periods}
\label{table::obs-statistical-results}
\centering
\begin{tabular}{lcccccc}
\hline
Region & station        & RMSE [TECU]   & $R^2$   & MAPE [\%] & QA03 [\%] & QA10 [\%] \\
\hline
\multirow{3}{*}{USA}    & CLRE & 1.3322 & 0.8894  & 14.23     & 23.50  & 63.93 \\
                        & KYTI & 1.4220 & 0.8890  & 11.13     & 24.36  & 62.53 \\
                        & ALAS & 1.3964 & 0.9073  & 9.33     & 23.21  & 61.80 \\
\hline
\multirow{3}{*}{Global} & TIXG & 1.4728 & 0.7584  & 16.20     & 18.65  & 55.76 \\
                        & GODE & 1.2460 & 0.9097  & 9.79     & 19.56  & 58.94 \\
                        & MAYG & 1.7371 & 0.8461  & 26.64     & 14.48  & 44.28 \\
\hline
\end{tabular}
\end{table*}
First of all, we present the DeepONet-STEC prediction results for the US CORS observation data. Fig.~\ref{fig:us-obs-rays}. represents the 30-day observation STEC value versus the 28-30-day model-predicted STEC value for each station in the single-satellite (G08) case based on the US CORS observation data. In Fig.~\ref{fig:usa-pred-obs}, the overall change of the observation STEC data at single station for all satellite rays within 30 days of single satellite varies periodically each day. The  unsmooth and fluctuating characteristics of the observation STEC data compared to the Nequick2 simulation STEC data are related to the quality of the station observation data, as well as  the complex real-time conditions of the ionosphere.  For different stations, the model predictions fluctuate differently, while for multiple days of data from the same station, there is a pattern of variation in the model predictions, but the model performs slightly worse on individual data with large values. The results in the USA region suggest that the model can learn the disparity between stations and the patterns of the change in the STEC data over time.

Secondly, we present the DeepONet-STEC prediction results for the Global observation data.
Fig.~\ref{fig:Global-obs-rays}. represents the 30-day observation STEC value versus the 28-30-day model-predicted STEC value for each station in the single-satellite (G08) case in the global region. From this scatterplot in Fig.~\ref{fig:global-pred-obs} , the observation STEC data, the overall change is generally periodic with large fluctuations, while the STEC value varies at different stations at the same day. The large latitude and longitude span and the uneven distribution of the global stations lead to a more complex dataset, which increases the difficulty of the model learning. The results show that in many test cases, especially for small values, the red dots overlap with the blue dots, indicating that the model-predicted STEC values do not differ much from the observed STEC values. The model still performs slightly worse for individual data with large values. For the three test stations at sparse distributed locations, the model predictions are still within acceptable limits.

The observation data prediction results for the two regions are summarized in Table III. Fig.~\ref{fig:usa-pred-obs}. present a comparison between the prediction value obtained from the model and the observation data in the USA region. In the USA region, the performance of the three stations is basically similar, with RMSE values around 1.3-1.4 TECU, $R^2$ around 0.89, QA03 around 23\%, and QA10 around 62\%. From the single-station scatter plots, it can be noted that the observation data are not as smooth as the simulation data, and there is a certain degree of fluctuation in STEC value from day to day. The global observation dataset MAPE up to $\sim 26.6\%$ is higher than that in the USA observation dataset $ 9.3\% \sim 14.2\%$.  

The observation date results demonstrate that the model's performance in the USA regional areas is stable. 
%% Figure 11 %%%%%%%%%%%%%%%%%%%
\begin{figure*}[!htbp]
	\centering
		\begin{minipage}[c]{0.32\linewidth}
			\centering
			\includegraphics[width=2.1in]{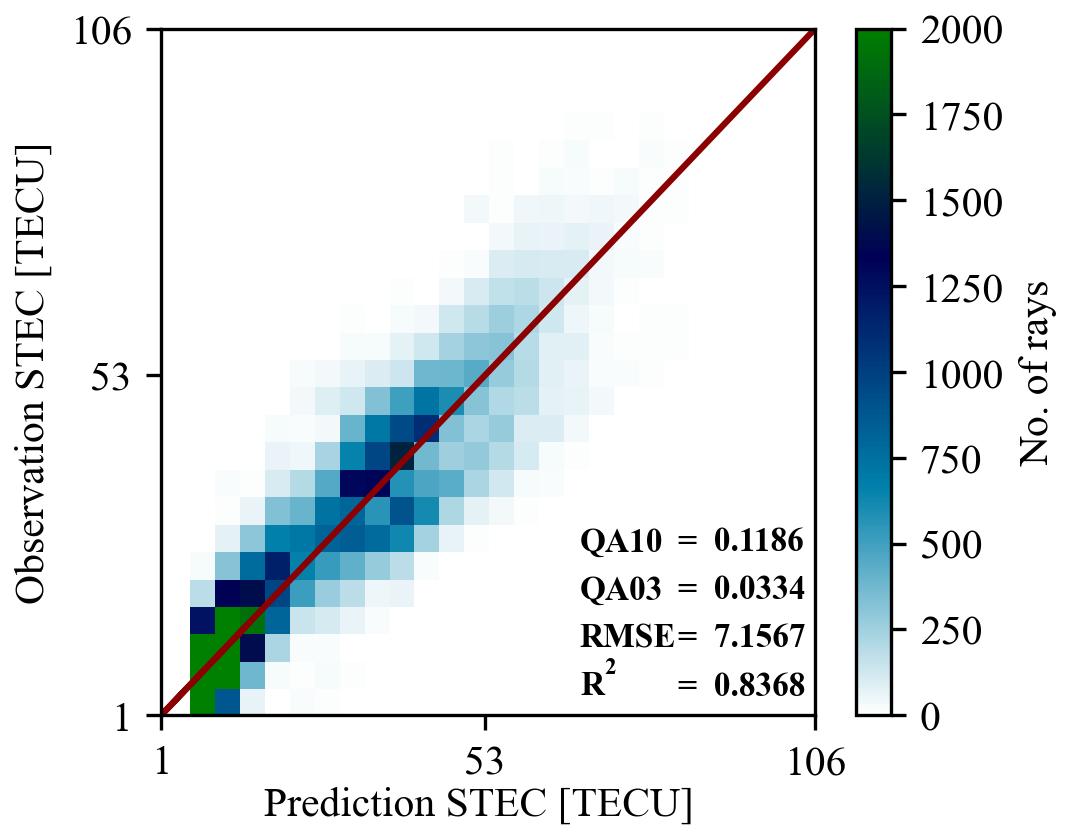}\\
			\vspace{0.02cm}
			\centering
			\includegraphics[width=2.1in]{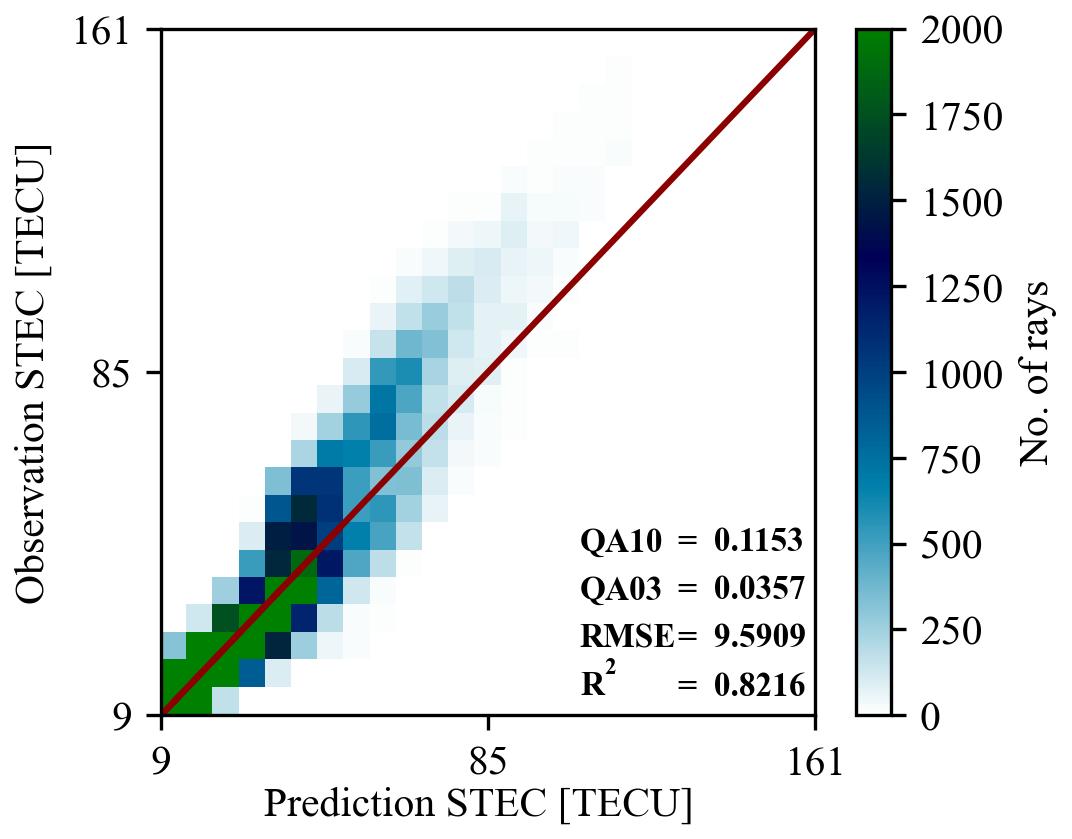}\\
			\vspace{0.02cm}
		\end{minipage}%
		\begin{minipage}[c]{0.68\linewidth}
			\centering
			\includegraphics[width=4.5in]{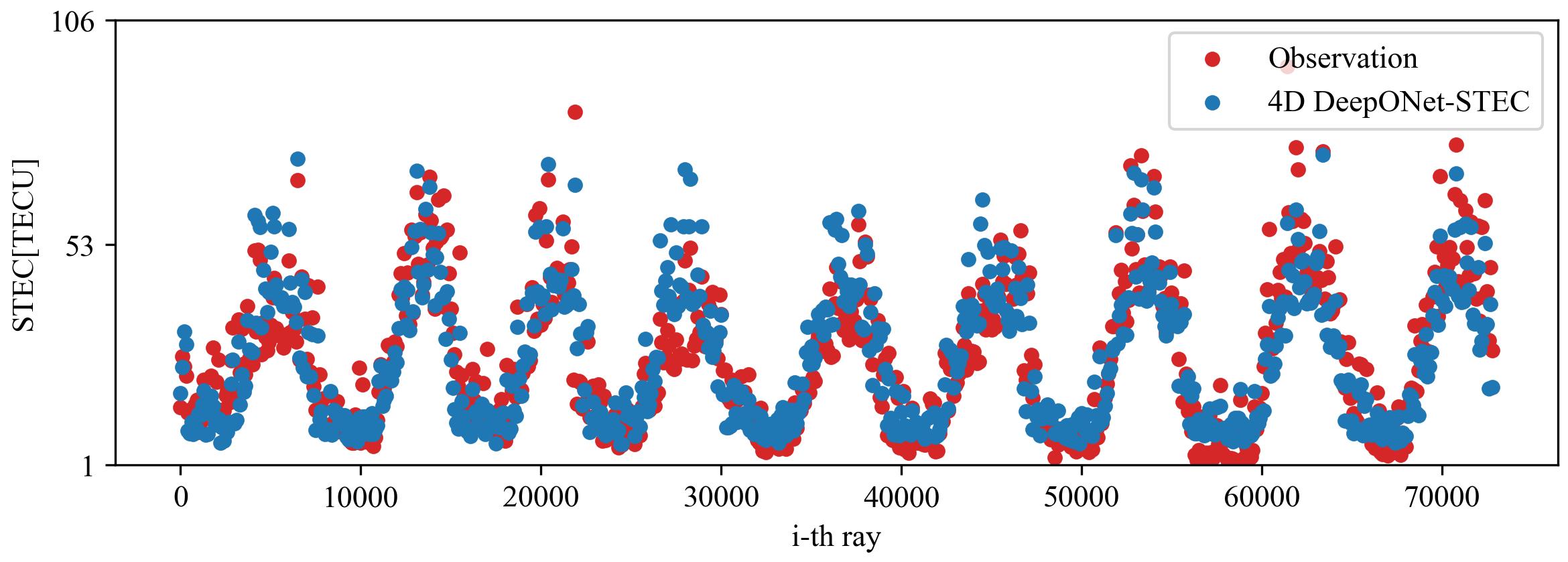}\\
			\vspace{0.02cm}
			\centering
			\includegraphics[width=4.5in]{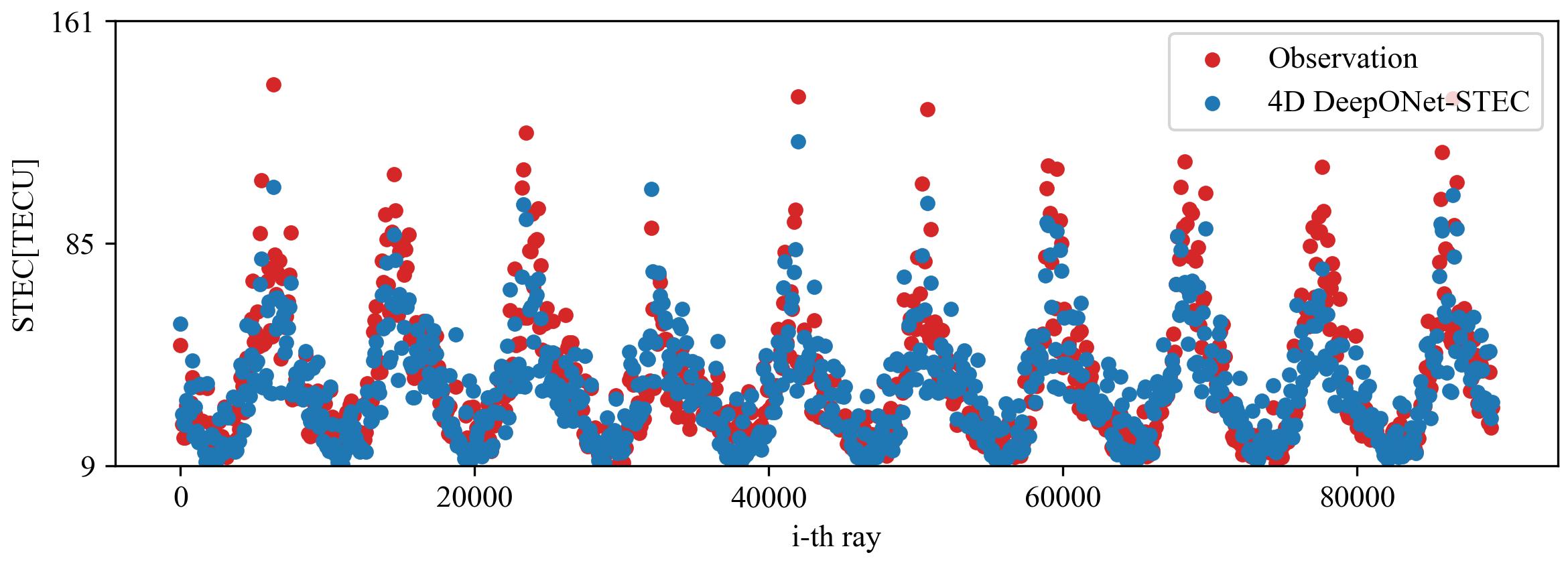}\\
			\vspace{0.02cm}
		\end{minipage}%
	\centering
	\vspace{-0.2cm}
    \caption{Global prediction results during the validation time span from March 12 to March 21, 2023 using the magnetic storm dataset at two validation stations. Up: SCH2 (Northern hemisphere); Down: FALK (Southern hemisphere).}
    \label{fig:valid}
\end{figure*}

%% Figure 12 %%%%%%%%%%%%%%%%%%%

\begin{figure*}[!htbp]
	\centering
			\includegraphics[width=6.2in]{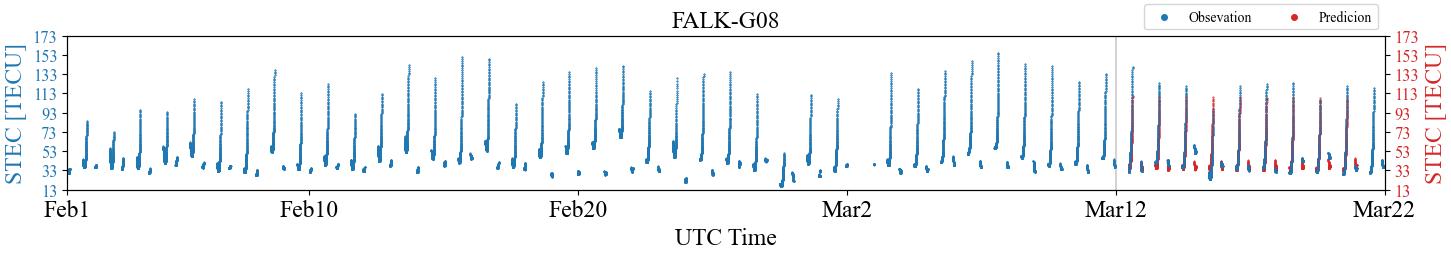}\\
            \includegraphics[width=6.2in]{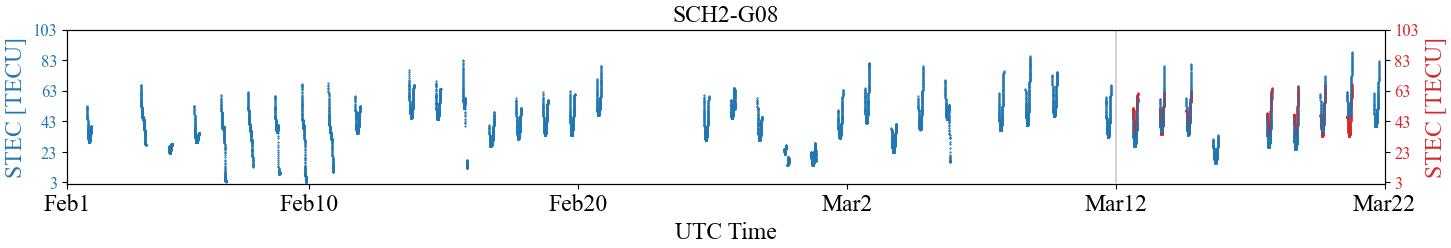}\\
	\centering
	\vspace{-0.2cm}
    \caption{Global prediction results from February 1 to March 21, 2023 at the validation station SCH2 and FALK using the magnetic storm dateset. The blue curve is the observation and the red curve is the predicted STEC value between March 12 to March 21, 2023.}
    \label{fig:valid_ray}
    \vspace{-0.6cm}
\end{figure*}

% CODE 
\begin{figure*}[!htbp]
	\centering
		\begin{minipage}[c]{0.32\linewidth}
			\centering
			\includegraphics[width=2.1in]{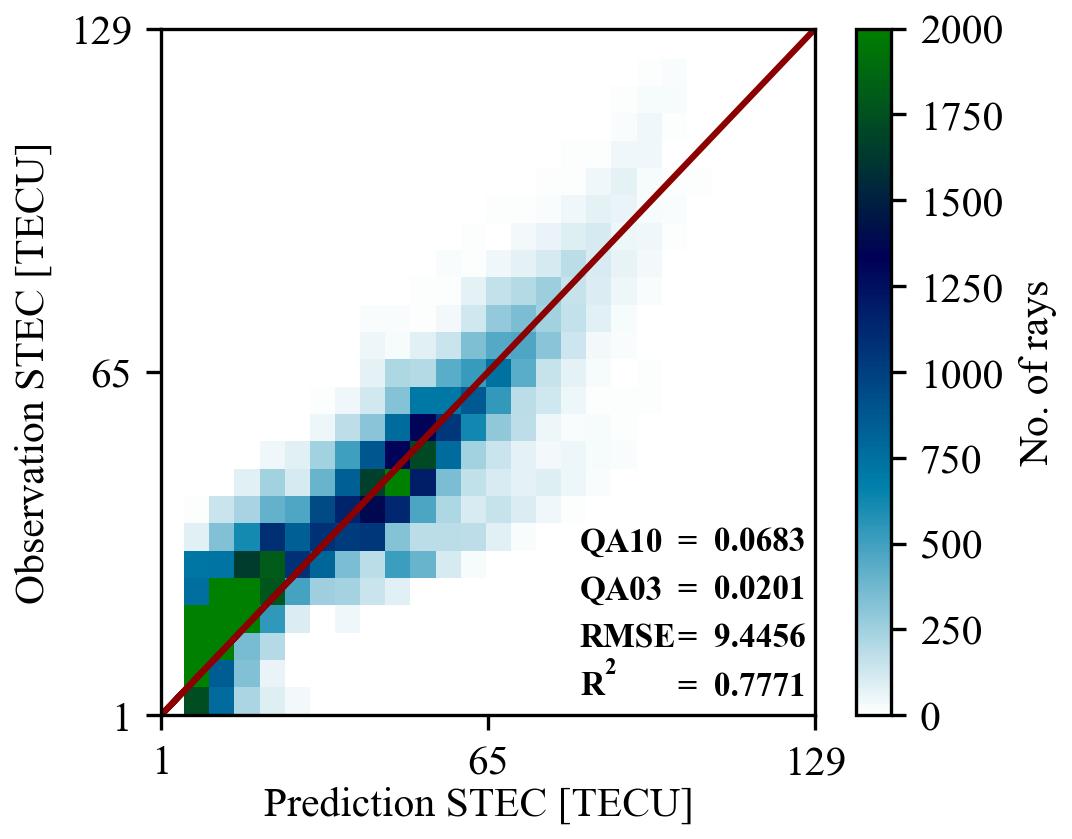}\\
			\vspace{0.02cm}
			\centering
			\includegraphics[width=2.1in]{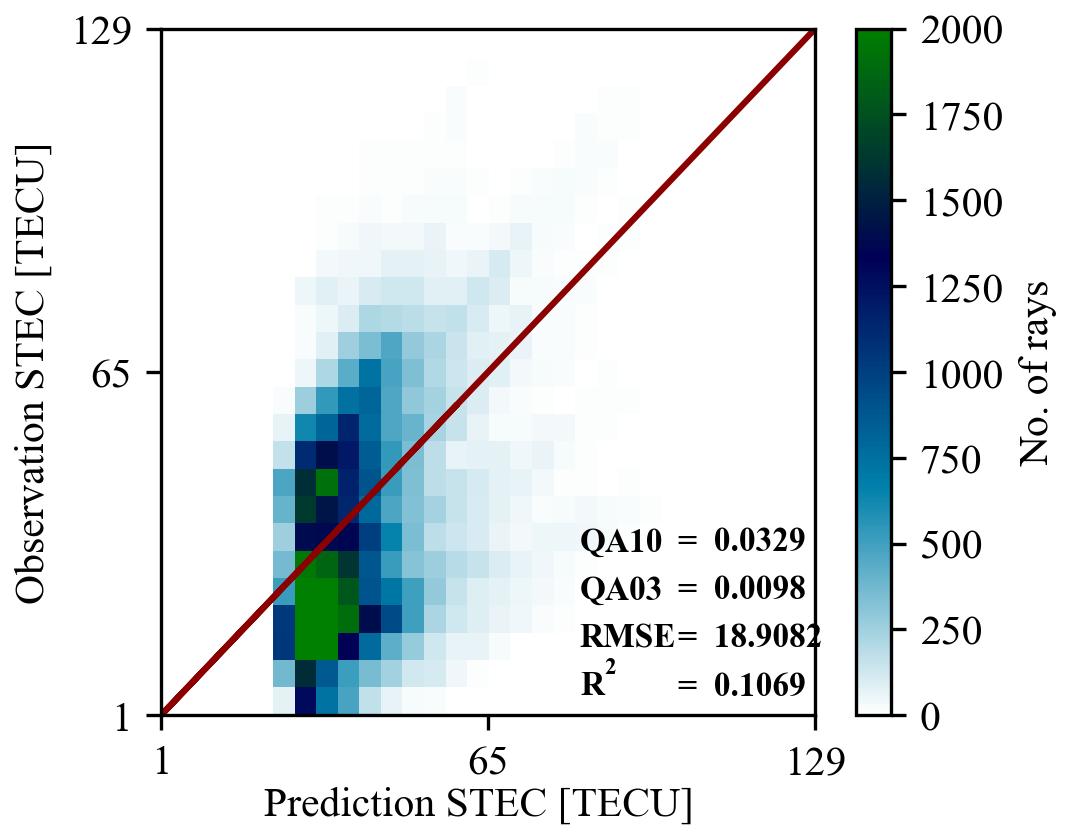}\\
			\vspace{0.02cm}
   		\centering
			\includegraphics[width=2.1in]{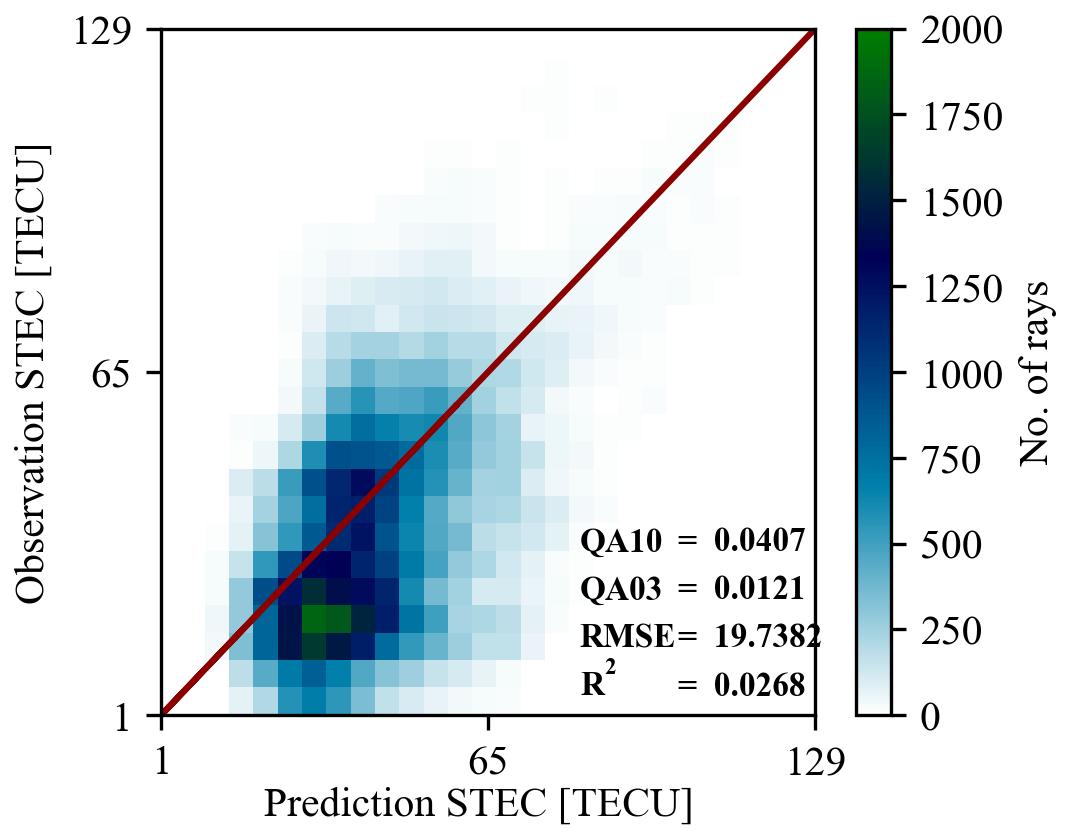}\\
			\vspace{0.02cm}
		\end{minipage}%
		\begin{minipage}[c]{0.68\linewidth}
			\centering
			\includegraphics[width=4.5in]{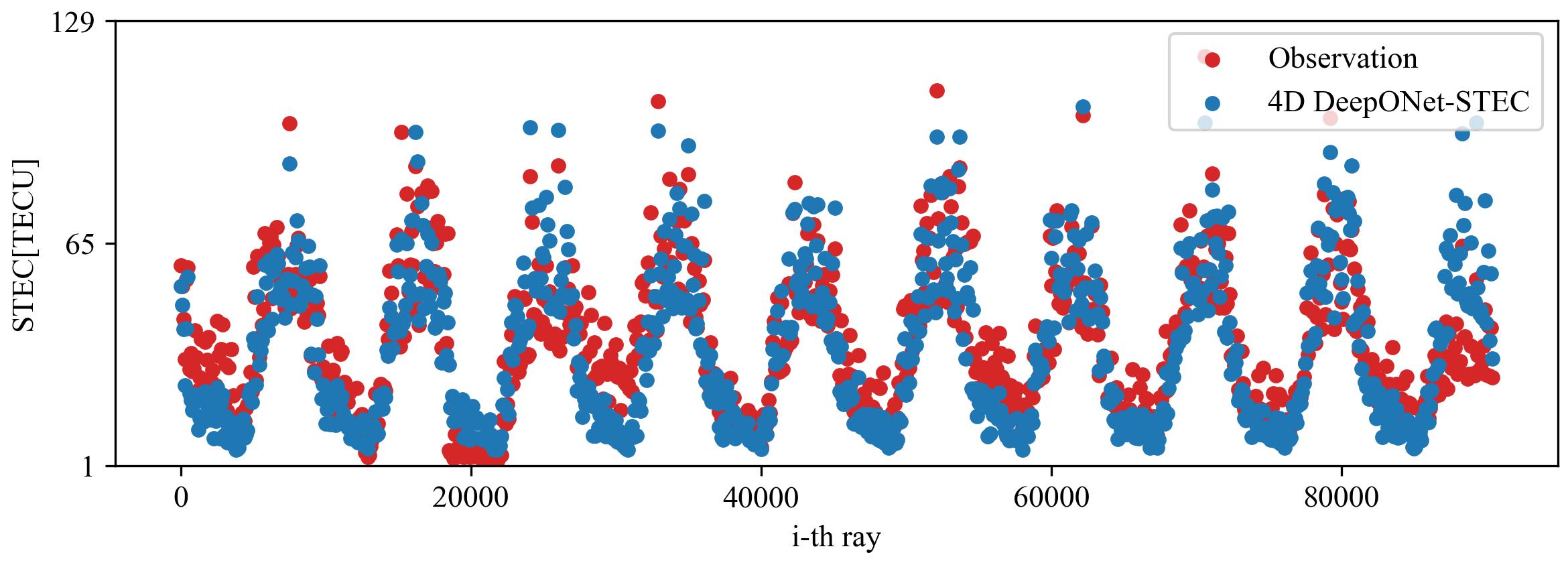}\\
			\vspace{0.02cm}
			\centering
			\includegraphics[width=4.5in]{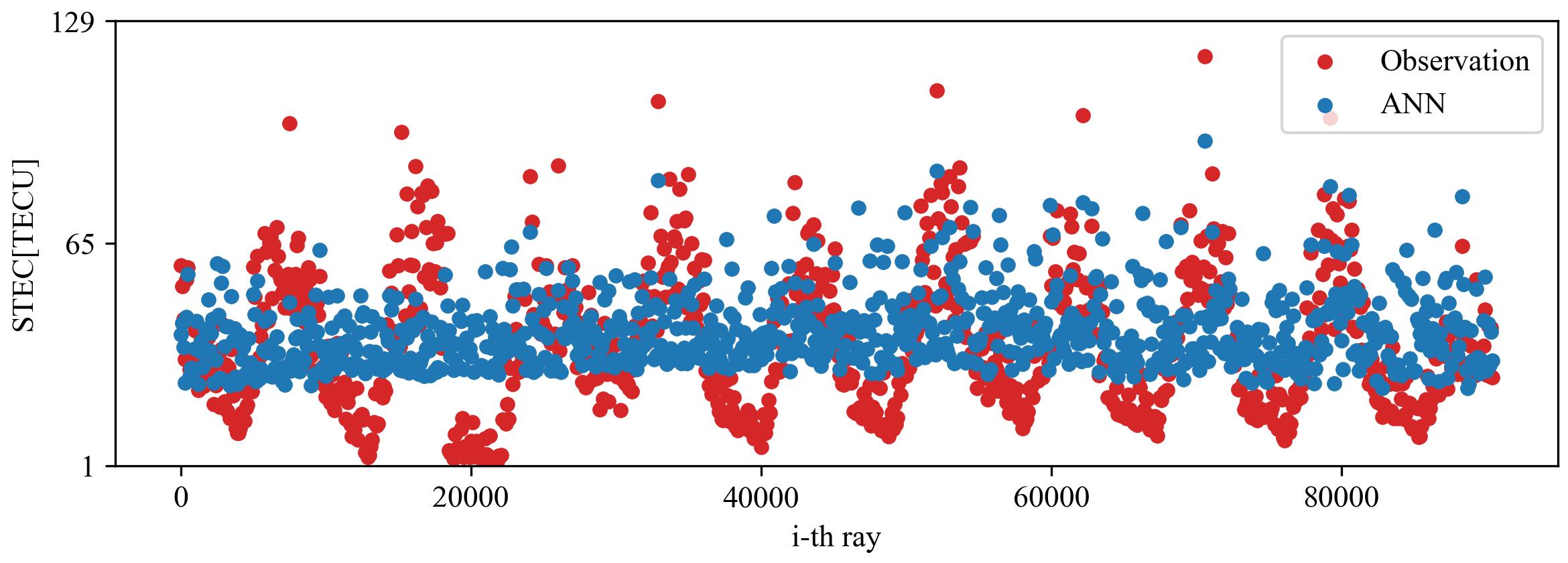}\\
			\vspace{0.02cm}
   		\centering
			\includegraphics[width=4.5in]{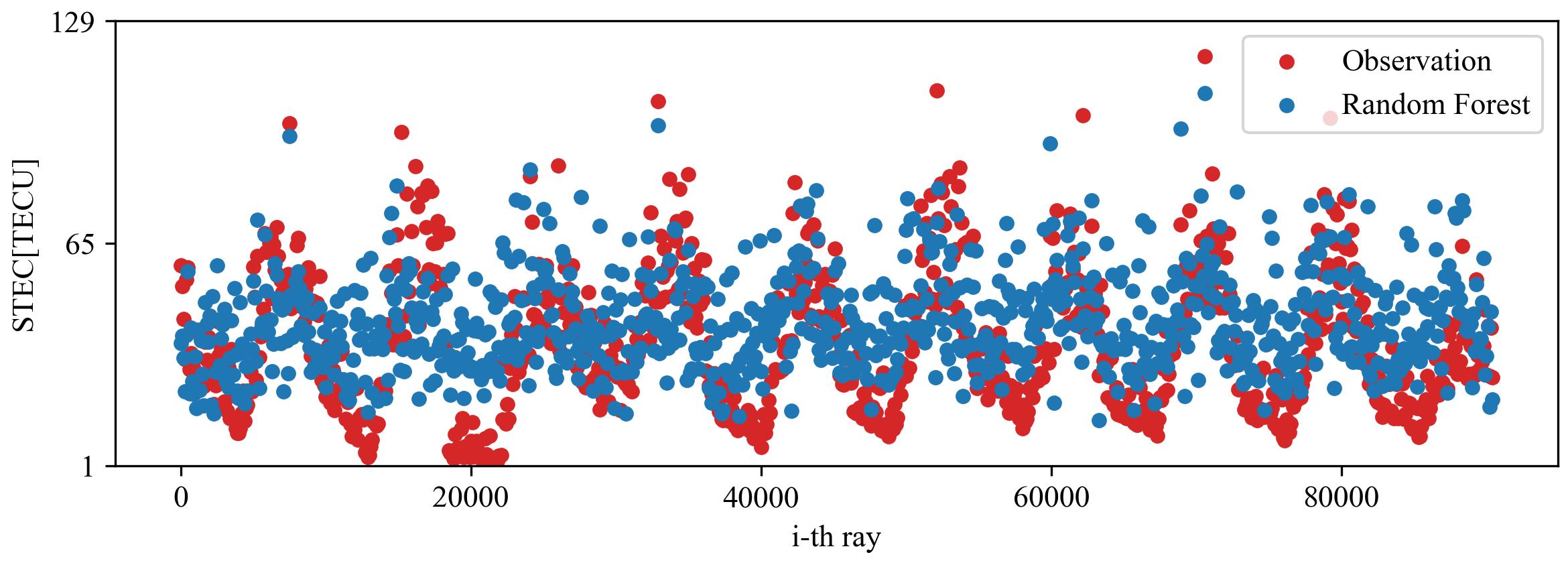}\\
			\vspace{0.02cm}
		\end{minipage}%
	\centering
	\vspace{-0.2cm}
    \caption{Global storm observation data prediction results by three methods during the time span March 12 to 21, 2023 at the test GODE station. Up: 4D DeepONet-STEC model, Middle: ANN model, Down: RF model.}
    \label{fig:gode3method}
\end{figure*}
\begin{figure*}[!htbp]
	\centering
			\includegraphics[width=7in]{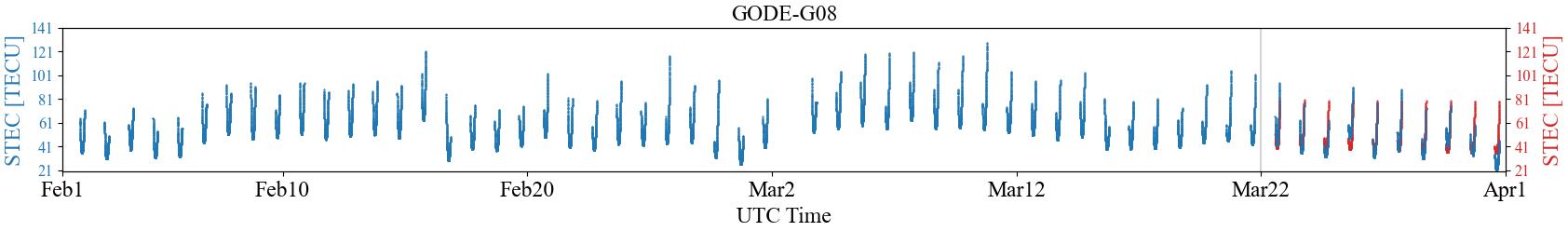}\\
			\vspace{0.02cm}
			\includegraphics[width=7in]{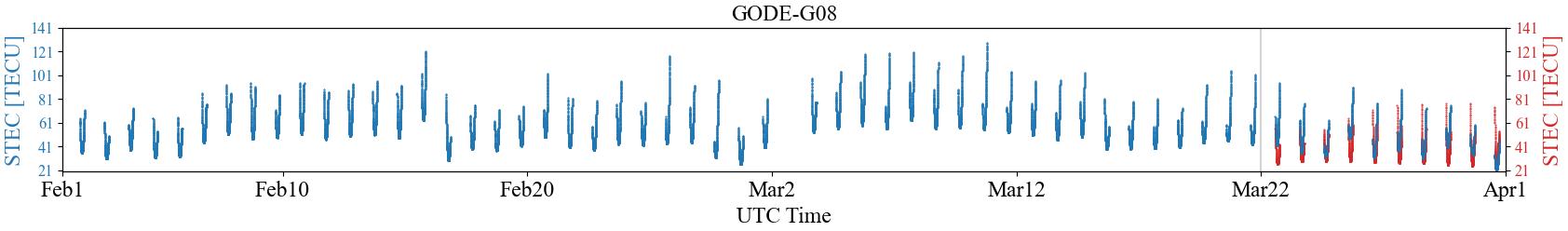}\\
			\vspace{0.02cm}
			\includegraphics[width=7in]{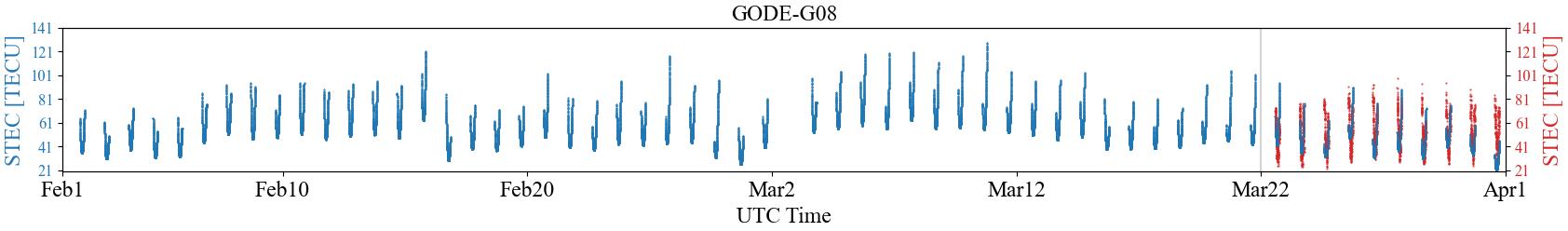}\\
			\vspace{0.02cm}
	\centering
	\vspace{-0.2cm}
    \caption{Predicted STEC for single satellite at the test GODE station using the magnetic storm dataset. Up: 4D DeepONet-STEC model. Middle: ANN model. Down: RF model. The blue curve is the observation and the red curve is the predicted STEC between March 22 and March 31, 2023.}
    \label{fig:3method-gode-rays}
    \vspace{-0.6cm}
\end{figure*}

Fig.~\ref{fig:global-pred-obs}. present a comparison between the predicted results obtained from the DeepONnet-STEC model and observation values of observation data in the global region. In the Global region, the GODE station in the Global region also exhibited high accuracy with an RMSE of 1.2460 TECU, $R^2$ value of 0.9097, even better than small region result. The TIXG station shows good result as small region result, where RMSE value is 1.4728 TECU and $R^2$ value is 0.7584.The MAYG station is slightly underperforming with RMSE values RMSE value of 1.7371 TECU and $R^2$ value of 0.8461. The single-site, single-satellite results can be noticed to show that there are different trends in STEC value changes at different location stations around the global area, and the data are all subject to fluctuations. This indicates that the predictive ability of the model is still credible for the observation data. For the MAYG station, it is noted that the simulated and observed data are different, which can indicate that there exist disparity between the model and observed data, which further validates the robustness of the model.

The prediction result of observation data of the 4D DeepONet-STEC model performance also shows high accuracy in STEC prediction problem. The selected stations within the North American and Global regions have consistently exhibited high accuracy, with RMSE values within 1.8 TECU and $R^2$ indices exceeding 0.75. Overall, due to the complexity of the observation data, the performance of the model on observation dataset is not as good as that on simulation dataset, but the value error of the model prediction results is generally small, which proves that the model can learn the whole range of STEC temporal and spatial patterns of change in a small observation dataset.

%In Fig.11, we counted the MAPE curves for the predicted three-day dataset at 30-minute intervals. Overall, for individual stations, the trend of MAPE is correlated with the trend of STEC values due to the fact STECs at different moments of the day come from different satellites. Our model is not time-series, but fits a function that outputs STEC given an arbitrary moment ray. At the same time, the quality of the observation data is fickle, MAPE does not show a clear temporal pattern.

% TIXG 
\begin{figure*}[!htbp]
	\centering
		\begin{minipage}[c]{0.32\linewidth}
			\centering
			\includegraphics[width=2.1in]{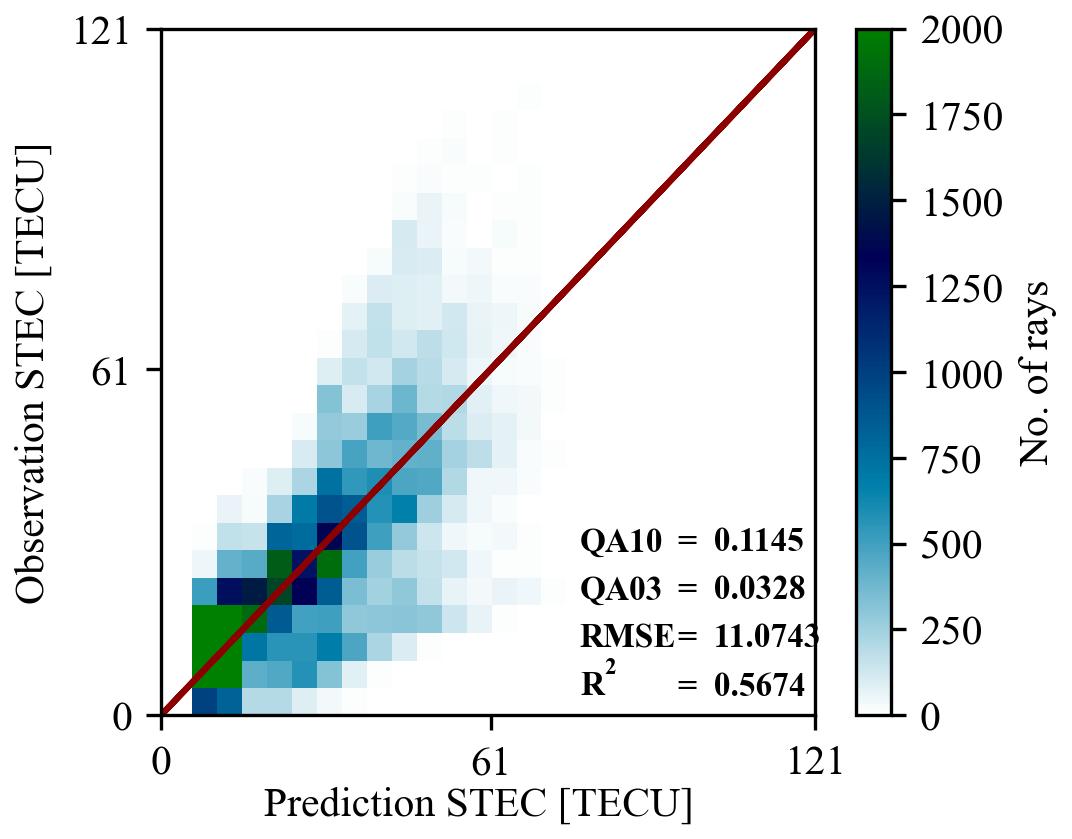}\\
			\vspace{0.02cm}
			\centering
			\includegraphics[width=2.1in]{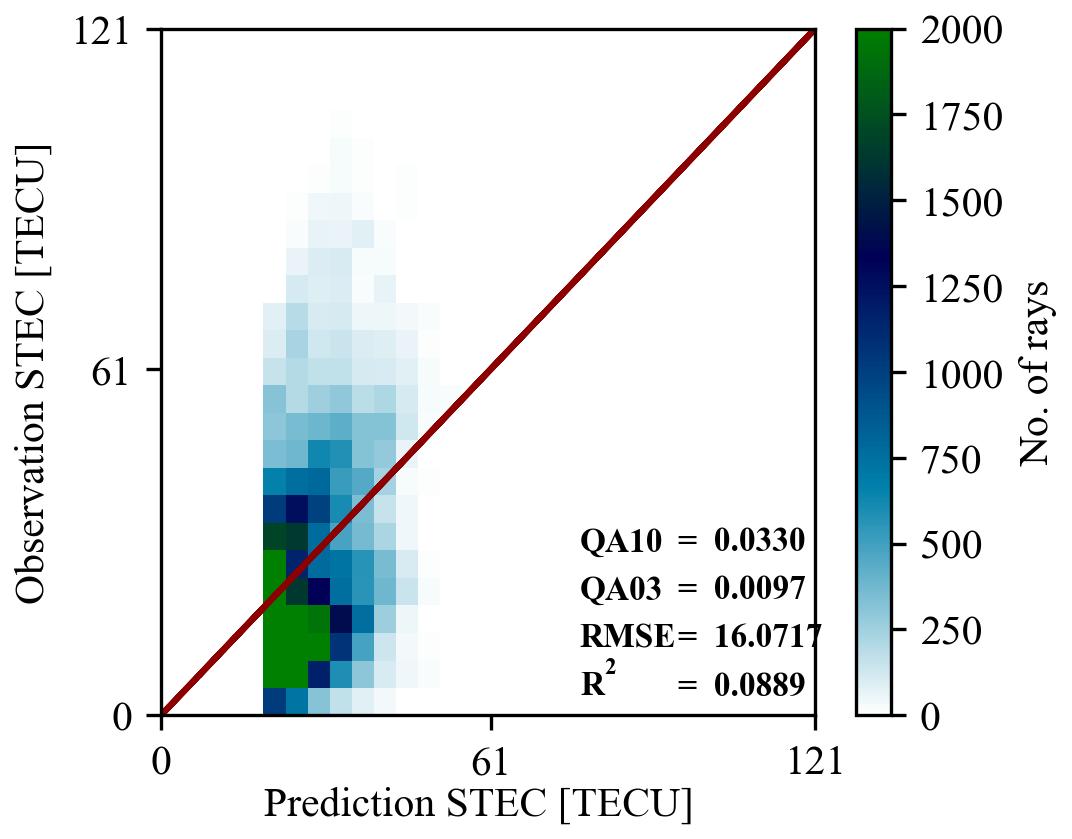}\\
			\vspace{0.02cm}
   		\centering
			\includegraphics[width=2.1in]{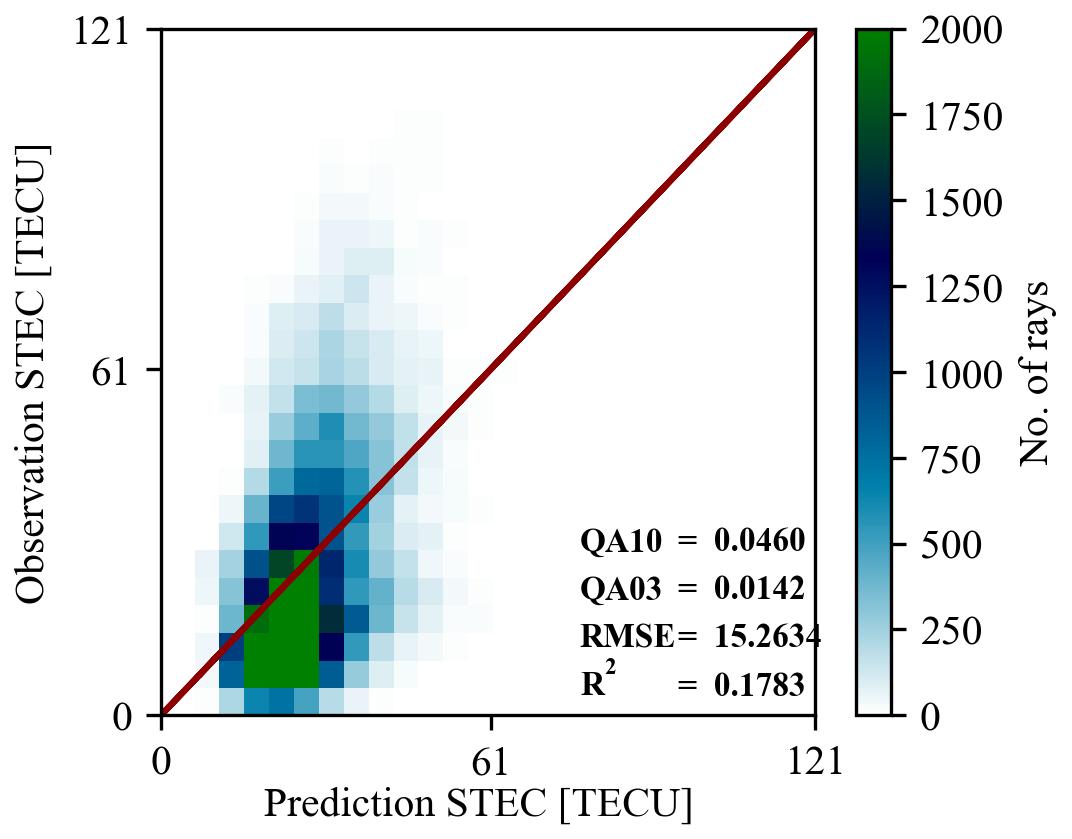}\\
			\vspace{0.02cm}
		\end{minipage}%
		\begin{minipage}[c]{0.67\linewidth}
			\centering
			\includegraphics[width=4.5in]{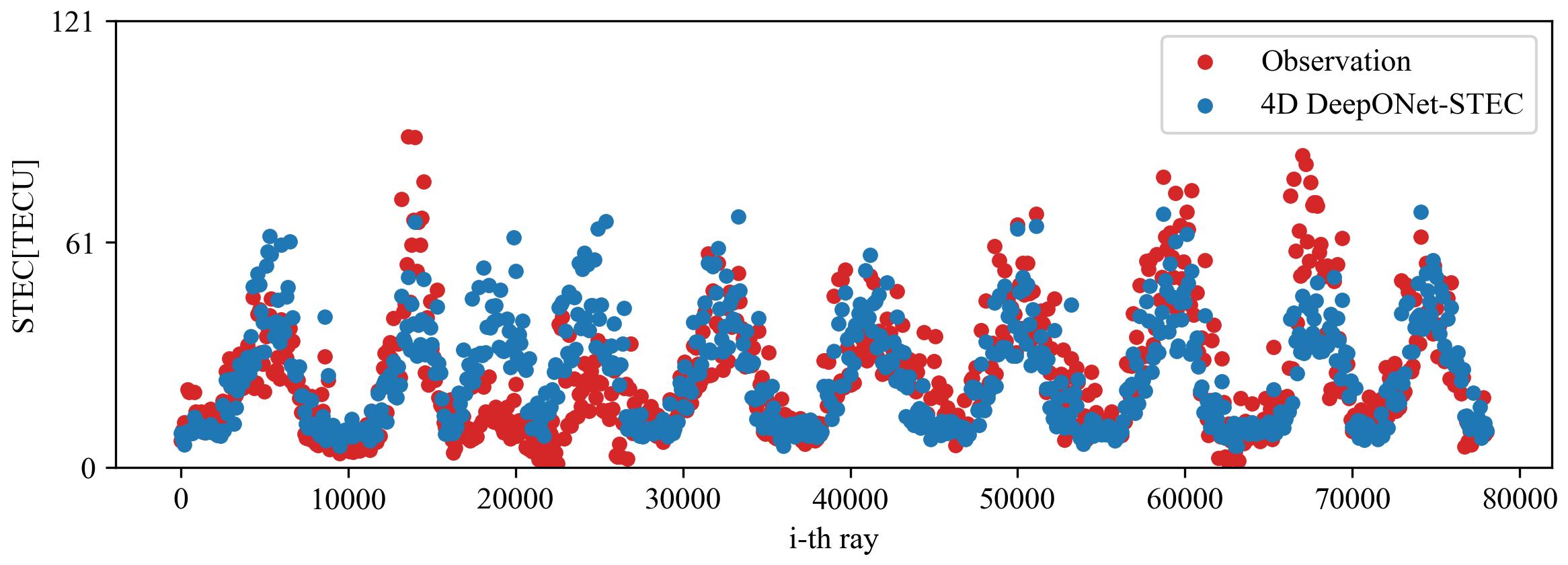}\\
			\vspace{0.02cm}
			\centering
			\includegraphics[width=4.5in]{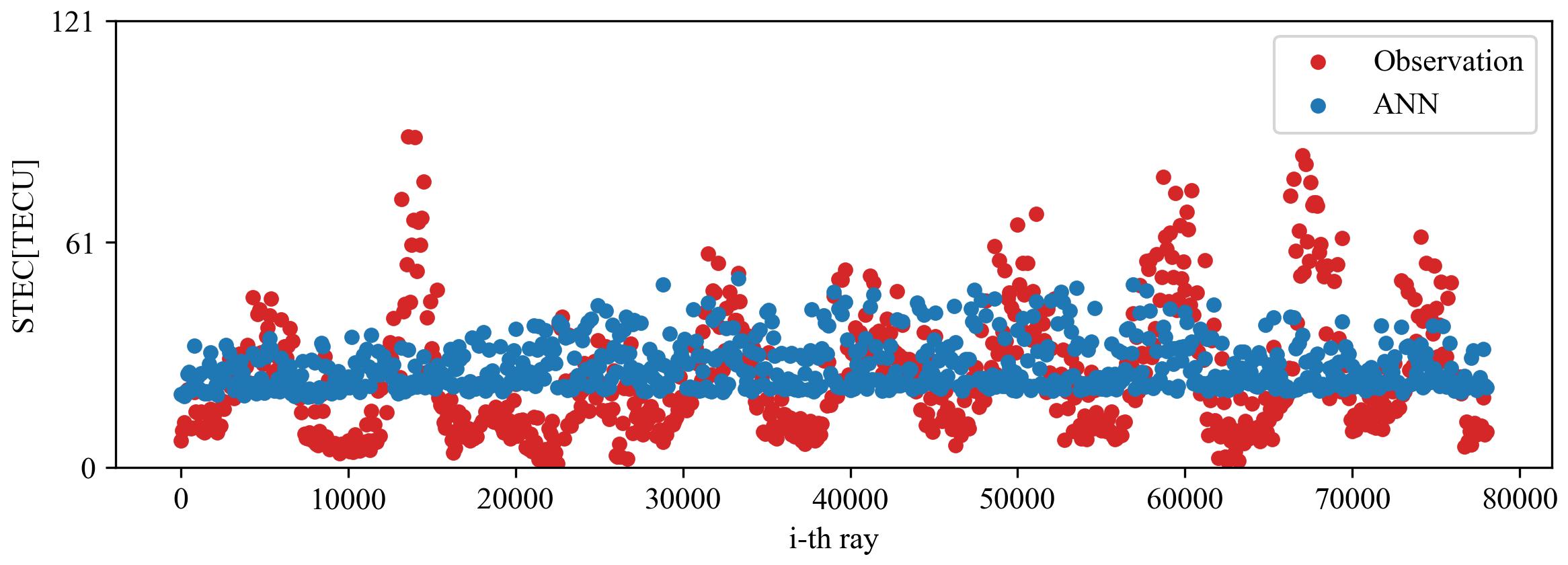}\\
			\vspace{0.02cm}
   		\centering
			\includegraphics[width=4.5in]{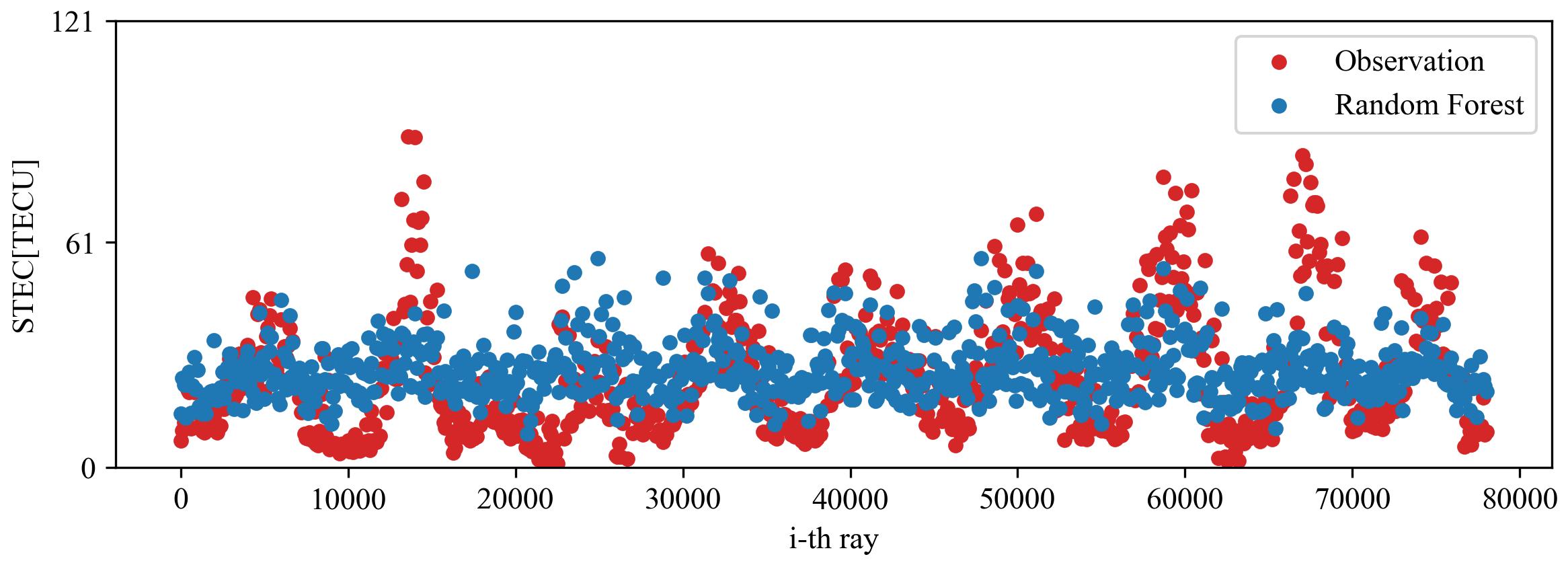}\\
			\vspace{0.02cm}
		\end{minipage}%
	\centering
	\vspace{-0.2cm}
    \caption{{Global storm observation data prediction result by three methods during the time span March 12 to 21, 2023 at the test TIXG station. Up: 4D DeepONet-STEC model. Middle: ANN model. Down: RF model.}}
    \label{fig:tixg3method}
\end{figure*}
\begin{figure*}[!htbp]
	\centering
			\includegraphics[width=7in]{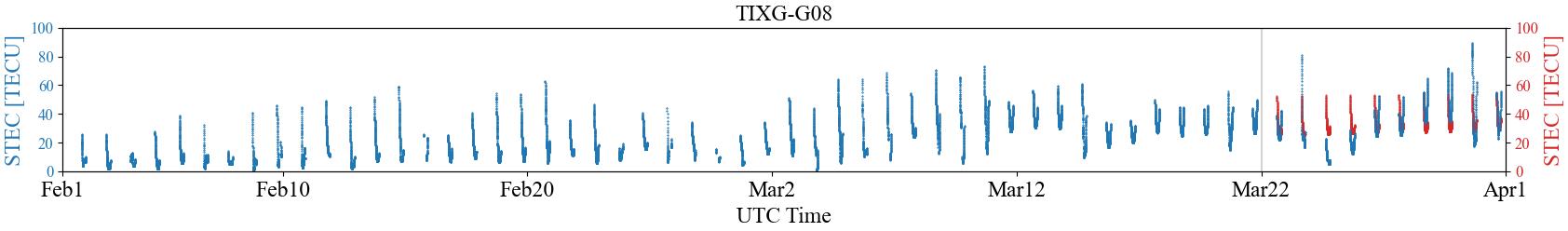}\\
			\vspace{0.02cm}
			\includegraphics[width=7in]{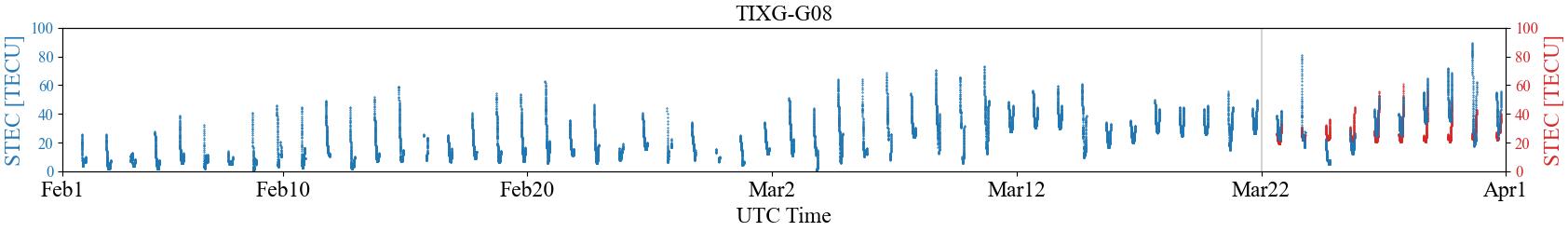}\\
			\vspace{0.02cm}
			\includegraphics[width=7in]{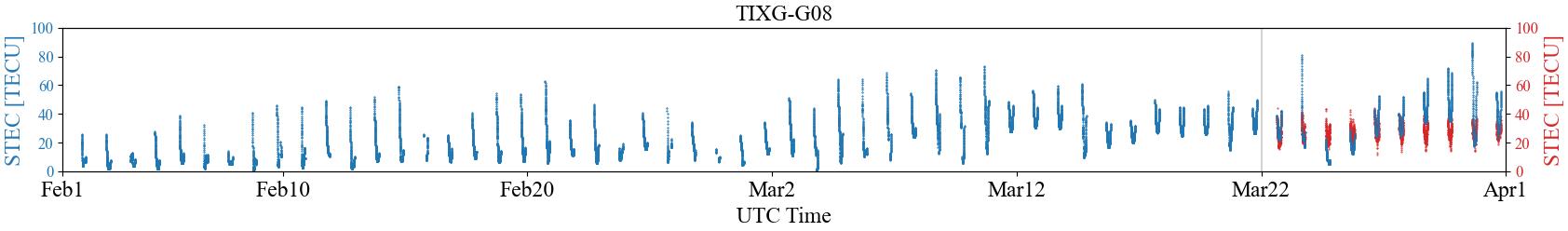}\\
			\vspace{0.02cm}
	\centering
	\vspace{-0.2cm}
 \caption{Predicted STEC for single satellite at the test TIXG station using the magnetic storm dataset. Up: 4D DeepONet-STEC model. Middle: ANN model. Down: RF model. The blue curve is the observation and the red curve is the predicted STEC between March 22 and March 31, 2023.}
    \label{fig:3method-tixg-rays}
    \vspace{-0.6cm}
\end{figure*}

%% MAYG 
\begin{figure*}[!htbp]
	\centering
		\begin{minipage}[c]{0.32\linewidth}
			\centering
			\includegraphics[width=2.1in]{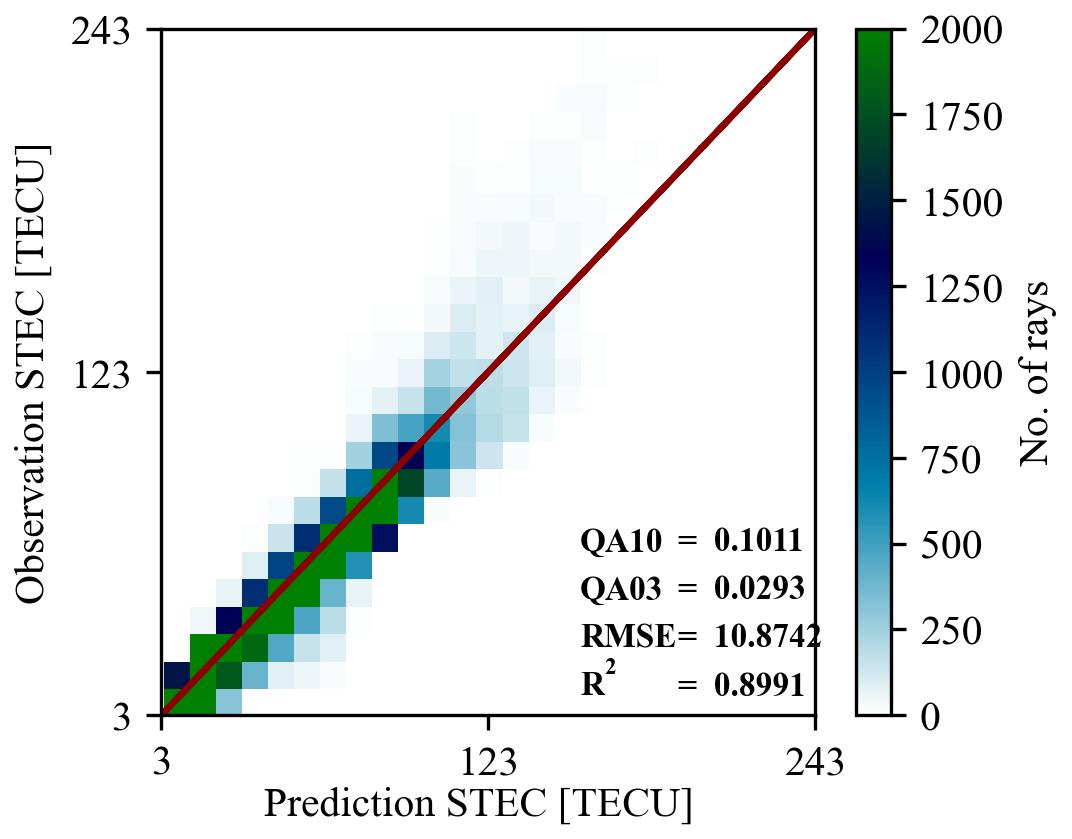}\\
			\vspace{0.02cm}
			\centering
			\includegraphics[width=2.1in]{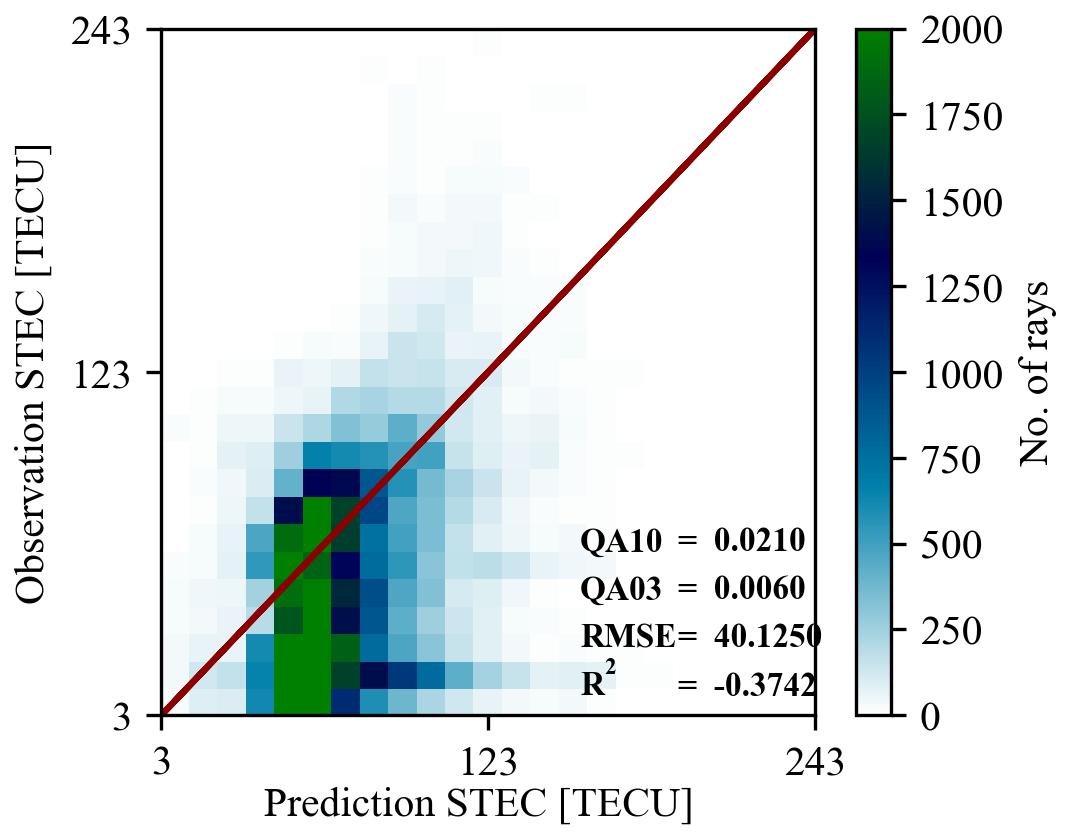}\\
			\vspace{0.02cm}
   		\centering
			\includegraphics[width=2.1in]{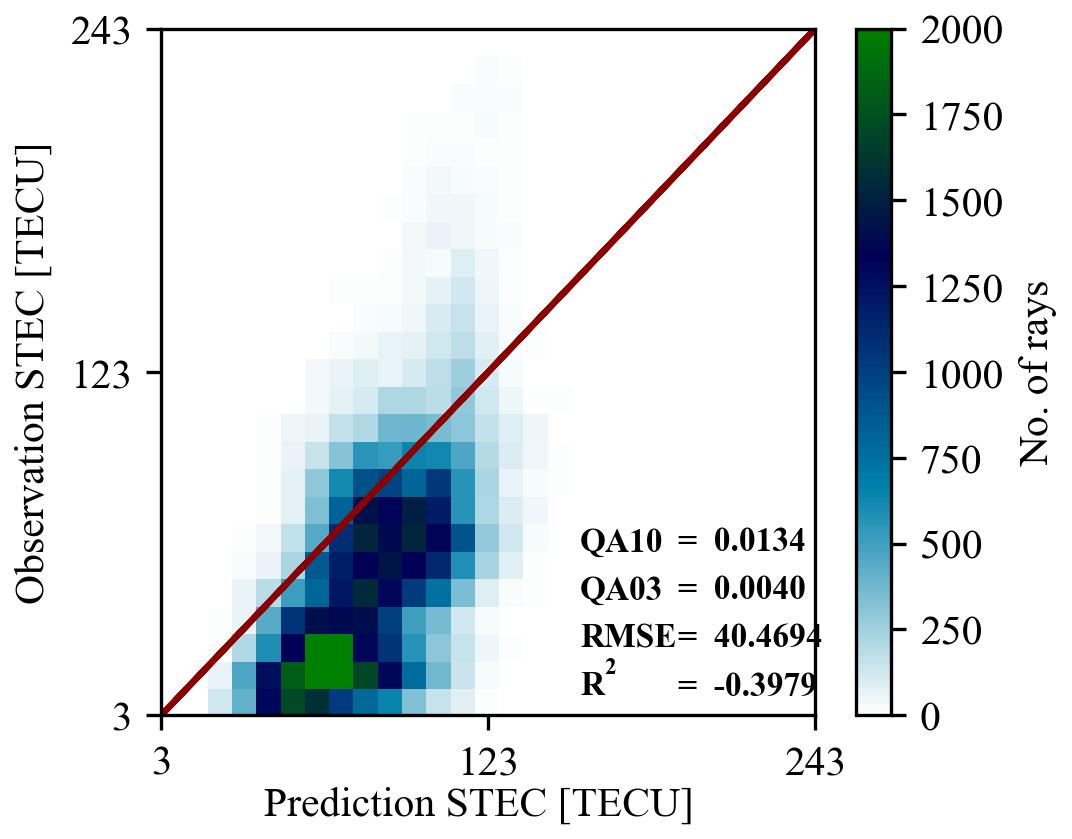}\\
			\vspace{0.02cm}
		\end{minipage}%
		\begin{minipage}[c]{0.68\linewidth}
			\centering
			\includegraphics[width=4.5in]{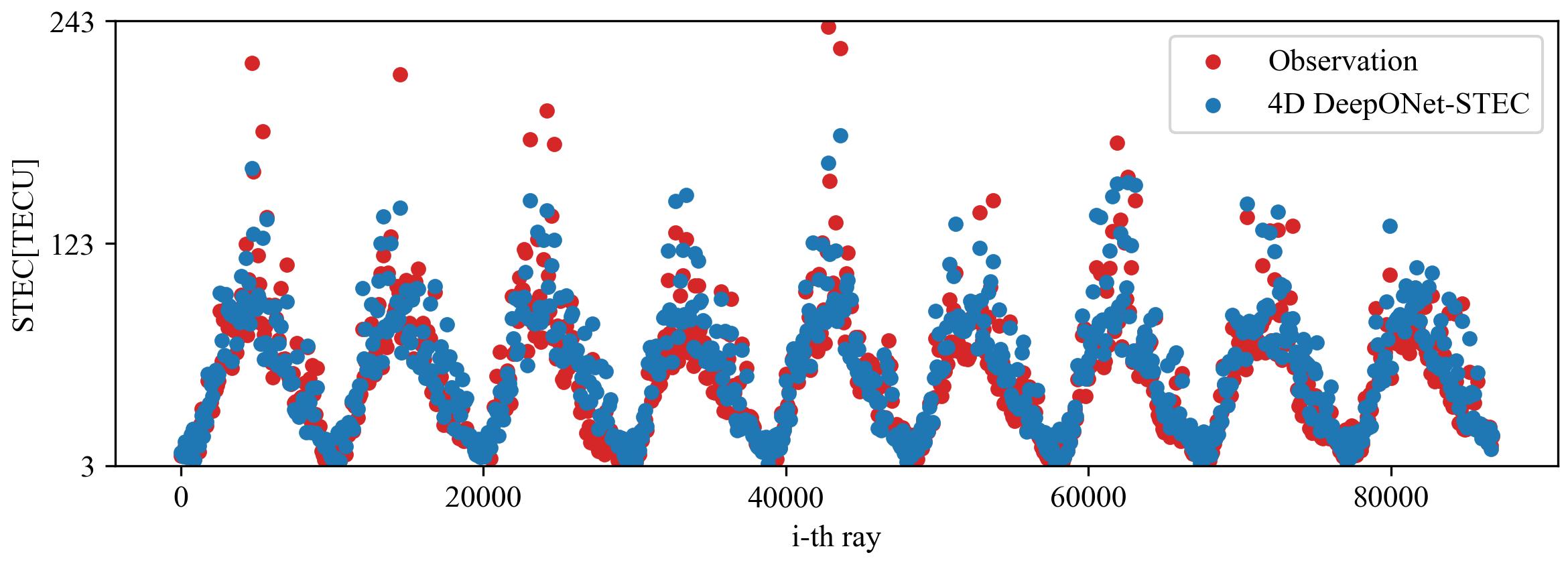}\\
			\vspace{0.02cm}
			\centering
			\includegraphics[width=4.5in]{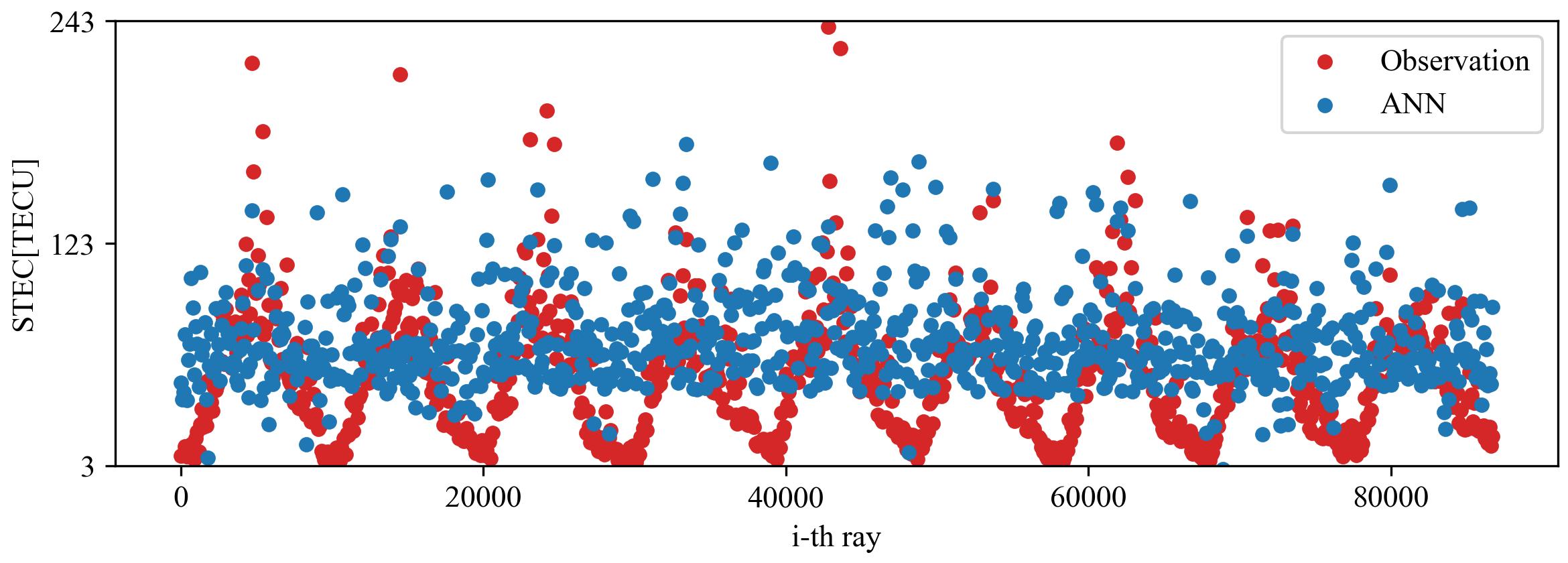}\\
			\vspace{0.02cm}
   		\centering
			\includegraphics[width=4.5in]{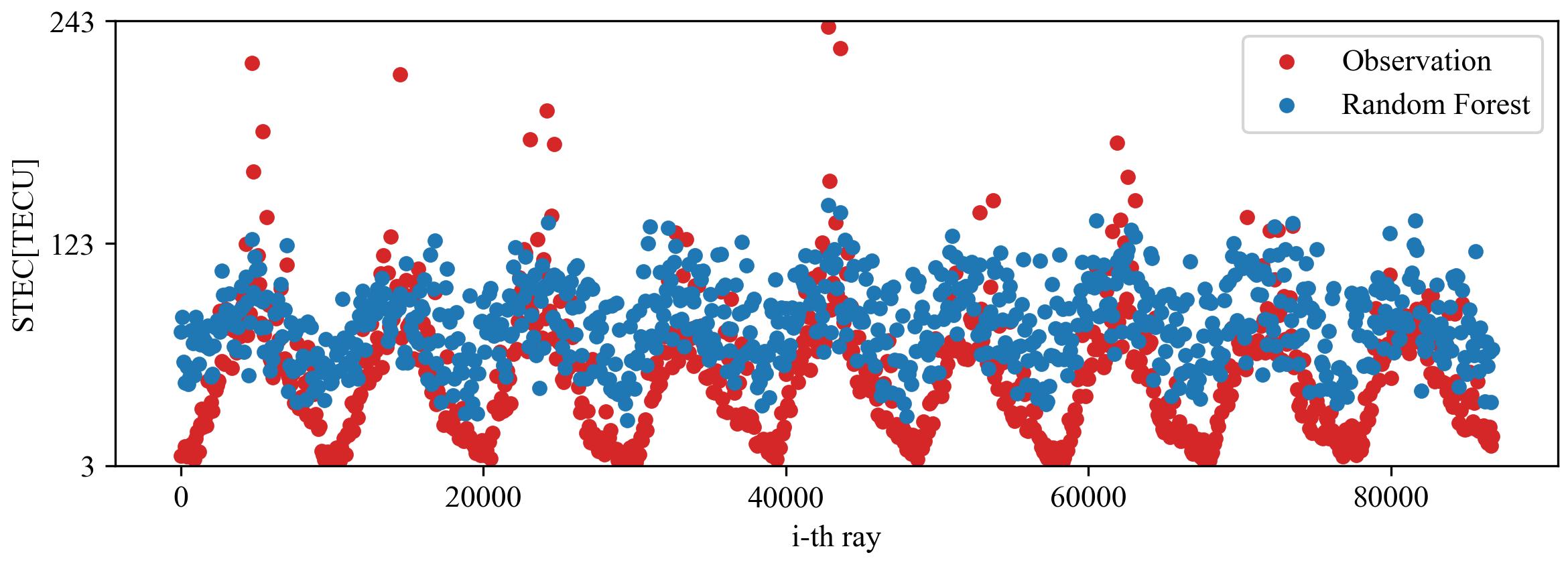}\\
			\vspace{0.02cm}
		\end{minipage}%
	\centering
	\vspace{-0.2cm}
    \caption{{Global storm observation data prediction result at the test MAYG station by three methods during the time span March 12 to 21, 2023 using the magnetic storm dataset . Up: 4D DeepONet-STEC model. Middle: ANN model. Down: RF model.}}
    \label{fig:mayg3method}
\end{figure*}

\begin{figure*}[!htbp]
	\centering
			\includegraphics[width=7in]{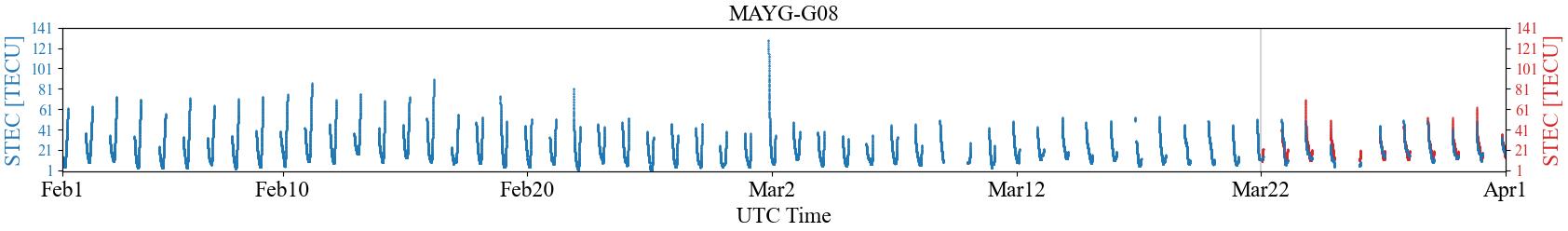}\\
			\vspace{0.02cm}
			\includegraphics[width=7in]{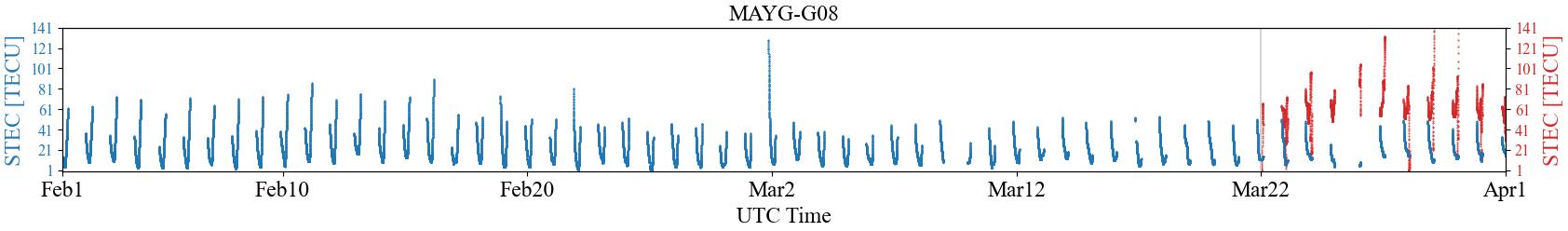}\\
			\vspace{0.02cm}
			\includegraphics[width=7in]{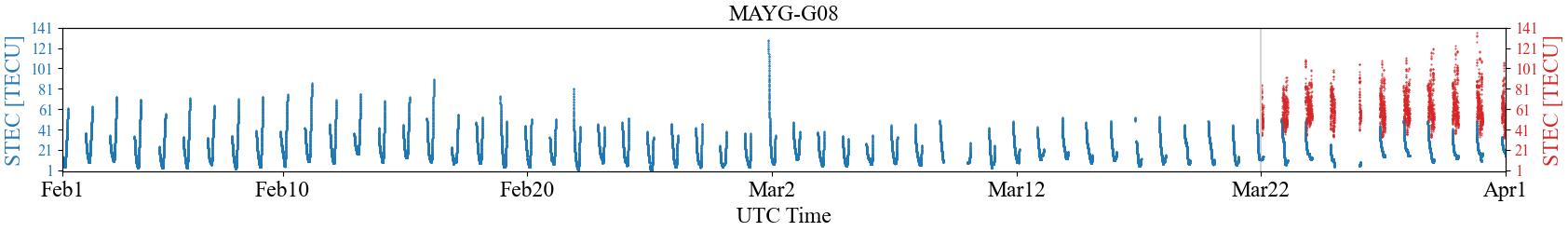}\\
			\vspace{0.02cm}
	\centering
	\vspace{-0.2cm}
    \caption{Predicted STEC for single satellite at the test MAYG station using the magnetic storm dataset. Up: 4D DeepONet-STEC model. Middle: ANN model. Down: RF model. The blue curve is the observation and the red curve is the predicted STEC between March 22 and March 31, 2023.}
    \label{fig:3method-mayg-rays}
    \vspace{-0.6cm}
\end{figure*}

\begin{figure*}[!htbp]
    \centering
    \subfigure[DeepONet-STEC result between March 22-31 in storm dataset]{
    \centering
    \begin{minipage}[b]{1\linewidth}
			\includegraphics[width=6.95in]{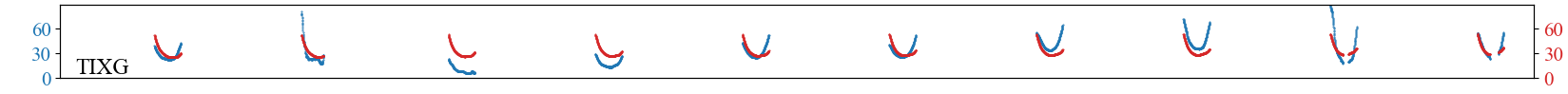}\\
			\includegraphics[width=7in]{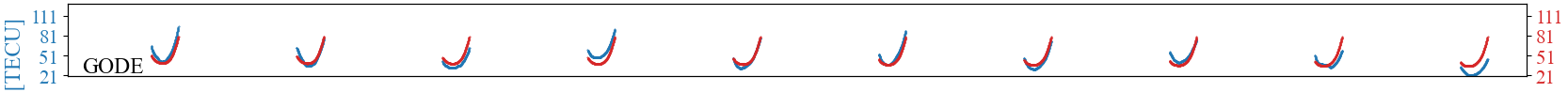}\\
			\includegraphics[width=7in]{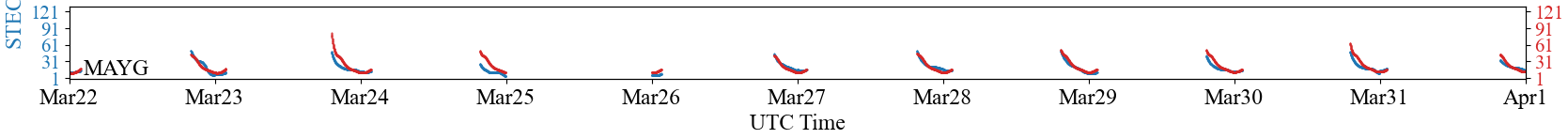}\\
			\vspace{-0.6cm}
    \end{minipage}
    }
    \subfigure[ANN result between March 22-31 in storm dataset]{
    \centering
    \begin{minipage}[b]{1\linewidth}
			\includegraphics[width=6.95in]{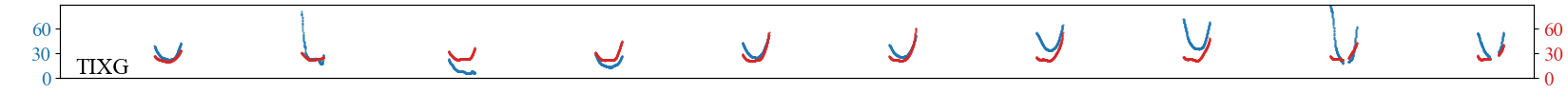}\\
			\includegraphics[width=7in]{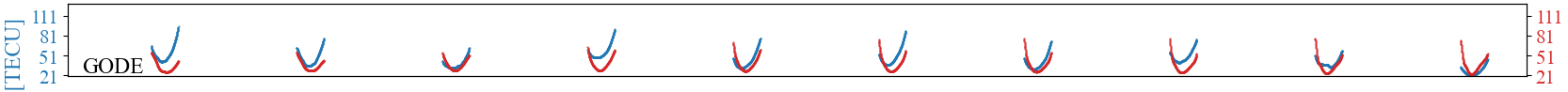}\\
			\includegraphics[width=7in]{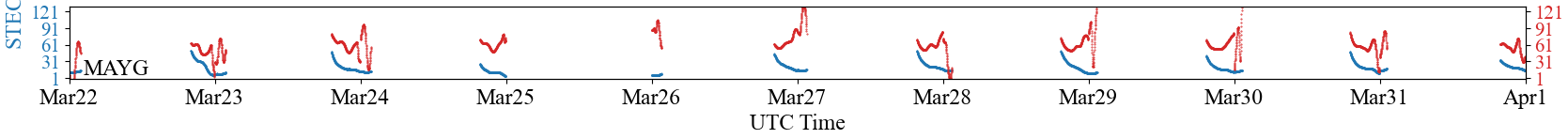}\\
			\vspace{-0.6cm}
    \end{minipage}
    }
    \subfigure[Random Forest result between March 22-31 in storm dataset]{
    \centering
    \begin{minipage}[b]{1\linewidth}
			\includegraphics[width=6.95in]{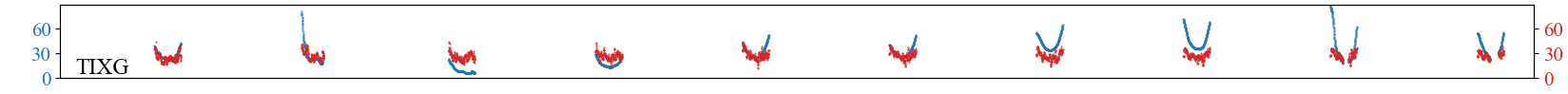}\\
			\includegraphics[width=7in]{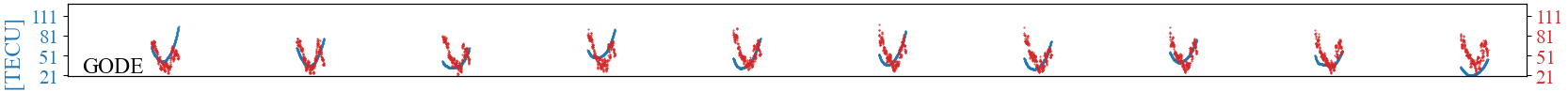}\\
			\includegraphics[width=7in]{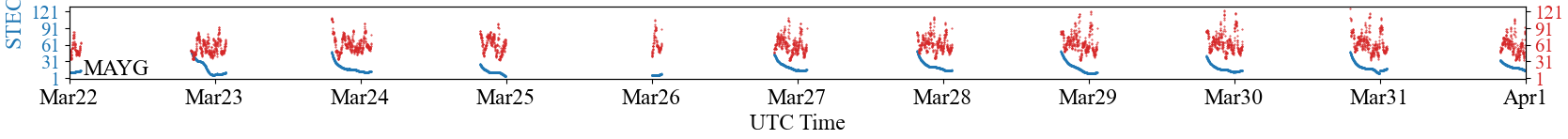}\\
			\vspace{-0.6cm}
    \end{minipage}
    }
	\vspace{-0.2cm}
    \caption{Global storm observation dataset predicted STEC of single-station, single-satellite (G08) for three test stations between March 22-31. The blue dots represents the observation, while the red dots represents the predicted STEC values by three models in storm periods.}
    \label{fig:storm-rays-magnify}
    \vspace{-0.4cm}
\end{figure*}

\subsection{Global Observation Data Results in Storm Periods}

In order to evaluate the performance of the DeepONet-STEC model, we compare the DeepONet-STEC with the other deep learning methods of random forest (RF) and ANN.  Both methods will directly learn the mapping function from the input value to the output value. Since the 4D DeepONet-STEC uses two networks with 16 and 30 layers respectively to learn, in order to compare with 4D DeepONet-STEC under relatively equal conditions, the network architecture of ANN is set to 46 layers, while the maximum depth of the tree of RF is limited to 46 layers. 

Here, we choose the two-month magnetic storm period data from February 1, 2023 to April 1, 2023 for validation as shown in Fig.~\ref{fig:kpdst30}(c)\&(d). Based on the the geomagnetic index Kp and Dst, the two storms occur from February 25 to March 14 with the minimum $\rm{Dst}=-138$, 2023 and March 22 to 31, 2023 with the minimum $\rm{Dst}=-184$. The training and validation data is from Feb 1 to March 21,2023 and the test data is from March 22 to April 1, 2023.

In Fig.~\ref{fig:valid} and Fig.~\ref{fig:valid_ray}, the prediction results at the validation stations are analyzed for the magnetic storm period data here. The training data does not contain validation stations, and the model prediction from validation stations from March 12 to March 21 will be recorded. The model results at two representative validation stations SCH2 in the northern hemisphere and FALK in the southern hemisphere, are presented in Fig.~\ref{fig:valid}. The results at two validation stations show that the $R^2$ coefficient is greater than 80\% , that captures the STEC variance well. For validation station SCH2, there are missing data for some dates because STEC data were not extracted for every station on every day, but still our model has good results. The results at the validation stations show that our model has exhibited good performance, therefore we further analyze the model performance at the test stations.

Figs. \ref{fig:gode3method}-\ref{fig:3method-mayg-rays} represent the two-month observation STEC value versus the model-predicted STEC value of the single-satellite (G08) at three test stations (TIXG, MAYG, GODE) based on the global observation storm data using DeepONet-STEC, ANN, and Random Forest. The DeepONet-STEC captures the trends much better than the RF and ANN. Table \ref{table:3method-results} compared the results of three method at TIXG, GODE and MAYG, respectively. The high latitude station (TIXG) in Fig.~\ref{fig:tixg3method} and Fig.~\ref{fig:3method-tixg-rays} performs the worst among the three sites. The $R^2$ coefficient at the GODE in Fig.~\ref{fig:gode3method} and Fig.~\ref{fig:3method-gode-rays} and MAYG station in Fig.~\ref{fig:mayg3method} and Fig.~\ref{fig:3method-mayg-rays} reaches to a good value in Table IV for DeepONet-STEC even during the storm period and degraded quickly at the high latitude TIXG site. The value of RMSE depends on the data retrieval accuracy for the PPP algorithm during the magnetic storm, which will be another subject to further investigate in the future.  

\begin{table*}[!htbp]
\caption{Three method prediction results comparison of the observation data based on the test stations in storm periods}
\label{table:3method-results}
\centering
\begin{tabular}{ccccccc}
\hline
Station               & Model             & RMSE[TECU]   & $R^2$   & MAPE[\%]  & QA03[\%]  & QA10[\%] \\ \hline
\multirow{3}{*}{TIXG} & DeepONet-STEC     & 11.0743      & 0.5674  & 46.72     & 3.28    & 11.45   \\
                      & ANN               & 16.0717      & 0.0889  & 98.36     & 0.97    & 3.30   \\
                      & RF                & 15.2634      & 0.1783  & 88.17     & 1.42    & 4.60   \\ \hline
\multirow{3}{*}{GODE} & DeepONet-STEC     & 9.4456       & 0.7771  & 29.93     & 2.01    & 6.83   \\
                      & ANN               & 18.9082      & 0.1069  & 88.76     & 0.98    & 3.29   \\
                      & RF                & 19.7382      & 0.0268  & 91.98     & 1.21    & 4.07   \\ \hline
\multirow{3}{*}{MAYG} & DeepONet-STEC     & 10.8742      & 0.8991  & 22.78     & 2.93    & 10.11   \\
                      & ANN               & 40.1250      & -0.3742 & 148.38    & 0.60    & 2.10   \\
                      & RF                & 40.4694      & -0.3979 & 152.76    & 0.40    & 1.34   \\ \hline

\end{tabular}
\end{table*}

It it noted that for active solar activities during magnetic storms, the STEC values are much larger and fluctuate more widely than that during quiet periods, reaching levels greater than 100 TECU. Because the magnetic storms in the predicted time period are stronger than that in the time period of the training set, there appear some extreme data  at individual stations such as MAYG. The prediction results by DeepONet for large value data shows relative smooth trend compared to the observation data.

The magnified results of the DeepONet-STEC during the prediction $10$ days period between March 22-31 are compared in Fig.~\ref{fig:storm-rays-magnify} for three test stations using the global observation storm dataset. The results show that due to the complexity of the training data during the magnetic storms, the prediction results of the Random Forest model fluctuate a lot, and the ANN model gives unstable prediction results. The DeepONet-STEC model maintains good stable predictions, which proves the robustness of our DeepONet-STEC model for global magnetic storm periods.

Overall, the 4D DeepONet-STEC model can relatively well learn the different STEC data characteristics during calm periods versus magnetic storms, whereas ANN and Random Forest methods cannot effectively capture the complex changing characteristics of STEC at different stations over long time periods.

%%%%%%%%%%%%%%%%%%Validation results is ingored here.
%\subsection{Validation Data Result}

%\begin{table}[!htbp]
%\caption{Statistical prediction results of the observation data based on the validation %stations in storm periods}
%\label{table:valid}
%\centering
%\begin{tabular}{lccccc}
%\hline
%Region & station     & RMSE [TECU]   & $R^2$ & QA03 [\%] & QA10 [\%] \\
%\hline
%\multirow{3}{*}{Global} & SCH2 & 7.1567 & 0.8368   & 3.34  & 11.86 \\
%                        & FALK & 9.5909 & 0.8216   & 3.57  & 11.53 \\
%\hline
%\end{tabular}
%\end{table}

% \begin{table}[!ht]
% \caption{
% Observation data
% }
% \label{table:obs-statistical-results}
% \centering
% \begin{tabular}{ccccc}
% \hline
% Region   & RMSE [TECU] & R2      & QA03 [\%] & QA10 [\%] \\ \hline
% USA      & 1.4617      & 0.8831  & 24.21     & 61.36  \\
% Global   & 1.4843      & 0.8605  & 17.91     & 54.00 \\ \hline
% \end{tabular}
% \end{table}

\section{Conclusion and Discussion}
In this paper, a deep learning-based framework, the 4D DeepONet-STEC model, has been proposed for predicting ionospheric STEC based on the deep operator network using global GNSS data. The 4D DeepONet-STEC  model first adopts the deep operator network techniques to achieve high 4D temporal-spatial resolution of the STEC values at any given satellite-station ray to enhance the accuracy and performance for providing accurate ionospheric corrections. The innovative Kernel method is adopted to construct the input function from the real observation data for the DeepONet-STEC model. 

The experimental results obtained from Nequick2 simulation data and observation data demonstrated the effectiveness and robustness of the 4D DeepONet-STEC model. The observation data has been extracted by the UCPPP with time resolution $30\,\rm{s}$. The model achieved a relatively low accuracy RMSE $\sim 1.5$ TECU and high $R^2 \sim 0.85 $ values for global observation data in quiet period data of one month, which indicates accurate prediction and good correlation between predicted and true values. The QA metrics, specifically the QA03 and QA10 metrics, revealed that the absolute errors decreases within acceptable percentages, which are specified to apply for providing reliable ionospheric corrections for positioning applications. In addition, the DeepONet-STEC prediction tested on data from two-month magnetic storm period data performs better for high solar activity scenarios in comparison with ANN and random forest model.

In conclusion, the 4D DeepONet-STEC model demonstrates strong potential for accurately predicting ionospheric TEC, enabling improved positioning accuracy in navigation applications.  The novelty of this work presents a framework for 4D estimation of ionospheric STEC based on the deep operator network. The model's performance evaluation, supported by comprehensive analysis and visualization, validates its effectiveness and reliability. Future work can focus on further refining and optimizing the 4D DeepONet-STEC model to enhance its accuracy and applicability across different geographical regions and varying ionospheric conditions. Additionally, incorporating real-time observation data into the model can improve its adaptability and responsiveness to dynamic ionospheric changes. This work based on machine learning pave the way for further advancements in ionospheric remote sensing and  space system applications by exploring advanced artificial intelligence.

\ifCLASSOPTIONcaptionsoff
  \newpage
\fi

\bibliographystyle{IEEEtran}
% \bibliography{IEEEabrv,../bib/paper}
% \bibliography{DeeponetV9Final}

\clearpage
\clearpage

\begin{IEEEbiography}[{\includegraphics[width=1in,height=1.25in,clip,keepaspectratio]{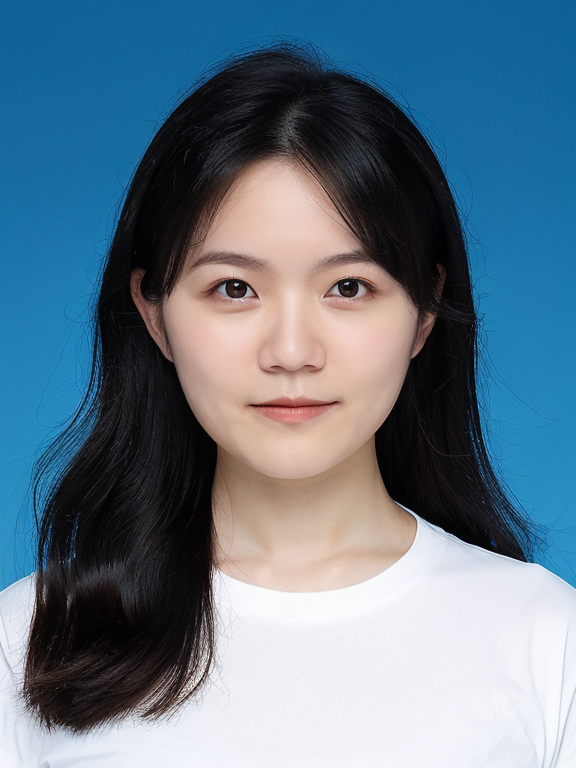}}]{Dijia Cai}
    received the Bachelor's degree in National University of Defense Technology in 2022. She is currently studying the Master's degree in electromagnetic field and microwave technology at Fudan University. Her research interests include ionospheric modeling and machine learning.

\end{IEEEbiography}
\begin{IEEEbiography}[{\includegraphics[width=1in,height=1.25in,clip,keepaspectratio]{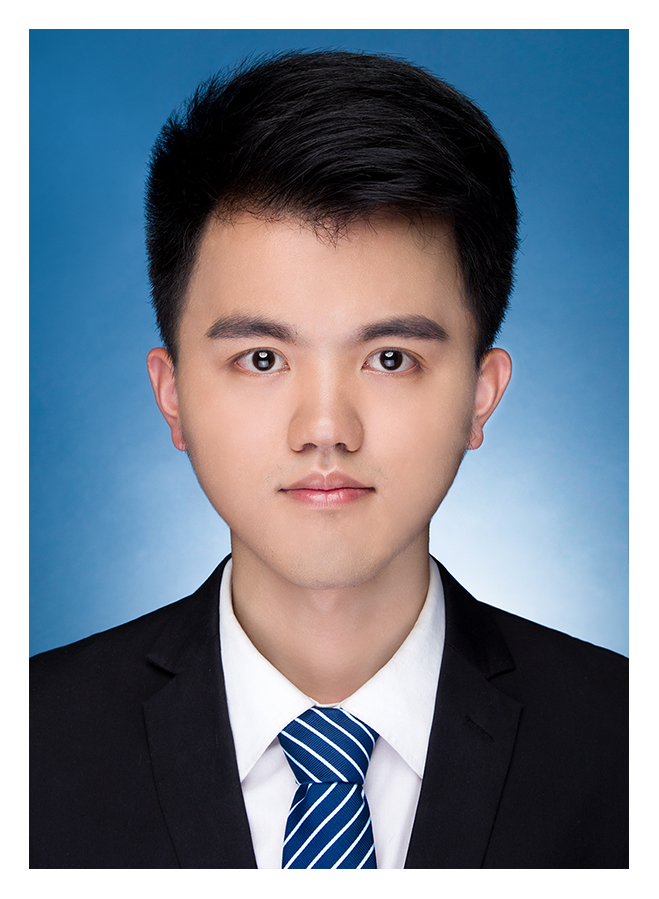}}]{Zenghui Shi}
    received his B.S. degree in Electronic Information Engineering from Xiamen University in 2020. He received his M.S. degree in electromagnetic field and microwave technology at Fudan University and Shanghai Innovation Center for BeiDou Intelligent Application in 2023. His research interests include ionospheric modeling and spatial and temporal prediction of ionospheric correction quantities.

\end{IEEEbiography}
\begin{IEEEbiography}[{\includegraphics[width=1in,height=1.25in,clip,keepaspectratio]{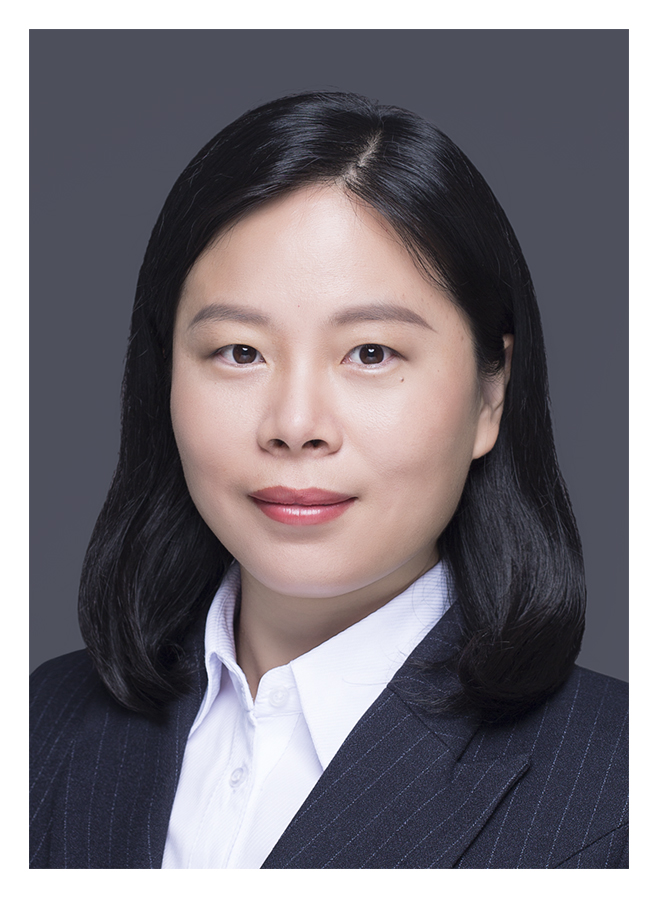}}]{Haiyang Fu}
	received the B.S. with Honor and M.S. degrees in power mechanics and engineering from Harbin Institute of Technology, Harbin, China, in 2006 and 2008, respectively and received the Ph.D. degree in electrical engineering from Virginia Polytechnic Institute and State University, Blacksburg, USA, in 2012. In 2013, she joined in school of information science and engineering at Fudan University as an assistant professor. Since 2023, she has been a professor in the Key Laboratory of Information Science of Electromagnetic Waves, Fudan University, Shanghai, China. She has been awarded as the URSI (International Union of Radio Science) Young Scientist Award in 2017 and currently served as the Chair of Woman in Radio Science Chapter of URSI-China (2024 $\sim$ ). She has served as the deputy session chair of Commission C4.1/D5.1 on Space Research COSPAR (2018-2022). Her current research focuses on ionospheric physics and remote sensing with artificial intelligence.
\end{IEEEbiography}

\begin{IEEEbiography}[{\includegraphics[width=1in,height=1.25in,clip,keepaspectratio]{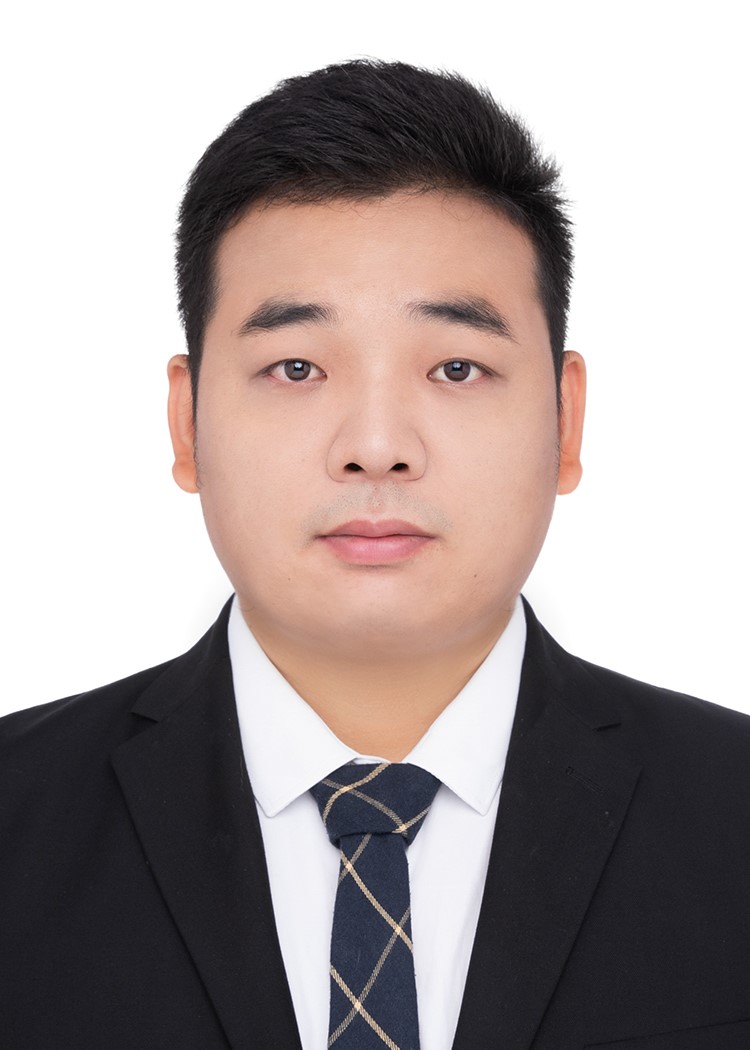}}]{Huan Liu}
is currently a research associate in the School of Mathematics, Shanghai University of Finance and Economics, Shanghai, China. He received the B.S. degree from the Central South University, Changsha, China in 2014, and Ph.D. degree in computational mathematics from Shandong University, Jinan, China in 2019. From 2020 to 2022, he was a postdoc research fellow in the School of Mathematical Science, Fudan University, Shanghai, China. His current research interests include computational methods for partial differential equations, scientific and engineering computing and inverse problems.

\end{IEEEbiography}

\begin{IEEEbiography}[{\includegraphics[width=1in,height=1.25in,clip,keepaspectratio]{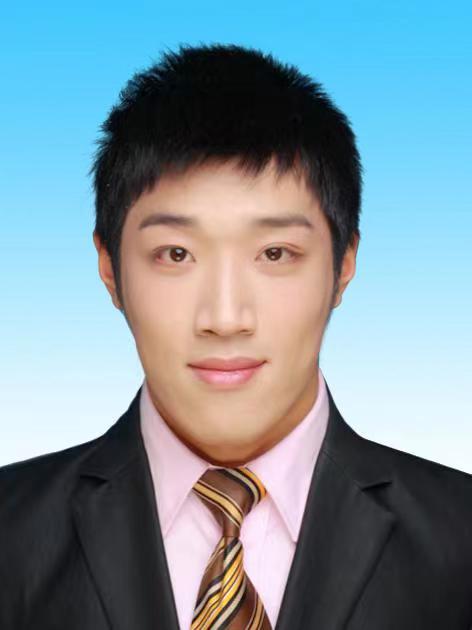}}]{Hongyi Qian}
received the B.Sc. from the School of Communication and Information Engineering, Shanghai University in 2014 and received the MS. from the School of information science and technology, Fudan University, Shanghai, in 2023. He is currently an Algorithm Engineer at China Mobile Research Institute, Shanghai, China, in the area of global navigation satellite system (GNSS) algorithms. Nowadays, he focuses on the ionospheric modelling in PPP-RTK algorithm.

\end{IEEEbiography}

\begin{IEEEbiography}[{\includegraphics[width=1in,height=1.25in,clip,keepaspectratio]{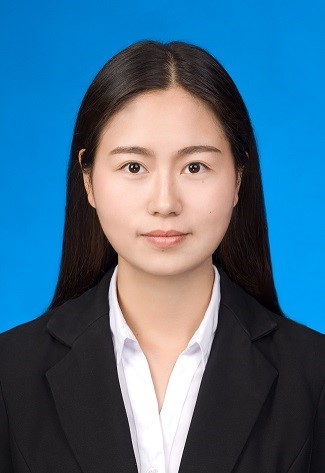}}]{Yun Sui}
received the Bachelor's degree in information Engineering from Nanjing University of Aeronautics and Astronautics, Jiangsu In 2013, and her Master's degree in Electromagnetic Field and Microwave Technology from Nanjing University of Aeronautics and Astronautics, Jiangsu in 2016. She is currently studying for her doctor's degree in electromagnetic field and microwave technology at Fudan University. From 2016 to 2018, she worked as an assistant engineer in Shanghai Key Laboratory of Electromagnetism. Her research interests include ionospheric modeling, GNSS positioning, and electromagnetic big data.

\end{IEEEbiography}

\begin{IEEEbiography}[{\includegraphics[width=1in,height=1.25in,clip,keepaspectratio]{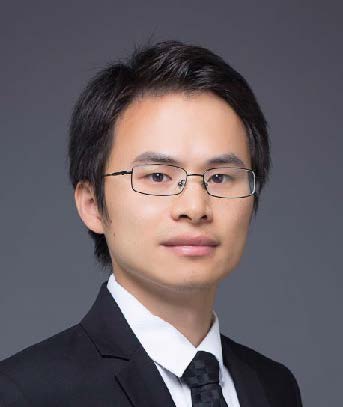}}]{Feng Xu}
(Senior Member, IEEE) received the B.E. degree (Hons.) in information engineering from Southeast University, Nanjing, China, in 2003, and the Ph.D. degree (Hons.) in electronic engineering from Fudan University, Shanghai, China, in 2008. He is currently a Professor and the Vice Dean of the School of Information Science and Technology and the Vice Director of the electromagnetic waves (MoE) Key Laboratory for Information Science of Electromagnetic Waves. His research interests include electromagnetic scattering modeling, SAR information retrieval, and radar system development.

Dr. Xu was a recipient of the second-class National Nature Science Award of China in 2011, the 2014 Early Career Award of the IEEE Geoscience and Remote Sensing Society, and the 2007 SUMMA Graduate Fellowship in the advanced electromagnetics area. He is a Topic Associate Editor of the IEEE TRANSACTIONS OF GEOSCIENCE AND REMOTE SENSING and was an Associate Editor of the IEEE GEOSCIENCE AND REMOTE SENSING LETTERS (2014–2021). He is the Founding Chair of the IEEE GRSS Shanghai Chapter and an IEEE GRSS AdCom Member.

\end{IEEEbiography}

\begin{IEEEbiography}[{\includegraphics[width=1in,height=1.25in,clip,keepaspectratio]{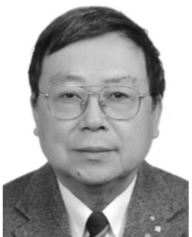}}]{Ya-Qiu Jin}
	received the bachelor’s degree in electrical engineering and computer science from Peking University, Beijing, China, in 1970, and the M.S., E.E., and Ph.D. degrees in electrical engineering and computer science from the Massachusetts Institute of Technology, Cambridge, MA, USA, in 1982, 1983, and 1985, respectively. In 1985, he joined Atmospheric Environmental Research, Inc., Cambridge, as a Research Scientist. From 1986 to 1987, he was a Research Associate Fellow at the City University of New York, New York, NY, USA. In 1993, he joined the University of York, York, U.K., as a Visiting Professor, sponsored by the U.K. Royal Society. He is currently the Te-Pin Professor and the Director of the Key Laboratory for Information Science of Electromagnetic Waves (MoE), Fudan University, Shanghai, China. He has authored more than 720 papers in refereed journals and conference proceedings and 14 books, including Polarimetric Scattering and SAR Information Retrieval (Wiley and IEEE, 2013), Theory and Approach of Information Retrievals From Electromagnetic Scattering and Remote Sensing (Springer, 2005), and Electromagnetic Scattering Modelling for Quantitative Remote Sensing (World Scientfic, 1994). His research interests include electromagnetic scattering and radiative transfer in complex natural media, microwave satellite-borne remote sensing, as well as theoretical modeling, information retrieval and applications in earth terrain and planetary surfaces, and computational electromagnetics. Dr. Jin was awarded by the Senior Research Associateship in NOAA/NESDIS by the USA National Research Council, in 1996. He was a recipient of the IEEE GRSS Distinguished Achievement Award (2015), the IEEE GRSS Education Award (2010), the China National Science Prize (1993 and 2011), the Shanghai Sci/Tech Gong-Cheng Award (2015), and the first-grade MoE Science Prizes (1992, 1996, and 2009) among many other prizes. He is the Academician of the Chinese Academy of Sciences, and a fellow of the The World Academy of Sciences, and the International Academy of Astronautics. He was a Co-Chair of TPC for IGARSS2011 in Vancouver, BC, Canada, and the Co-General Chair of IGARSS2016 in Beijing, China. He was the Associate Editor of the IEEE TRANSACTIONSON GEOSCIENCE AND REMOTE SENSING from 2005 to 2012, the member of the IEEE GRSS AdCom, and the Chair of the IEEE Fellow Evaluation of GRSS from 2009 to 2011. He is an IEEE GRSS Distinguished Speaker and an Associate Editor of the IEEE ACCESS.
\end{IEEEbiography}

\end{document}